\documentclass[a4paper,fleqn,usenatbib]{mnras}

\usepackage{mathptmx}
\usepackage{verbatim}

\usepackage[T1]{fontenc}
\usepackage{ae,aecompl,pdflscape}

\usepackage{graphicx}	
\usepackage{amsmath}	
\usepackage{amssymb}	
\usepackage{subfig}
\usepackage{float}
\usepackage{epstopdf}
\usepackage[colorinlistoftodos]{todonotes}  %

\title[The first 62 AGN in MaNGA - IV: Ionized gas]{The first 62 AGN observed with SDSS-IV MaNGA - IV: gas excitation and star-formation rate distributions}

\author[J. C. do Nascimento et al.]{Jana\'ina C. do Nascimento$^{1,3}$\thanks{E-mail:},
Thaisa Storchi-Bergmann$^{1,3}$,
N\'icolas D. Mallmann$^{1,3}$, 
\newauthor Rog\'erio Riffel$^{1,3}$,
Gabriele S. Ilha$^{2,3}$,
Rogemar A. Riffel$^{2,3}$,
Sandro B. Rembold$^{2,3}$,
\newauthor J\'aderson Schimoia$^{3,5}$,
Luiz Nicolaci da Costa$^{3,4}$,
Marcio A.G. Maia$^{3,4}$, and
\newauthor Alice D. Machado$^{2,3}$
\\
$^{1}$Departamento de Astronomia, IF, Universidade Federal do Rio Grande do Sul, CP 15051, 91501-970, Porto Alegre, RS, Brazil\\
$^{2}$Departamento de F\'isica, CCNE, Universidade Federal de Santa Maria, 97105-900, Santa Maria, RS, Brazil\\
$^{3}$Laborat\'orio Interinstitucional de e-Astronomia - LIneA, Rua General Jos\'e Cristino 77, Rio de Janeiro, RJ - 20921-400, Brazil\\
$^{4}$Observat\'orio Nacional, Rua General Jos\'e Cristino, 77, Rio de Janeiro, RJ, 20921-400, Brazil\\
$^{5}$Departamento de F\'isica, Universidade Federal de Santa Catarina, CP 88040-900, Florian\'opolis, SC, Brazil\\
}

\date{Accepted XXX. Received YYY; in original form ZZZ}

\pubyear{2018}
\hypersetup{draft}
\begin{document}
\label{firstpage}
\pagerange{\pageref{firstpage}--\pageref{lastpage}}
\maketitle

\begin{abstract}

We present maps of the ionized gas flux distributions, excitation,  star-formation rate SFR, surface mass density $\Sigma_{H+}$, and obtain total values of SFR and ionized gas masses {\it M} for 62 Active Galactic Nuclei (AGN) observed with SDSS-IV MaNGA and compare them with those of a control sample of 112 non-active galaxies. The most luminous AGN -- with $L(\rm{[OIII]}\lambda 5007) \ge 3.8\times 10^{40}\,\mbox{erg}\,\mbox{s}^{-1}$, 
and those hosted by earlier-type galaxies are dominated by Seyfert excitation within 0.2 effective radius $R_e$ from the nucleus, surrounded by LINER excitation or transition regions, while the less luminous and hosted by later-type galaxies show equally frequent LINER and Seyfert excitation within $0.2\,R_e$. The extent $R$ of the region ionized by the AGN follows the relation $R\propto\,L(\rm{[OIII]})^{0.5}$ -- as in the case of the Broad-Line Region. The SFR distribution over the region ionized by hot stars is similar for AGN and controls, while the integrated SFR -- in the range $10^{-3}-10$\,M$_\odot$\,yr$^{-1}$ is also similar for the late-type sub-sample, but higher in the AGN for 75\% of the early-type sub-sample. We thus conclude that there is no signature of AGN quenching star formation in the body of the galaxy in our sample. We also find that 66\% of the AGN have higher ionized gas masses $M$ than the controls -- in the range 10$^5-3\times10^7$\,M$_\odot$ -- while 75\% of the AGN have higher $\Sigma_{H+}$ within $0.2\,R_e$ than the control galaxies.

\end{abstract}




\section{Introduction}

Active Galactic Nuclei (AGN), that are triggered when supermassive black holes (SMBH) accrete gas at the center of galaxies \citep{SB19}, play a major role in their evolution. This occurs as a result of radiative and mechanical feedback from the AGN: the generation of large regions of hot, photoionized gas by the AGN, as well as associated outflows, may affect the evolution of the galaxy over time by quenching their star formation \citep{H14, Brusa+16, Carniani+16}. This quenching, as the galaxy evolves, ends up setting limits to the total mass of the host galaxy \citep{S+08}. 

bf The investigation of the effect of AGN on its host galaxy has been facilitated in recent years via the use of Integral Field Spectroscopy. One project that does just that, for $\sim $10\,000 nearby galaxies, is the MaNGA (Mapping Nearby Galaxies at the Apache Point Observatory) survey, described in \citet{bundy15}, which is part of the fourth-generation Sloan Digital Sky Survey (SDSS-IV) \citep{Law+16,Yan+16,Blanton+17}. The Integral Field Units design and performance are characterized in \citet{Drory+15}.
During the survey's operation time -- from 2014 to 2020 -- a total of 300 active galaxies -- the MaNGA AGN -- are expected to be observed.

Our group AGNIFS (AGN Integral Field Spectroscopy) has joined the SDSS-IV collaboration via the LIneA laboratoy (Laborat\'orio Interinstitucional de e-Astronomia) with the goal of investigating the relation between MaNGA AGN and their host galaxies comparing the properties of the host galaxies with those of a control sample of non-active galaxies. This is the fourth paper of a series in which we aim at comparing the resolved stellar and gas properties of the AGN observed with MaNGA with those of the control sample. In Paper I \citep{Rembold17} we have reported the selection of the first 62 AGN observed with MaNGA, released in its fifth Product Launch (MPL-5) of the MaNGA data reduction pipeline \citep{Law+16} as well as  that of a control sample of 112 non-active galaxies, and compared the nuclear stellar population of the two samples. The cubes have been processed using the version 2.0.1 of the MaNGA Data Reduction Pipeline \citep{Law+16}. In Paper II \citep{Mallmann18} we have studied the spatial distribution of the stellar population properties including the resolved star formation history, while in Paper III \citet{Ilha18} we have compared the gas and stellar kinematics.

In the present paper, our goal is to map the gas excitation, the extent of the region ionized by the AGN (the Narrow-Line Region - NLR), quantify and map the star-formation rate (hereafter SFR) as well as to obtain the total ionized gas masses of the AGN host galaxies. In order to verify which properties are related to the AGN, we have compared them with those of a sample of control galaxies, investigating also if and how these properties depend on the host galaxy type and AGN luminosity.

In evaluating the effect of the AGN on its host galaxy, it is important to distinguish regions that have been ionised by the central AGN from other ionization sources. We have used for this the BPT diagnostic diagrams \citep{bpt81,VB87}, in particular [OIII]$\lambda$5007/H$\beta$ vs. [NII]$\lambda6584$/H$\alpha$. \citet{kauf03} and \citet{kewley06} used this diagrams to empirically improve the definition of the boundary separating AGN from star-forming galaxies, introducing the so-called transition region, with objects of these regions also called composite AGN-Starburst, most frequently due to the inclusion in the same observation aperture of both an AGN and nearby star forming regions.

Besides AGN and young stars, hot evolved stars can ionize the local gas and produce a spectrum similar to that of a LINER, as has been shown by \citet{Cid+10} and references therein. In order to identify these regions, \citet{Cid+10} proposed the WHAN diagram, a relation between the equivalent width of H$\alpha$ W(H$\alpha$) versus the [NII]$\lambda6584$/H$\alpha$ line ratio. When W(H$\alpha$) is lower than 3\AA, the source of gas ionisation can be post-AGB stars, and not necessarily an AGN or a starburst. Galaxies with W(H$\alpha$)$\le$3\AA\ are thus considered ``retired", in the sense that the origin of the gas emission is due neither to nuclear nor to starburst activity. As the [NII]/H$\alpha$ ratio of these galaxies are similar to those of LINERs, these objects have been dubbed ``LIERs". We have thus used the WHAN diagram to locate these regions.

By being able to map the gas excitation over most of the galaxy, we have measured the extent $R$ of the region excited by the AGN -- the Narrow-Line Region (NLR). We investigate the relation between $R$ and L([OIII]) that has been addressed over the years by a number of authors \citep{Schmitt03a,bennert02,greene11,liu+13,hainline+13,sb18}, using small samples, sometimes combining in-homogeneous data and finding varying slopes for the relation. With our MaNGA sample we  contribute to this investigation in the low-luminosity end with an homogeneous and larger sample than in previous studies.

The triggering of the AGN depends on the availability of gas to feed the SMBH at the center. Some studies argue that AGN host galaxies have more gas in the inner few kpc than non-active galaxies, from which the nuclear activity is triggered, what seems indeed to be the case at least for early-type galaxies \citep{SL07,Martini03}, while \citet{Hicks13} found higher molecular gas surface mass densities within the inner $\approx50$ pc of nearby Seyfert galaxies than in a matched control sample. Other studies point out that the Star-Formation Rate (SFR) seems to be enhanced in the nuclear region ($\sim$ inner kpc) of AGN \citep{ds12,Esquej+14,mushotzky+14}, but these studies have been done in the infrared, using the PAH (Polycyclic Aromatic Hydrocarbon) feature at 11.25$\mu$m or continuum fluxes at 70$\mu$m and 160$\mu$m as indicators of star formation, presumably not affected by the presence of an AGN. In the optical spectra, there is no such feature, and we must first use a diagnostic  -- e.g. the BPT diagram -- to find the regions ionized by star formation where we can then calculate the SFR. We will thus not be able to calculate the SFR at the nucleus of the AGN  but we have calculated and mapped the SFR in the regions of the galaxies ionized by hot stars. In the AGN nuclear region we have calculated the mass of ionized gas instead. For completeness, we have also mapped the distribution of ionized gas over the body of all the galaxies.

The present paper is organized as follows. We present a brief description of the sample chosen for this work in Section 2. We then describe the measurements from the MaNGA data and present the corresponding maps in Section 3. These results are discussed in Section 4 and the conclusions presented in Section 5.

\section{Sample and data}

The sample studied in this paper comprises the first 62 AGN observed with MaNGA and 112 control galaxies listed in Table\,\ref{table1}, and has been described in Paper I \citep{Rembold17}. In summary, in order to investigate the relation between the AGN and their host galaxy properties, we have identified in the fourteenth data release \citep[DR14; ][]{dr14}  all galaxies observed by MaNGA whose emission-line ratios in the SDSS-III \citep{Gunn+06} single nuclear spectrum were dominated by ionization by an active nucleus, as indicated in the BPT \citep{bpt81} diagnostic diagram. We have also used the WHAN \citep{Cid+10} diagnostic diagram in the selection of the AGN, in order to avoid ``LIERs", as discussed in the Introduction. For each AGN, \citet{Rembold17} have chosen two control non-active galaxies matching the AGN host stellar mass, morphology, distance and inclination. 

The data thus comprise the MaNGA datacubes of the 62 AGN and 112 control sample galaxies of Paper I and listed in Table\,\ref{table1}, resampled to square pixels of $0\farcs5\times0\farcs5$.

In our sample of 62 AGNs, there are 20 early-type hosts, 38 late-type hosts, with 2 in interaction, according to the classification in the Galaxy Zoo GZ1 ~\citep{L11}. There are 2 galaxies that are undefined, corresponding to galaxies whose probabilities of being elliptical and spiral in GZ1 are rigorously the same. These galaxies were excluded from the part of the analysis in which we separate early and late-type galaxies.

 Regarding the AGN luminosity, our sample comprises  17 luminous AGN -- as we have called in Paper I -- and  45 low-luminosity AGN, with the division between the two set at L(\rm{[OIII]})$\lambda$ 5007 = 3.8$\times 10^{40}$\,\mbox{erg}\ \mbox {s}$^{-1}$.

\section{Measurements and Maps}
\label{measurements}
In this section, we describe the measurements we have obtained from the data and present their corresponding maps.

The emission-line fluxes were obtained by fitting the line profiles with Gaussian curves using the Gas AND Absorption Line Fitting \citep[GANDALF; ][]{Marc06} routine, written in IDL (Interactive Data Language). This routine was adapted for the analysis of the MaNGA data cubes and details about the measurements can be found in \citet{Ilha18}.
The choice of this routine was due to the fact that it fits both the stellar population spectra and the profiles of the emission lines, after the subtraction of the stellar population contribution. 

Results are presented in Figs.~\ref{fig:elliptical} to ~\ref{fig:l_luminosity} for four representative trios of AGN hosts and their respective control galaxies. These trios were selected to sample, respectively, an early-type AGN, a late-type AGN, a luminous AGN and a low-luminosity AGN. The maps for the remaining objects are shown in Figs.~\ref{fig:first_ap}~\ref{fig:last_ap} of the Appendix.

\subsection{Line fluxes}

Line fluxes were obtained from the integrated line-profiles, from which the luminosities were calculated. Uncertainties in the fluxes range from less then 1\% to a few percent in the central region up to $\approx$10\% at the borders of the FoV.
Resulting luminosity maps are shown for the [OIII]$\lambda$5007\AA\ emission line in units of 10$^{38}$\,ergs\,s$^{-1}$ per spaxel for each active galaxy and its two controls in the fourth row of the left set of panels of Figs.~\ref{fig:elliptical} to~\ref{fig:l_luminosity}, for the four representative trios and in the Appendix for the remaining galaxies. 

An average one-dimensional profile was also built and is shown in the corresponding right panel of the figures above. These profiles were obtained following the method described in \citet{Mallmann18}:  averages of 30 radial profiles equally spaced in the azimuthal direction of the galaxy, limited to angular distances from the major axis of $\theta_{max} = \tan^{-1} (b/a)$ degrees, where $a$ is the galaxy semi-major axis and $b$ is the semi-minor axis,
extracted from the MaNGA's {\it drpall} table. This choice was made because profiles closest to the minor axis, when corrected for projection, introduced too much noise in the average profiles, probably as a result of obscuration effects when the galaxies have high inclinations.

\subsection{Gas Excitation}

In order to map the excitation of the gas, emission-line ratio maps were obtained and used to build the diagnostic diagram \citep{bpt81} [OIII]$\lambda$5007/H$\beta$ vs. [NII]$\lambda$6583/H$\alpha$ (hereafter called BPT diagram) as well as the WHAN diagram \citep{Cid+10} for each galaxy. 

From the above diagrams, excitation maps were obtained, according to the following procedure: we first built the BPT diagram and corresponding excitation map, ``painting" the regions of the galaxy with different colors corresponding to the different excitation regions: green for Seyfert excitation, magenta for LINER excitation, blue for starburst excitation and grey for composite (or transition region). 

After building the BPT excitation map, we built the WHAN diagram in order to check if the regions with AGN excitation (Seyfert and LINER) were ``LIERs'', in which case we painted the corresponding regions orange. 

\begin{figure*}
    \includegraphics[width=1.\columnwidth]{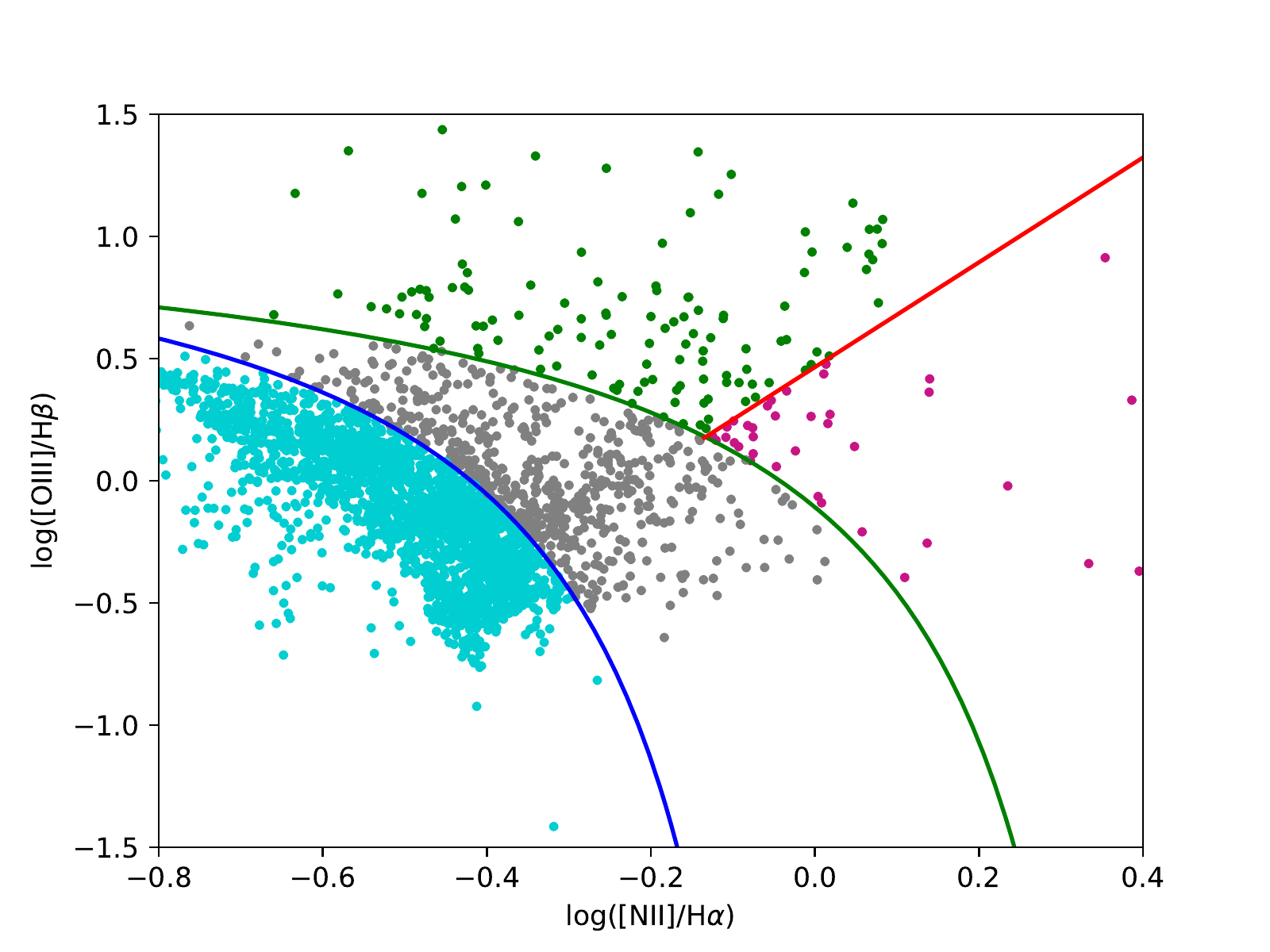}
    \includegraphics[width=0.7\columnwidth]{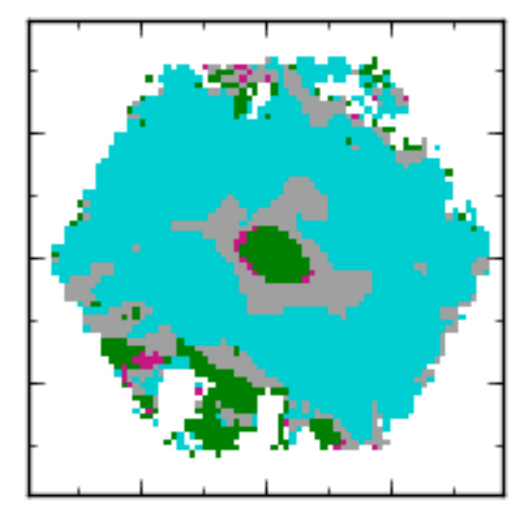}
    \caption{BPT diagram (left) for the control galaxy plateifu 8613-12702 where each point represent each spaxel, different colors identifying each type of excitation and the corresponding excitation maps to the right, where the tick markers are separated by 5$^{\prime\prime}$. The blue lines separates the starburst excitation and composite (or transition region), the green line separates the composite and AGN excitation, while the red line separates Seyfert(top) and LINER(bottom) excitation. Typical uncertainties in logarithmic in the range 0.02--0.1.} 
    \label{fig:bpt}
\end{figure*}

Figure~\ref{fig:bpt} shows an example of BPT diagram in which each point is a spaxel of the datacube of the control galaxy ``plateifu 8613-12702" in the left panel and the corresponding excitation map is shown in the right panel. The red line -- marking the separation between Seyfert and LINER excitation and the blue line -- marking the separation between the starburst and transition region, were obtained from \citep{kauf03}. The green line -- marking the separation between the transition region and AGN  was obtained from \citep{Kewley01}. 

\begin{figure*}
    \includegraphics[width=1.\columnwidth]{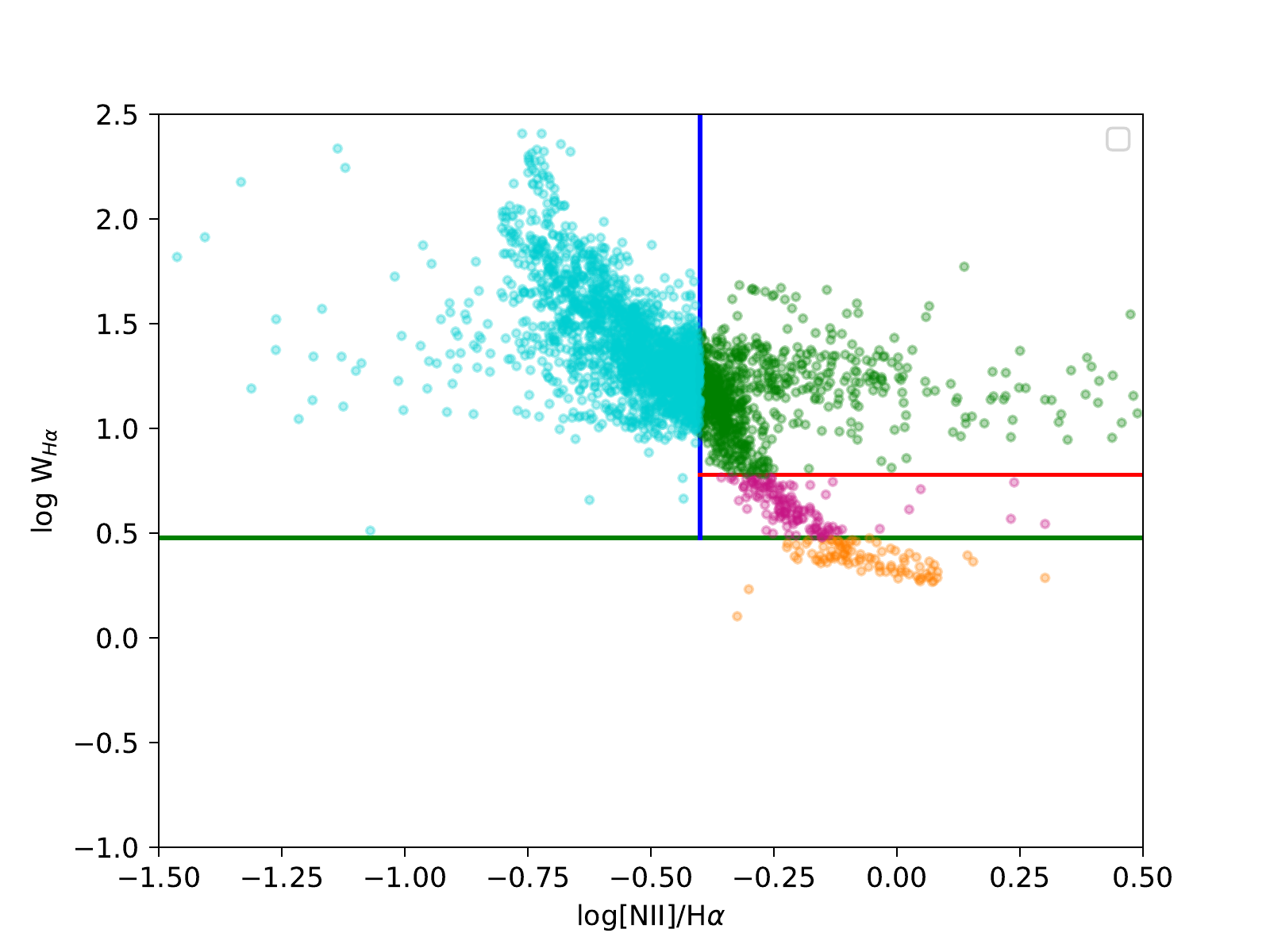}
    \includegraphics[width=0.7\columnwidth]{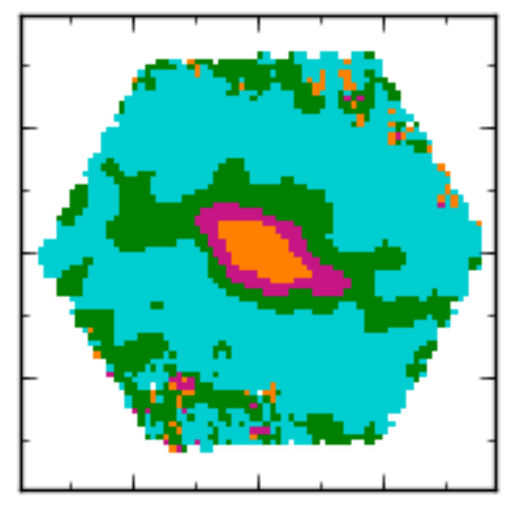}
    \caption{WHAN diagram (left) and excitation map (right) of the control galaxy identified as plateifu 8613-12702. In the excitation map the tick marks are separated by 5$^{\prime\prime}$. Blue lines separate staburst excitation to the left, while the red line separates Seyfert (top) and LINER (bottom) excitation. The green line shows the limit below which the excitation corresponds to a LIER. The dashed vertical line shows an alternative (proposed by us) division between starburst and AGN excitation that gives an excitation classification in better agreement with that of the BPT diagram. Typical uncertainties in logarithm are in the range 0.02--0.1.}
    \label{fig:whan}
\end{figure*}

Figure~\ref{fig:whan} shows the WHAN diagram \citep{Cid+10} on the left panel, and the corresponding excitation map on the right panel. The blue line separates starburst and AGNs and the red line separates Seyferts and LINERs as in the BPT diagram, but the green line here separates the LINERs (above the line) from the LIERs (below the line).

Uncertainties in the line ratios range from a few percent in the central regions, up to 10-30\% at the borders of the FOV.

We note that the excitation maps obtained using the BPT diagram shows some differences when compared to the one obtained using the WHAN diagram: many regions classified as composite and even starburst in the BPT diagrams appear as Seyfert or LINER in the WHAN diagram. The reason for this can be understood through a comparison between Figs.\ref{fig:bpt} and \ref{fig:whan}: while the division line between starburst and AGN in the WHAN diagram  corresponds to log([NII]/H$\alpha)=-0.4$, in the BPT diagram this division is curved and has the transition region between the starburst and AGN, while this transition region is not defined in the WHAN diagram. In the BPT diagram, the log([NII]/H$\alpha)$ ratios can reach a value of 0.1 for LINERs and up to -0.1 for Seyferts, higher than the fixed value of -0.4 of the WHAN diagram. 

Considering the above, we have adopted the excitation classification of the BPT diagram and only changed it to LIER in the cases where the excitation in the BPT diagram is Seyfert or LINER but is LIER in the WHAN diagram, keeping the transition region and starburst classifications of the BPT diagram.

We note that in order to improve the consistency of the two excitation maps for our sample, the division line in the WHAN diagram should correspond to somewhat a larger value than proposed in \citet{Cid+10} of  log([NII]/H$\alpha)\approx-0.25$. We have drawn a vertical blue dashed line in Fig.~\ref{fig:bpt} at this value in the WHAN diagram in Fig.~\ref{fig:bpt}.

We have used the procedure above to map the different type of excitation over the body of the galaxy and build the excitation maps shown in the second line of Figs.~\ref{fig:elliptical} to \ref{fig:l_luminosity}
and Figs.~\ref{fig:first_ap} -- \ref{fig:last_ap} of the Appendix.

\subsection{Star Formation Rate}

The excitation maps of both the AGN and controls show many locations in which the excitation is typical of starbursts or HII regions, in which the line emission is due to ionization of the gas by young stars, and the star formation rate (SFR) can be obtained from the H$\alpha$ luminosity L(H$\alpha$). We decided to include in the calculation of the SFR also the transition regions, which, although have some contribution from AGN excitation, do include a large contribution of a starburst component. As there may be some additional ``hiden" star-formation contribution in the region dominated by AGN excitation, the assumption that all H$\alpha$ emission is due to starbursts in the transition region may compensate for this ``hidden"  star formation contribution.

We have used the following expression to obtain the SFR, from \citet{Kennicutt}: 

\begin{equation}
\mathrm{SFR}=7.9\times10^{-42} \times \mathrm{L(H{\alpha})}
\end{equation}

In order to calculate the H$\alpha$ luminosity, we have corrected the H$\alpha$ flux, F(\mbox{H}$\alpha$), for reddening, using the extinction law of \citet {cardelli98}:

\begin{equation}
\mathrm{ L(H\alpha)}= \mathrm{4\pi d^{2}F(\mbox{H}\alpha)10^{C(\mbox{H}\alpha)}}
\end{equation}
where $d$ is the galaxy distance obtained from the redshift values listed in Paper I, F(H$\alpha$) is the H$\alpha$ flux and $C(\mbox{H}\alpha)$ is the interstellar extinction coefficient at the H$\alpha$ wavelength calculated as described in the next subsection. 

We have obtained both the integrated SFR, shown in the last column of Table\,\ref{table1} as well as the SFR surface density $\Sigma_{SFR}$ dividing the SFR at each spaxel by the corresponding area (0\farcs5$\times$0\farcs5) in kpc$^2$, obtaining its value in units of $\mathrm{M_{\odot}\,yr^{-1}\, kpc^{-2}}$.

The $\Sigma_{SFR}$ maps for the AGN and their control galaxies are shown in the third row of Figs.~\ref{fig:elliptical} to~\ref{fig:l_luminosity}
and Figs.~\ref{fig:first_ap} -- \ref{fig:last_ap} of the Appendix.

\subsection{Extinction}
The gas extinction was calculated from the observed H$\alpha$/H$\beta$ line ratio adopting case B recombination \citep{of06} and the \citet{cardelli98} reddening law:

\begin{equation}
\mathrm{A_{V}}=\mathrm{7.23\times\log \left[\frac{F(\mbox{H}\alpha)}{F(\mbox{H}\beta)}\times \frac{1}{2.87}\right]}  
\end{equation}

\noindent where $F(\mbox{H}\alpha)$ and $F(\mbox{H}\beta)$ are the observed fluxes.

The extinction maps for the AGN and their control galaxies are shown in the fifth row of the panels of Figs.~\ref{fig:elliptical} to~\ref{fig:l_luminosity}
and Figs.~\ref{fig:first_ap} -- \ref{fig:last_ap} of the Appendix.

\subsection{Ionized gas masses}
As we cannot calculate the SFR at the regions that are not ionized by hot stars, we have calculated the mass of ionized gas, that can be obtained at all locations with H$\alpha$ or H$\beta$ emission using \citep{of06}:

\begin{equation}
\mathrm{M} = Vfn_{e} m_{p}
\label{eq:mass}
 \end{equation}

\noindent where $V$ is the volume of the emitting region, $f$ is the filling factor, $n_{e}$ is the electron density and $m_{p}$ the proton mass.

Based on \citet{of06}, assuming case B recombination, we obtain:

\begin{equation}
\mathrm{V}f=\mathrm{8.1\times 10^{59} \frac {L_{41}(\mbox{H}_\beta)}{n_{3}^{2}}~~{\rm  cm^{-3} }} 
\label{eq:volum}
\end{equation}

\noindent where $L_{41}(\mbox{H}\beta)$ is the $\mbox{H}\beta$ luminosity in units of $10^{41}$ erg s$^{-1}$ and $n_{3}$ is the electron density in units of $10^{3}$ cm$^{-3}$. 


The emitting gas mass can be calculated by combining equations~(\ref{eq:mass}) and (\ref{eq:volum})\citep{p97} 
where we replaced L(H$\beta$) for L(H$\alpha$)/2.87:

\begin{equation}
\mathrm{M \approx 2.4\times 10^{5}\frac{L_{41}(\mbox{H}\alpha)}{n_{3}}}~~{\rm M_{\odot}} 
\end{equation}

In order to obtain the gas density $n$, we have used the ratio between the doublet lines [SII]$\lambda$6717 and [SII]$\lambda$6731\AA, as the corresponding two levels have different collision forces and radiative transition probabilities. The flux observed in each component of the doublet depends on the relative population at each energy level, which is sensitive to the electron density \citep{of06}. The calculation was done using the task {\it temden} of the package {\it stsdas.nebular} of IRAF and the adopted gas temperature was  10,000\,K. In many locations the line ratio was above the limit [SII]$\lambda$6717/ [SII]$\lambda$6731\AA=1.45, which corresponds to densities of $\approx$\,100\,cm$^{3}$ or lower. In these cases, we have adopted the value of 100\,cm$^{-3}$. Due to low signal-to-noise ratio, some measurements of the [SII] lines were not possible, and, in order to be able to calculate the gas masses at these locations, we decided to adopt also a density value of $100$ cm$^{-3}$. This is justified by the fact that these regions were surrounded by line ratios indicating this density or lower. 

Considering that the mass of the gas is inversely proportional to $n$, the resulting ionized gas masses are actually lower limits. In order to evaluate the effect of the density on the estimated gas masses, we have repeated the calculations for gas densities of 10\,cm$^{-3}$ and 1\,cm$^{-3}$. The result is an increase in the gas masses in the range 10--40\%, depending on the galaxy. This small increase is due to the fact that most of the gas emission comes from the regions with the highest line fluxes, for which the density value could be obtained. 
 
As for the $\Sigma_{SFR}$, we have also obtained the ionised gas surface mass density $\Sigma_{H^+}$ by dividing the mass at each pixel by its area. The corresponding maps are shown in the bottom row of  Figs.~\ref{fig:elliptical} to~\ref{fig:l_luminosity} and Figs.~\ref{fig:first_ap} -~\ref{fig:last_ap} of the Appendix.

\begin{figure*}

    \includegraphics[width=2.1\columnwidth]{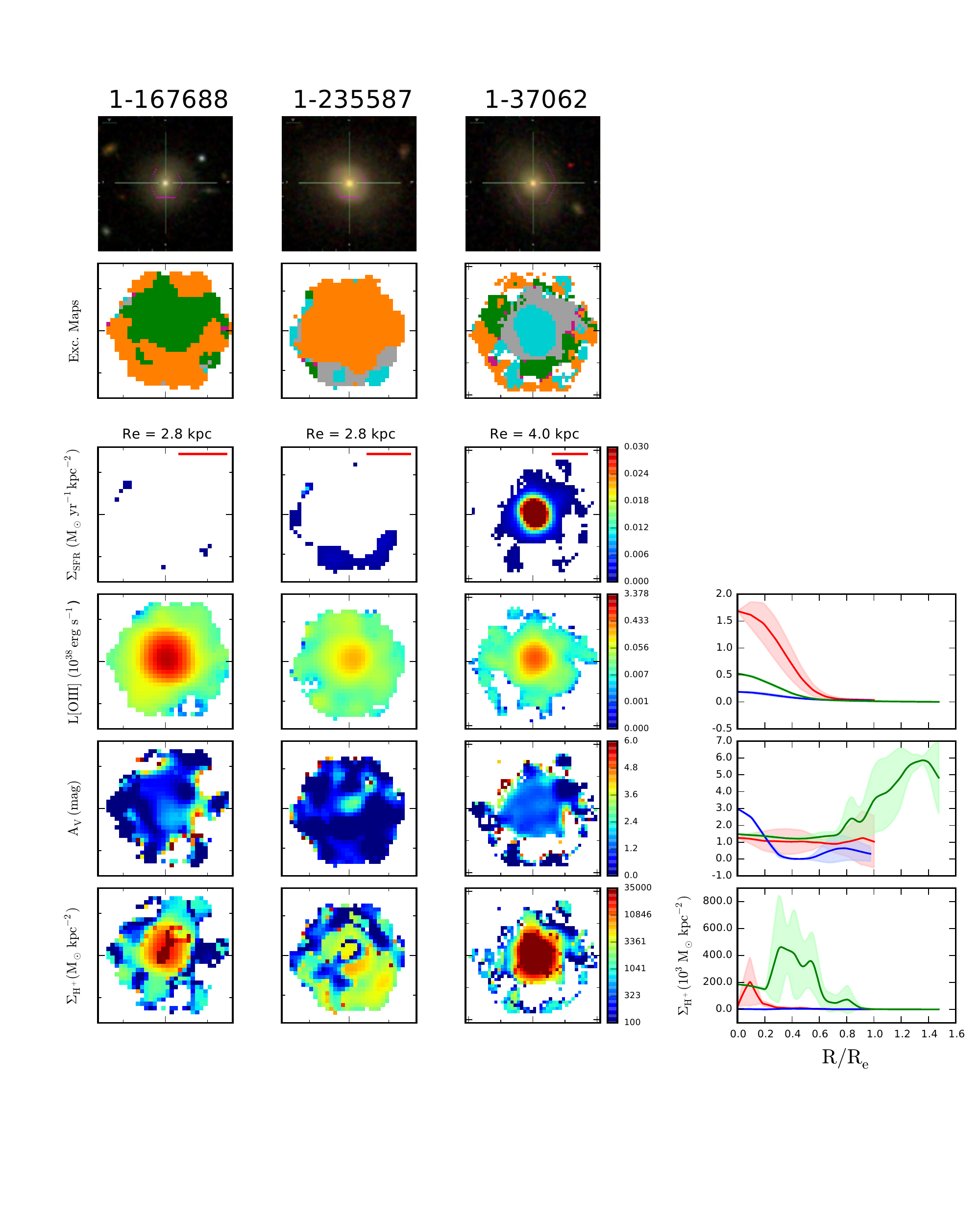}
    \vspace*{-30mm}
    \caption{Surface distribution of the properties measured for a typical early-type AGN and its two controls, showing the AGN on the left and the two control galaxies at the center and right. Top row: SDSS-IV images with the MaNGA footprint over-plotted in pink; second row: excitation maps where green corresponds to Seyfert, magenta to LINER, blue to starburst, grey to transition region and orange to LIER excitation; third row: $\Sigma_{SFR}$ maps; fourth row: L{\rm[OIII]} in units $(10^{38}$ erg\,$\mbox{s}^{-1}$\,spaxel$^{-1}$); fifth row: extinction maps; bottom row: $\Sigma_{H^+}$ maps. Bottom right: azimuthally averaged properties of the bottom three left panels, where the AGN is shown in red, the first control in blue and second control in green. The  tick marks are separated by 5$^{\prime\prime}$, and the scale at the galaxy is given by the red horizontal line shown in the $\Sigma_{SFR}$  map panels. This line corresponds to the galaxy effective radius R$_e$ with extent in kpc given above the panel.}
    \label{fig:elliptical}
\end{figure*}

\begin{figure*}
    \includegraphics[width=2.1\columnwidth]{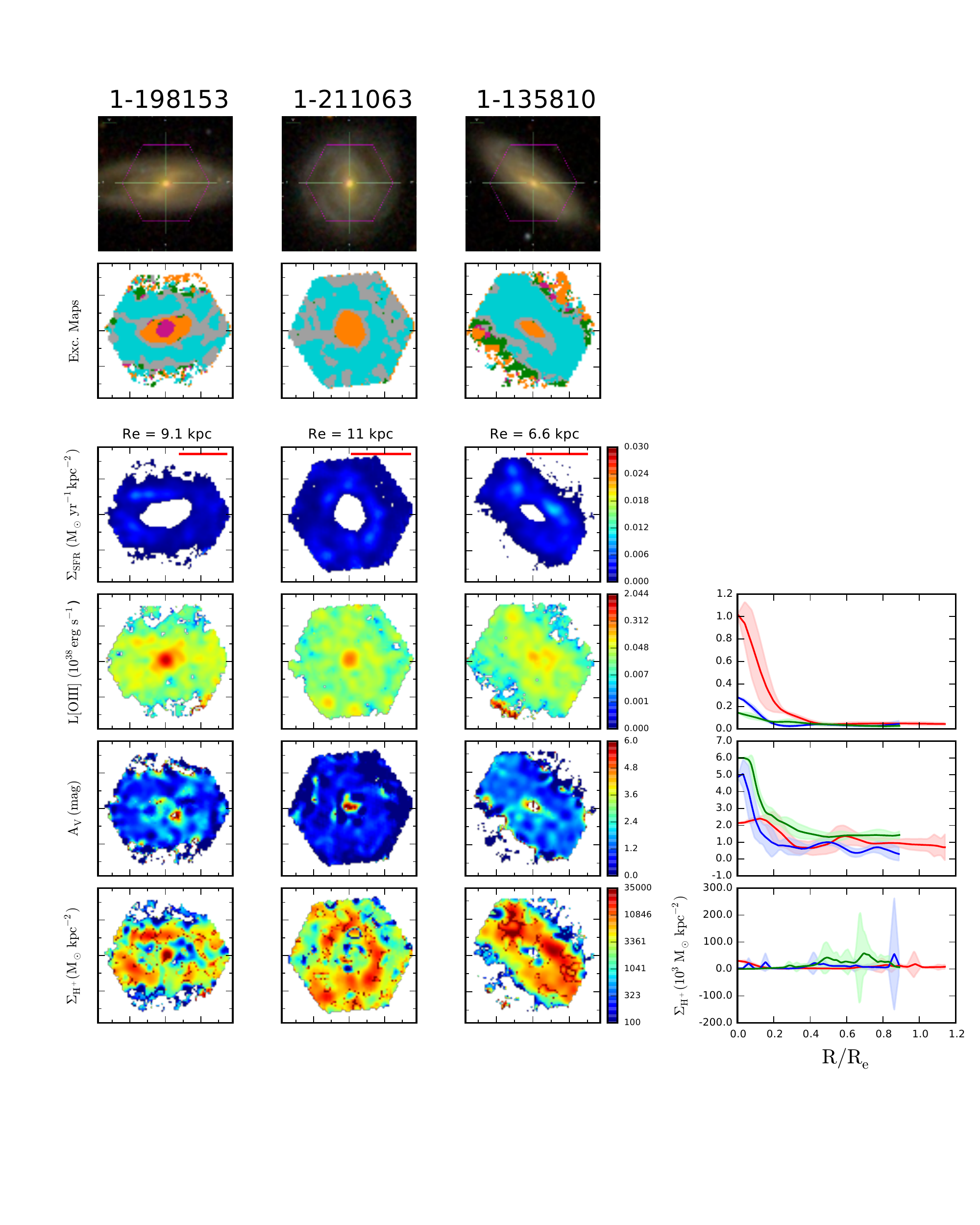}
    \vspace*{-30mm}
    \caption{Same as Fig.~\ref{fig:elliptical} for a typical late-type AGN.}
    \label{fig:spiral}
\end{figure*}

\begin{figure*}
    \includegraphics[width=2.1\columnwidth]{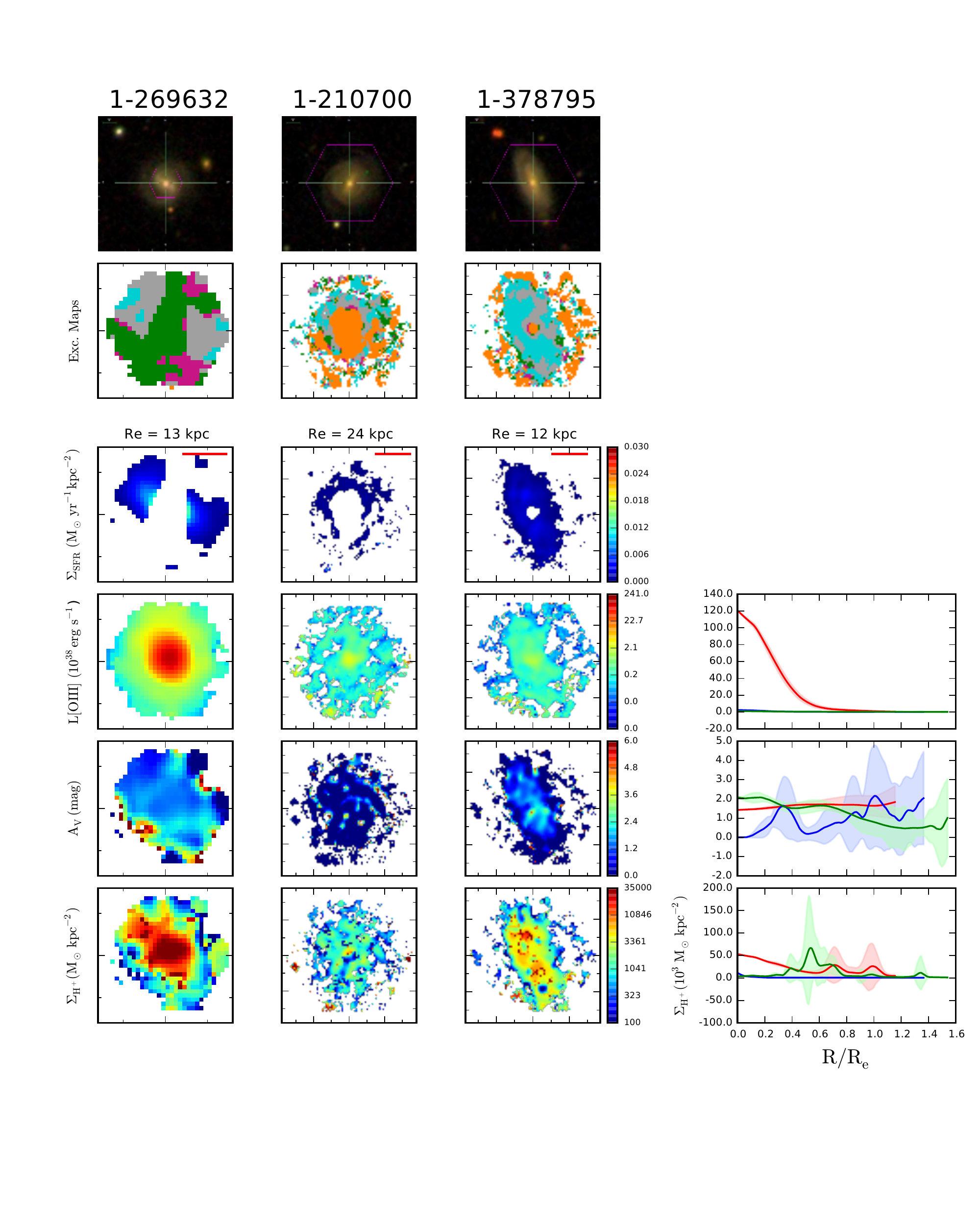}
    \vspace*{-30mm}
    \caption{Same as Fig.~\ref{fig:elliptical} for a high-luminosity AGN.}
    \label{fig:h_luminosity}
\end{figure*}

\begin{figure*}
    \includegraphics[width=2.1\columnwidth]{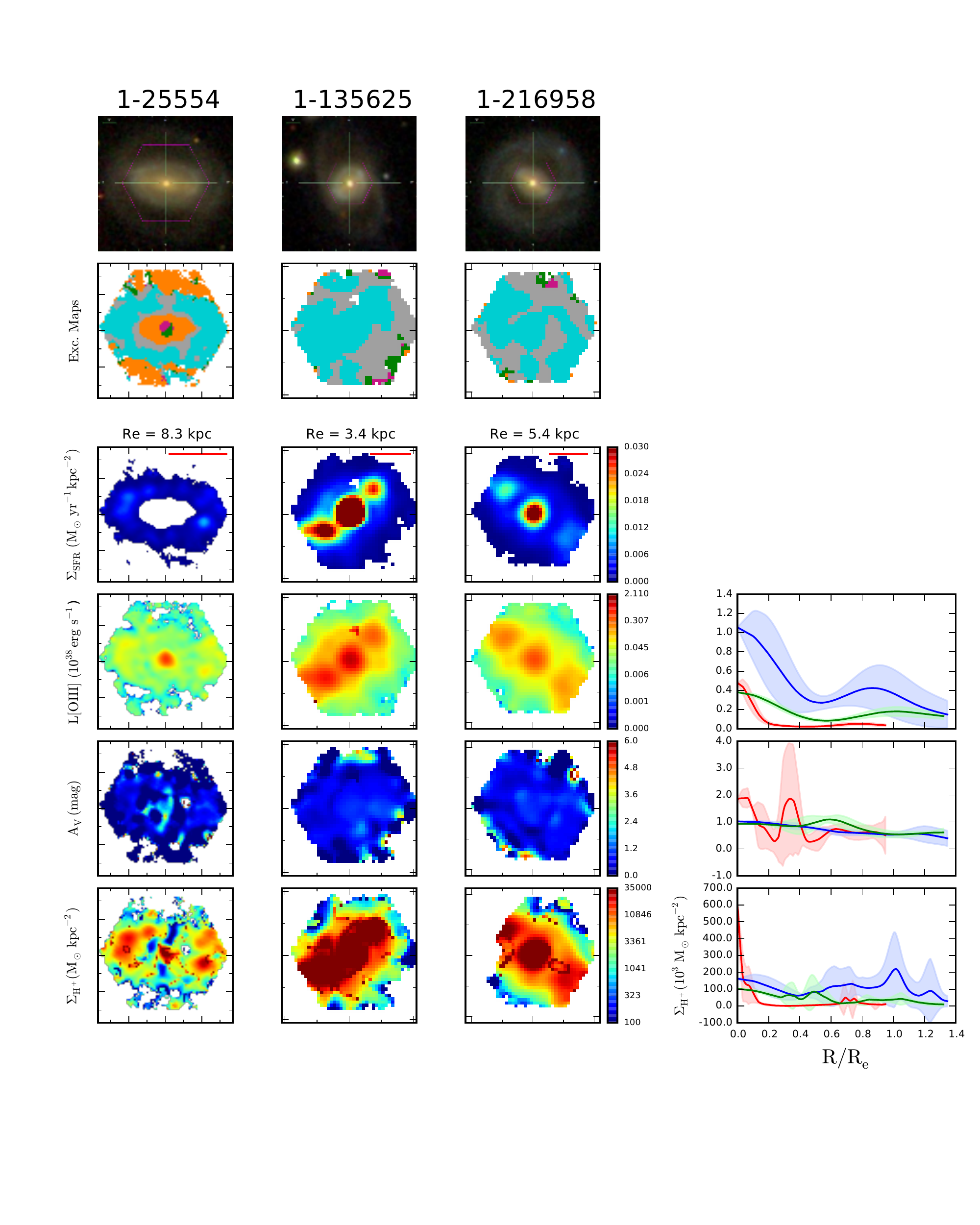}
    \vspace*{-30mm}
    \caption{Same as Fig.~\ref{fig:elliptical} for a low-luminosity AGN.}
    \label{fig:l_luminosity}
\end{figure*}

\begin{table*}
\caption{I - Parameters of the AGN hosts (columns 1-5) and control galaxies (columns 6-10): (1) galaxy identification in the MaNGA survey; (2) log$_{10}$(L[OIII]) for L in units of 10$^{40}$ erg s$^{-1}$; (3) log$_{10}$ of the extent of AGN excitation region in kpc; (4) total ionized gas mass in units of log$_{10}$($M_\odot$); (5) mean nuclear (within 0.2\,$R_{e}$) ionized gas surface mass density in units of $(M_{\odot}$ kpc$^{-2}$; (6) total SFR in logarithm units of M$_\odot$\,yr$^{-1}$; 6--10: same properties for the control sample.
}
\label{table1}
\begin{tabular}{cccccccccccccccc}
\hline
AGN ID & L$\mbox{[OIII]}$ & R & Gas mass & $\Sigma_{nuc}$ & SFR & CS ID & L$\mbox{[OIII]}$ & Gas mass  & $\Sigma_{nuc}$  & SFR \\
 (1)    & (2)  & (3)    & (4)   & (5)   & (6)   & (7)   & (8)   & (9)     & (10)  & (11)     \\
\hline
1-558912 & 56.82$\pm$1.25 & 1.12 & 6.75 & 3.99  & -0.42 & 1-71481 & 0.10$\pm$0.20 & 6.10  & 3.38 & -0.93\\
         &                &      &      &       &       & 1-72928 & 0.09$\pm$0.23 & 5.80  & 3.56 & -1.54 \\
1-269632 & 30.08$\pm$1.69 & 1.25 & 6.88 & 4.68  &-0.23 & 1-210700 & 1.55$\pm$0.44 & 6.75 & 3.65 & -0.51\\
         &                &      &      &       &      & 1-378795 &0.72$\pm$0.31& 6.94 & 3.66 & 0.14\smallskip\\
1-258599 & 20.95$\pm$0.67 &1.41  &7.39 & 5.32  &-0.41 & 1-93876 &0.46$\pm$0.36&5.94 & 3.07 &-1.59 \\
         &                &      &      &       & 		& 1-166691 &0.09$\pm$0.49&5.69 & 3.31 &-1.67 \\
1-72322 & 20.66$\pm$0.43 & 1.43  & 6.92 & 4.07  &-0.28 & 1-121717 & 1.40$\pm$0.57&7.09 &3.81 & 0.67\\
         &                &      &      &       & 	    & 1-43721 &1.91$\pm$0.52&5.95 & 3.12 & -1.06\smallskip\\
1-121532 & 11.68$\pm$0.96 & 1.18 & 6.68 & 4.59  &0.30  & 1-218427 & 0.72$\pm$0.62&5.49 & 1.63 & -2.13\\
         &                &      &      &       &   	& 1-177493 & 2.29$\pm$0.28&6.08 & 3.63 & -1.22\\
1-209980 & 11.01$\pm$0.17 &1.14  &6.23 & 4.77  &-0.69	& 1-295095 & 0.15$\pm$0.03&5.99 & 2.92 & -0.54\\
         &                &      &      &       &  	& 1-92626 &  0.76$\pm$0.07&5.98 & 3.17 & -0.53 \smallskip\\
1-44379 & 8.94$\pm$0.14 &  0.98  & 7.10 &  4.25 &-0.03 & 1-211082 & 0.19$\pm$0.04&6.05 &3.27 & -0.35\\
         &               &       &       &       &  	& 1-135371 & 0.25$\pm$0.07&6.59 &4.31 & -0.26\\
1-149211 & 7.88$\pm$0.14 & 0.61  &  6.50 & 4.46  &-0.93	& 1-377321 & 4.53$\pm$0.13&6.72 &5.31 & -0.32\\
         &               &       &       &       &  	& 1-491233 & 0.25$\pm$0.03&6.78& 6.33 & -0.28 \smallskip\\
1-173958 & 6.79$\pm$0.30 &  0.74 & 7.32 & 4.24  &0.83 	& 1-247456 & 0.57$\pm$0.16&6.87 & 4.47 & 0.23 \\
         &               &       &       &       & 	    & 1-24246 &  0.11$\pm$0.06&6.20 &2.85 & -0.23 \\
1-338922 & 6.77$\pm$0.90 & 1.49  & 7.84 & 5.08  &0.89  & 1-286804 & 2.23$\pm$0.43&7.34 & 3.67 & 0.25 \\
         &               &       &       &       & 	    & 1-109493 & 0.15$\pm$0.18&5.62 & 2.62 & -1.64 \smallskip\\
1-279147 & 6.77$\pm$0.20 & 0.90  & 5.72 & 4.19  &-0.76 & 1-283246 & 0.23$\pm$0.06&5.00 & 2.36 & -1.31\\
         &               &       &       &       &  	& 1-351538 & 0.46$\pm$0.13&6.36 & 4.14 & -0.23\\
1-460812 & 6.46$\pm$0.31 & 0.86  & 6.33  & 4.98  &-1.12	& 1-270160 & 0.70$\pm$0.39&5.71 & 3.45 & -1.85\\
         &               &       &       &       & 	    & 1-258455 & 0.49$\pm$0.14&5.33 & 2.79 & -2.14\smallskip\\
1-92866 & 6.12$\pm$0.30 & 0.81   &6.12  & 4.14  &-1.71	& 1-94514 & -- & 4.79 & 3.23 & -4.04 \\
        &               &        &        &       & 	    & 1-210614 & 0.40$\pm$0.14&5.10 & 2.84 & -1.92 \\
1-94784 & 5.96$\pm$0.12 & 0.33   &6.34  & 4.57  &-0.49	& 1-211063 & 0.20$\pm$0.04&6.45 & 3.64 & -0.29 \\
        &               &        &        &       & 	    & 1-135502 & 0.50$\pm$0.05&5.70 & 3.37 & -0.90\smallskip\\
1-44303 & 5.56$\pm$0.12 & 0.91   & 6.57  & 3.99  &-0.21	& 1-339028 & 0.44$\pm$0.08&5.44 &3.35 & -1.31 \\
         &              &        &        &       & 	    & 1-379087 & 0.72$\pm$0.13&6.76 & 4.72 & 0.12\\
1-339094 & 5.29$\pm$0.09 & 0.57  & 5.99  & 4.57  &2.53 	& 1-274646 & 0.35$\pm$0.04&5.01 & 3.76 & -3.28 \\
         &              &        &        &       & 	    & 1-24099 & 0.06$\pm$0.03&4.80 & 3.10 & -2.99 \smallskip\\
1-137883 & 3.87$\pm$0.12 & 0.66  & 6.42  & 4.86  &-1.54	& 1-178838 & 0.10$\pm$0.02&5.44 &1.92 & -1.29\\
         &              &        &        &       &      & 1-36878 & 0.28$\pm$0.04&6.68 & 5.48 & -0.15\\
1-48116 & 3.79$\pm$0.08 & 0.62   & 6.49   & 4.90  &-0.26 & 1-386452 & 0.32$\pm$0.04&6.31 &5.40 & -0.50\\
         &              &        &        &       &      & 1-24416 & 0.22$\pm$0.03&5.58 & 3.43 & -1.01\smallskip\\
1-256446 & 3.74$\pm$0.15 & 0.82  & 5.83 & 3.97   &-1.25 & 1-322671 &  --&4.28 & 1.63 & -3.38\\
         &               &       &      &        &      & 1-256465 & 0.59$\pm$0.11&5.68 & 4.27 & -1.40\\
1-95585 & 3.58$\pm$0.16 & 0.96   & 7.03  & 3.74   &0.23  & 1-166947 & 0.13$\pm$0.08&5.68 & 3.08 & -0.88\\
         &              &        &        &       &      & 1-210593 & 0.43$\pm$0.14&6.01 & 2.87 & -0.56\smallskip\\
1-135641 & 3.52$\pm$0.09 & 1.01  & 6.75  & 4.93  &-0.09 & 1-635503 & 0.15$\pm$0.06&7.35 & 5.41 & 0.37\\
         &              &        &        &       &      & 1-235398 & 0.16$\pm$0.05&6.39 &5.26 & -0.34\\
1-259142 & 3.47$\pm$0.20 & 0.59  & 6.63 &  4.28  &-0.35 & 1-55572 & 0.12$\pm$0.04&6.49 & 3.14 & -0.40\\
         &              &        &        &       &      & 1-489649 & 0.30$\pm$0.08&5.61 & 3.35 & -1.34\smallskip\\
1-109056 & 3.24$\pm$0.08 & 0.72  & 6.35  &  4.64 &-0.37 & 1-73005 & 0.20$\pm$0.06&6.37 &3.27 & -0.33\\
         &               &       &       &       &      & 1-43009 & 0.12$\pm$0.04&6.53 &4.20 & -0.26\\
1-24148 & 3.17$\pm$0.05 & 0.18   &  5.16  &  4.32 &-2.81 & 1-285031 & 0.26$\pm$0.04&6.61 &5.67 & -0.61\\
         &              &        &        &       &      & 1-236099 &  0.07$\pm$0.01&5.24 & 4.53 & -2.19\smallskip\\
1-166919 & 2.64$\pm$0.25 &0.62   & 6.78  & 3.81  &-0.01 & 12-129446 & 0.28$\pm$0.09&6.99 & 4.43 & 0.21\\
         &              &        &        &       &      & 1-90849 & 0.28$\pm$0.05&6.49 & 3.71 & -0.10\\
1-248389 & 2.55$\pm$0.09 &0.15   & 4.99 &  3.79  &-2.96 & 1-94554 & 0.22$\pm$0.04&5.09 &3.75 & -3.08\\
         &               &       &      &        &      & 1-245774 & 0.29$\pm$0.07&6.67 & 4.14 & -0.13\smallskip\\
1-321739 &2.24$\pm$0.10 & 0.97  & 6.93  & 4.69   &0.06  & 1-247417 & 0.16$\pm$0.04&7.27 & 4.94 & 0.18\\
         &              &        &       &        &      & 1-633994 & 0.36$\pm$0.09&5.90 & 3.14 & -0.68\\
1-234618 & 2.23$\pm$0.23 & 1.17   & 7.09 & 4.60   &-0.10 & 1-282144 & 0.10$\pm$0.02&7.83 & 4.46 & 0.17\\
         &               &       &     &         &      & 1-339125 & 0.45$\pm$0.23&5.62 & 2.90 & -1.32\smallskip\\
\hline
\end{tabular}
\label{table1}
\end{table*}

\begin{table*}
\contcaption{}
\begin{tabular}{ccccccccccccccc}
\hline
AGN ID & L\mbox{[OIII]} & R & Gas mass & $\Sigma_{nuc}$ & SFR &CS ID & L\mbox{[OIII]} & Gas mass  & $\Sigma_{nuc}$ & SFR\\
 (1)    & (2)  & (3)    & (4)   & (5)   & (6)   & (7)   & (8)   & (9)     & (10)  & (11)     \\
\hline
1-229010 & 2.11$\pm$0.09 & 0.38  & 6.31 & 3.51   &-0.46 &1-210962 & 0.35$\pm$0.06&5.60 & 3.11 & -1.16\\
         &               &        &       &        &      &1-613211 & 0.16$\pm$0.06&5.90 & 3.64 & -3.15\\
1-211311 & 1.99$\pm$0.06 & 0.58   & 5.68 &  3.76  &-0.95 &1-25688 & 0.10$\pm$0.02&5.66 &4.03 & -0.70\\
         &               &        &      &        &      &1-94422 & 0.24$\pm$0.03&5.88 & 3.82 & -0.96\smallskip\\
1-373161 & 1.87$\pm$0.11 & 0.54   & 5.90 &  3.98  &-2.34 &1-259650 & 0.67$\pm$0.20&5.53 & 3.72 & -1.81\\
         &               &        &      &        &      &1-289865 &  0.11$\pm$0.09&7.17 &3.11 & -2.47\\
1-210646 & 1.80$\pm$0.10 & 0.55   & 7.09 & 3.48   &0.22  &1-114306 & 0.33$\pm$0.16&6.92 & 4.09 & 0.22\\
         &               &        &      &        &      &1-487130 & 0.27$\pm$0.10&4.32 & 4.72 & 0.15\smallskip\\
1-351790 & 1.72$\pm$0.03 & 0.46   & 5.40 & 4.01   & -inf  &1-23731 & 0.02$\pm$0.01&4.76 & 3.76 & -3.87\\
         &               &        &      &        &      &1-167334 & 0.47$\pm$0.05&6.87 & 1.63 & -2.72\\
1-163831 & 1.67$\pm$0.13 & 0.68   & 6.82 & 3.60   &0.12  &1-247456 & 0.57$\pm$0.16&6.01 & 4.47 & -0.03\\
         &               &        &      &        &      &1-210593 & 0.43$\pm$0.14&6.81 & 2.87 & -0.56\smallskip\\
1-22301 & 1.67$\pm$0.23  & 0.69   &  7.05 &  4.16  &0.31  &1-251871 & 0.24$\pm$0.18&8.23 & 3.11 & 0.26\\
         &               &        &       &        &      &1-72914 & 0.13$\pm$0.07&6.45 & 3.35 & 0.46\\
1-248420 & 1.66$\pm$0.06 & 0.56   & 6.92  & 3.96  &-0.11 &1-211063 & 0.20$\pm$0.04&6.15 & 3.64 & -0.25\\
         &               &        &      &        &      &1-211074 & 0.20$\pm$0.04&4.53 & 3.26 & -0.46\smallskip\\
1-23979 & 1.60$\pm$0.05  & 0.51   &  5.34 &  4.62  &-2.37 &1-320681 & 0.09$\pm$0.07&5.25 & 3.25 & -inf\\
         &               &        &       &        &      &1-519738 & 0.11$\pm$0.04&6.46 &  2.60 & -4.21\\
1-542318 & 1.58$\pm$0.07 & 0.48   & 5.64 & 3.94   &-1.62 &1-285052 & 0.11$\pm$0.03&5.30 & 3.37 & -0.49\\
         &               &        &      &        &      &1-377125 & 0.57$\pm$0.14&5.60 & 1.63 & -0.54\smallskip\\
1-95092 &  1.54$\pm$0.07 & 0.33   & 6.49 &  4.65  &-0.12 &1-210962 & 0.35$\pm$0.06&6.35 & 3.11 & -1.14\\
         &               &        &      &        &      &1-251279 & 0.37$\pm$0.06&5.43 & 4.31 & -0.32\\
1-279676 & 1.52$\pm$0.14 &1.12    & 6.07 & 3.35   &-0.55 &1-44789 & 0.32$\pm$0.09&5.39 & 2.95 & -1.72\\
         &               &        &      &        &      &1-378401 & 0.57$\pm$0.14&6.20 & 2.86 & -2.16\smallskip\\
1-201561 & 1.37$\pm$0.15 & 0.88   & 6.34 & 4.55   &-0.69 &1-24246 &  0.11$\pm$0.06&6.46 & 2.85 & -0.54\\
         &               &        &      &        &      &1-285052 & 0.11$\pm$0.03&5.87 & 3.37 & -0.50\\
1-198182 & 1.34$\pm$0.11 & 0.33   & 5.74 & 4.30   &-2.44 &1-256185 & 0.25$\pm$0.04&5.12 & 3.32 & -1.60\\
         &               &        &      &        &      &1-48053 & --&5.68 & 3.70 & -1.82\smallskip\\
1-96075 & 1.26$\pm$0.13 &  0.73   & 7.10 & 3.66    &0.39  &1-166947 & 0.13$\pm$0.08&5.70 & 3.08 & -0.78\\
         &              &         &       &       &       & 1-52259 & 0.30$\pm$0.09&6.87 & 4.36 & 0.22\\
1-519742 & 1.19$\pm$0.03 & 0.39   & 5.68 & 4.14   &-1.00 & 1-37079 & 0.02$\pm$0.01&4.97 & 3.15 & -1.55\\
         &               &       &      &        &      &1-276679 & 0.05$\pm$0.01&6.17 & 4.58 & -0.61\smallskip\\
1-491229 & 1.14$\pm$0.11 & 0.32   & 5.69 & 4.35   &-2.38 &1-94554 & 0.22$\pm$0.04&5.09 & 3.75 & -3.21\\
         &               &       &      &        &      &1-604048 &  0.39$\pm$0.08&6.80 & 4.32 & -0.03\\
1-604761 & 1.00$\pm$0.13 & 0.97  & 6.86 & 3.49   &-0.11 &1-210173 &  0.52$\pm$0.13&7.12 & 3.01 & 0.0\\
         &               &       &      &     	 &      &1-71525 & 0.19$\pm$0.06&6.84 & 3.27 & 0.07\smallskip\\
1-25725 & 0.92$\pm$0.05 &0.70   & 5.91  & 4.69   &-1.01 &1-211079 & 0.03$\pm$0.04&4.48 & 3.58 & -inf\\
         &              &        &       &        &      &1-322074 & --&4.43 & 2.95 & -2.64\\
1-94604 & 0.86$\pm$0.07 &0.55    & 6.20  & 4.07   &-0.49 &1-295095 & 0.15$\pm$0.03&5.99 & 2.92 & -0.30\\
         &              &       &       &        &      &1-134239 & 0.23$\pm$0.06&6.53 & 3.89 & -0.29\smallskip\\
1-37036 & 0.84$\pm$0.06 & 0.16   & 5.36  &  4.07  &-2.16 &1-210785 &  --&4.37 & 1.97 & -4.06\\
         &              &       &       &        &      &1-25680 & 0.34$\pm$0.04&5.65 & 4.42 & -2.08\\
1-167688 & 0.84$\pm$0.02 & 0.54  & 5.54 & 5.42   &-3.59 &1-235587 & 0.08$\pm$0.02&4.97 & 3.36 & -2.04\\
         &               &      &      &        &      &1-37062 & 0.27$\pm$0.03&6.45 & 5.53 & -0.60\smallskip\\
1-279666 & 0.84$\pm$0.07 &0.89   & 5.57 & 3.92   &-1.54 &1-392976 &  0.10$\pm$0.03&5.16 & 3.59 & -2.17\\
         &               &      &      &        &      &1-47499 & 0.15$\pm$0.06&4.63 & 2.51 & -2.67\\
1-339163 & 0.82$\pm$0.07 &  0.94 & 6.74 & 4.02   &-0.26 &1-136125 & 0.08$\pm$0.02&6.42 & 2.78 & -0.41\\
         &               &       &      &        &      &1-626830 & 0.15$\pm$0.04&5.93 & 3.67 & -0.80\smallskip\\
1-258774 & 0.77$\pm$0.10 & 0.77  & 5.75 & 4.98   &-1.85 &1-379660 & 0.37$\pm$0.07&5.45 & 3.88 & -2.02\\
         &               &       &      &        &      &1-48208 &  0.12$\pm$0.04&5.06 & 2.88 & -2.41\\
1-198153 & 0.76$\pm$0.08 &0.32   & 6.42 & 4.24   &-0.50 &1-211063 &  0.20$\pm$0.04&6.45 & 3.64 & -0.31\\
         &               &       &      &        &      &1-135810 & 0.08$\pm$0.02&6.47 & 2.95 & -0.34\smallskip\\
1-91016 &  0.76$\pm$0.09 & 0.73  & 6.50 & 4.44   &-0.52 &1-338828 & 0.43$\pm$0.05&6.62 & 5.15 & -0.01\\
         &               &       &      &        &      &1-386695 &  0.81$\pm$0.09&6.81 & 5.37 & -0.03\\
1-279073 & 0.63$\pm$0.06 & 0.15   & 5.62 & 4.21   &-2.10 &1-211100 &  --&6.70 & 2.36 & -4.88\\
         &               &       &      &        &      &1-210784 & 0.15$\pm$0.05&4.42 & 3.62 & -3.86\smallskip\\
1-135044 &  0.61$\pm$0.04 & 0.43 & 6.32 & 4.49  &-0.29 &1-218280 & 0.12$\pm$0.03&6.40 & 3.78 & -0.10\\
         &                &      &     &        &      &1-211063 & 0.20$\pm$0.04&6.45 & 3.64 & -0.22\\
\hline
\end{tabular}
\end{table*}

\begin{table*}
\contcaption{}
\begin{tabular}{ccccccccccccccc}
\hline
AGN mangaID & L\mbox{[OIII]} & R$_{exc}$ & Gas mass & log$_{10}\Sigma_{nuc}$ & SFR &CS mangaID & L\mbox{[OIII]} & Gas mass  & log$_{10}\Sigma_{nuc}$ & SFR\\
 (1)    & (2)  & (3)    & (4)   & (5)   & (6)   & (7)   & (8)   & (9)     & (10)  & (11)     \\
\hline
1-148068 & 0.45$\pm$0.15 & 0.64  & 6.72  & 2.87  &-0.18 &1-166947 & 0.13$\pm$0.08&5.68 &3.08 & -1.13 \\
         &               &       &      &        &      &1-55572 & 0.12$\pm$0.04&6.49 & 3.14 & -0.30\\
1-277552 & 0.44$\pm$0.05 &0.39   & 7.46 & 3.78   &0.20  &1-264513 & 0.33$\pm$0.05&7.54 & 5.17 & 0.61\\
         &               &       &       &       &      &1-136125 & 0.08$\pm$0.02&6.42 & 2.78 & -0.31\\
1-217050 & 0.43$\pm$0.03 &  -    & 5.64  & 4.64  &-2.43 &1-135372 & 0.01$\pm$0.23&5.04 & 3.16 & -2.71\\
         &               &       &      &       &      &1-274663 & 0.08$\pm$0.02&5.34 & 4.29 & -3.64\smallskip\\
1-25554 &  0.24$\pm$0.03 & 0.09  & 6.11  & 4.75  &-0.61 &1-135625 & 0.56$\pm$0.04&6.55 & 5.15 & -0.10\\
         &               &       &       &       &      &1-216958 & 0.23$\pm$0.02&6.12 &4.89 & -0.50\\
1-135285 &  0.20$\pm$0.04 & 0.20     & 6.41 & 4.23  &-0.25 &1-633990& 0.25$\pm$0.03&5.82 & 4.68 & -0.83\\
         &               &       &       &       &      &1-25688 & 0.10$\pm$0.02&5.90& 4.03 & -0.77\\
\hline
\end{tabular} 
\end{table*}

\section{Discussion}

In this section, we discuss the properties mapped in the previous section in Figs.~\ref{fig:elliptical} to~\ref{fig:l_luminosity}, respectively for an early-type host, a late-type host, a high-luminosity and a low-luminosity AGN host. Figs.~\ref{fig:first_ap} -~\ref{fig:last_ap} of the Appendix show the maps for the remaining galaxies.

 We present and discuss also here the extents of the regions ionized by the AGN and their relation with L\mbox{[OIII]}, as well as histograms of the total SFRs, ionized gas masses and nuclear surface mass densities, comparing the results obtained for the AGN with those of the control sample.

\subsection{Excitation}

\subsubsection{Early and late-type AGN hosts}

In the excitation maps of the 20 early-type AGN hosts, 9 (45\%) present Seyfert type excitation in their nuclear region (which we define as the region within 0.2\,R$_e$) and up to distances ranging from $\approx$ 3 to 25 kpc. Beyond this region, the excitation varies, including types HII, transition, LIER or LINER. 

The control galaxies, in their majority, present LIER excitation at the nucleus and surroundings, while HII excitation, as seen in the control 2 of Fig.~\ref{fig:elliptical} is less frequently seen in early-type hosts.

Four galaxies (20\% of the early-type hosts) have nuclear LINER excitation, that can also be observed beyond the nucleus, reaching $\approx$ 1.5 - 3.5 kpc, and extra nuclear LIER excitation. The controls in turn have mostly LIER cores. Another 6 galaxies exhibit both LINER and Seyfert type excitation in the nucleus (within R$_e$), indicating that the line ratios are borderline between the Seyfert and LINER regions of the BPT diagram. Their controls usually have LIER excitation in the nucleus.

Regarding the group of 38 late type AGN, as the one in Fig.~\ref{fig:spiral}: 16 (42\%) show LINER excitation from the nucleus until distances in the range 1.6 - 27 kpc; 13 (34\%) present Seyfert excitation extending to 2.4 - 18 kpc); the remaining 9 (24\%) show LINER/Seyfert nuclear excitation. The controls all show similar maps, with LIER, HII or transition excitation in the nucleus. 

\subsubsection{High and low-luminosity AGN hosts}

Regarding the 17 most luminous AGN, 12 (70\%) have Seyfert excitation at the nucleus and throughout most of the galaxy, up to distances from the nucleus in the range 4 - 26 kpc. 
Three (20\%) galaxies show LINER excitation at the nucleus usually surrounded by Seyfert excitation, and extended to distances in the range 5.5 - 27\,kpc, and 2 (12\%) present Seyfert excitation at the nucleus surrounded by LINER excitation. The nuclear excitation of the control galaxies include type HII, transition and LIER.

The 45 low-luminosity AGN show varied types of nuclear excitation: $\approx$ 30\% have mixed LINER/Seyfert excitation, $\approx$ 40\% have LINER excitation at the nucleus and up to 1.4 -0 9.4 kpc, while the other $\approx$ 30\% have Seyfert excitation at the nucleus and up to 2.4 - 15 kpc. At the nucleus, the controls show LIER or starburst excitation, while, outside the nucleus, AGN and controls present similar maps, dominated by star-forming disks in the late types and LIER excitation in the early types, what can be attributed to the low level of activity of the AGN.

Finally, there are 5 controls that present Seyfert or LINER excitation at the nucleus and 1 case of one AGN (1-217050) which presents LIER excitation. This result indicates disagreement between the line ratios measured in the SDSS-III spectra (that we have used to select the sample) and in the nuclear MaNGA spectrum, which could be due to aperture size differences between SDSS-III and IFU MaNGA and/or measurement uncertainties and/or intrinsinc variation of the AGN.

In the case of the  control galaxy 1-286804, which are actually two galaxies in interaction, one galaxy nucleus present Seyfert and the other LINER excitation, what was also not seen in the SDSS-III spectrum.

\subsubsection{Extra-nuclear AGN excitation}

We have measured the extent $R$ of the region ionized by the AGN in the excitation map -- that can be identified with the narrow-line region (NLR) of the AGN -- in order to verify if it increases with the AGN luminosity, as expected \citep{Schmitt03a,Schmitt03b}. These values are listed in column 2 of Table \ref{table1}, and range from 1.2\,kpc to 27\,kpc, with estimated measurement uncertainties of about 15\%. Due to the relatively large uncertainty, we preferred not to correct the data for the galaxy inclination. 

\begin{figure}
    \includegraphics[width=1.1\columnwidth]{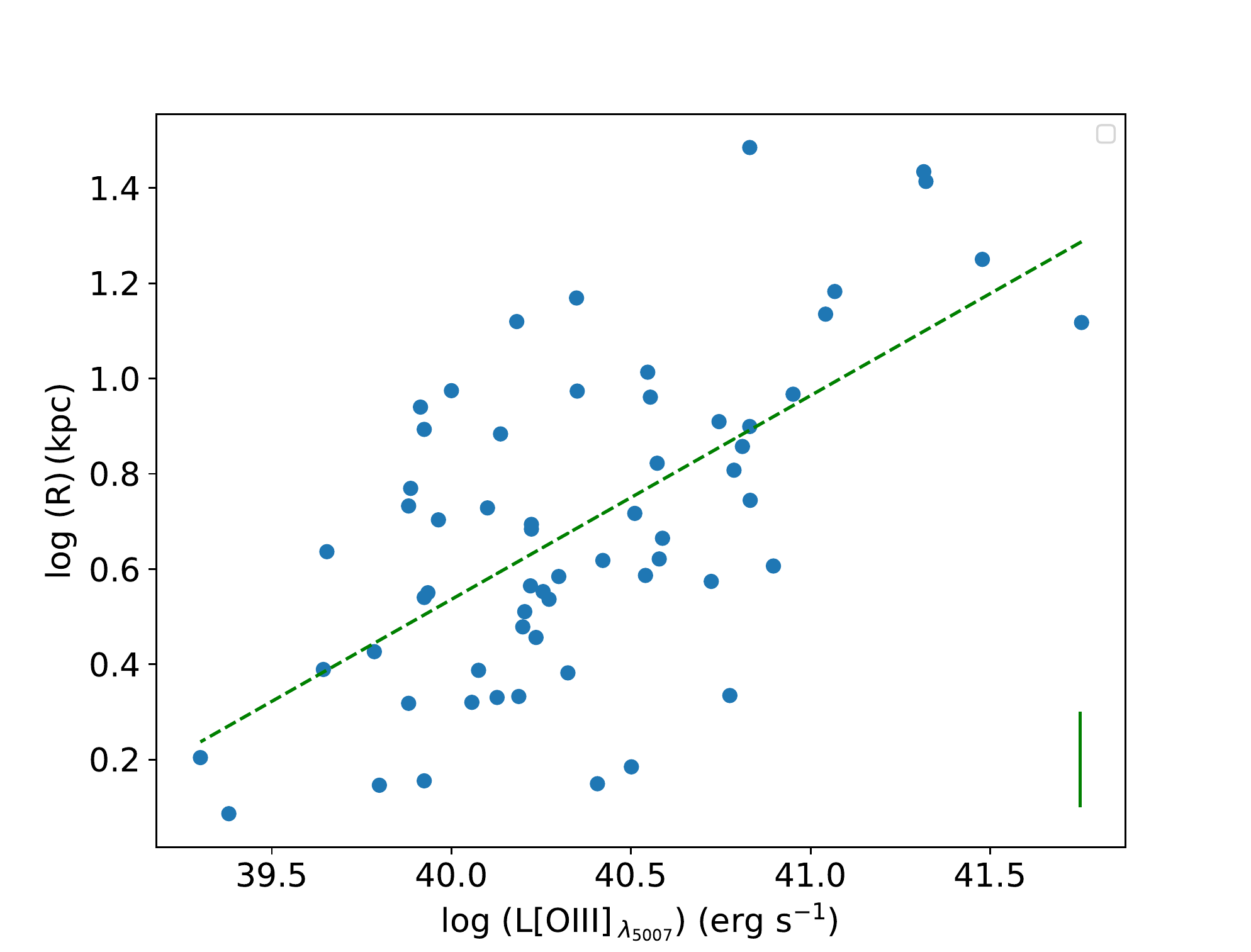}
    \caption{Relation between the extent of the region ionized by the AGN $R$ and $L\mbox{[OIII]}$, showing also a linear regression to the data (see text). The bar in the bottom corner shows a typical uncertainty in $R$ of about 10\%.}
    \label{fig:L_R}
\end{figure}

We have plotted the $R$ values against L[OIII] in Fig.~\ref{fig:L_R}, showing that, on average, the extent of the region ionized by the AGN do increase with the AGN luminosity. A linear least squares regression applied to the data gives the following relation:

\begin{equation}
\mathrm \log(R) = \mathrm{(0.43\pm0.07)\,\log L\mbox{[OIII]} - (16.58\pm2.78)} 
\end{equation}

\noindent with a correlation coefficient of $r=0.62$. 

It is interesting to compare the above relation with a similar one obtained between the extent of the NLR and $L\mbox{[OIII]}$ by \citet{bennert02} and more recently by \citet{sb18}. We note that the regression coefficients above agree with those of the latter authors within the uncertainties. There is a vertical shift to larger values in our relation compared to that of \citet{sb18}, which we attribute to the poorer resolution of the MaNGA data, that can barely resolve 1\,kpc in the closest galaxies of our sample, while \citet{sb18} can resolve down to $\approx$\,100\,pc.

The above relation supports the proportionality $R\,\propto$\,L$^{0.5}$, known to be valid for the Broad-Line Region (BLR) \citep{kaspi05,peterson14}. Given the expression for the ionization parameter $U\propto(L_{AGN}$/($4\pi\,R^2\,n_e\,\,c$), where $c$ is the light speed and $n_e$ is the gas density, $R$ is $\propto\,L^{0.5}$  if the product $U\,n_e$ is a constant. 

bf In many galaxies, both AGN and controls, there are some external spaxels, detached from the inner AGN excitation region, that show Seyfert or LINER excitation , according to the BPT diagram. This finding in the MaNGA data has been already pointed out by previous authors such as \citet{W18}. One possible interpretation is that this gas has been previously ionized by an AGN, but which has presently `` turned off", thus characterizing a ``relic AGN". 

We have nevertheless noticed, that in many cases these spaxels are surrounded by LIER excitation. When we consider that our adopted ``division" between AGN and LIERs is a sharp limit of the equivalent width W(H$\alpha$)$=$3\AA, typical uncertainties of 10--30\% in such small W(H$\alpha$) values will move the spaxel from AGN to LIER excitation and vice-versa. We thus do not favor the ``relic AGN" hypothesis in these cases, and attribute to possible uncertainties in W(H$\alpha$) values.

We have also noted that in many cases the extranuclear spaxels with AGN excitation correspond to high latitude locations in a galaxy seen close to edge-on, suggesting an origin in a warm ionized medium (WIM) \citep[e.g.][]{sb96}, that shows [NII]/H$\alpha$ ratios similar to those of AGN \citep{Br+08}, such as the case of Fig.\,A2 in the appendix. 

There are a few cases that may be indeed ``relic AGN", and should be further investigated for having more extended detached regions with AGN excitation. The following objects are candidates for being relic AGN: MaNGA ID 1-121717, 1-43721, 1-178838, 1-37062, 1-377125, 1-386695 and 1-25688.

\subsection{Star Formation Rate SFR}

The SFR surface density $\Sigma_{SFR}$ maps are presented in the third row of Figs.~\ref{fig:elliptical} to ~\ref{fig:l_luminosity} for the four selected AGN and in the Appendix for the other galaxies.

In the case of the early-type AGN of Fig.~\ref{fig:elliptical}, there is almost no star formation, while the first control galaxy presents some star formation at its border with low $\Sigma_{SFR}$ (bellow 0.006\,M$_\odot$\,yr$^{-1}$\,kpc$^{-2}$) and the second shows a high $\Sigma_{SFR}$ (above 0.03\,M$_\odot$\,yr$^{-1}$\,kpc$^{-2}$) in the nuclear region. The other early-type AGN hosts exhibit either more extranuclear star formation than the controls (11 cases) or similar values (8 cases), with only 3 cases with lower SFR in the AGN than in the controls. The $\Sigma_{SFR}$ values are usually low, as above. A minority of control galaxies present higher values of $\Sigma_{SFR}$ than the AGN, and when this happens, the $\Sigma_{SFR}$ is highest in the central region of the galaxies.

In the case of the late type AGN, as illustrated in Fig.~\ref{fig:spiral}, the $\Sigma_{SFR}$ maps of AGN and controls are similar to each other, with typical values that do not vary much and are in the range $\approx$ 0.003--0.01\,M$_\odot$\,yr$^{-1}$\,kpc$^{-2}$. There is a large extra nuclear region with star formation, with the $\Sigma_{SFR}$ values being usually larger in the inner radii and decreasing outwards, although in some cases there are enhancements associated with spiral arms. Some control galaxies show star formation in the nuclear region, in which case $\Sigma_{SFR}$ peaks at the nucleus. In these cases, the $\Sigma_{SFR}$ values are $\approx$\,30 times larger than that in the AGN.

In the high-luminosity AGN of fig.~\ref{fig:h_luminosity}, there is star formation both in the AGN and controls only in the disk of the galaxies (thus beyond the nuclear region). The $\Sigma_{SFR}$ values do not vary much; only in this AGN we notice an increase in the values towards the center but this could be due also to the increase in the H$\alpha$ flux due to the AGN proximity, as most of the region is a transition one (AGN combined with star formation, that we included in the calculation of the SFR). The remaining trios in this group show similar behavior, with $\Sigma_{SFR}$ maps for the AGN similar to those of the controls, in which star formation is only observed outside the nucleus. There are a few cases with almost no star formation, and a few others in which the controls show high $\Sigma_{SFR}$ values in the center.

In Figure~\ref{fig:l_luminosity}, for a low luminosity AGN, we observe that the AGN has star formation only in the galaxy disk -- with an enhancement in $\Sigma_{SFR}$ in what appears to be a star-forming ring. Both controls present similar $\Sigma_{SFR}$ values at the galaxy disks but larger values mostly at the nucleus, but also in an elongated structure beyond the nucleus that seems to be a bar. The values at the nucleus are at least 30 times higher than the typical values in the disk.

For the other low-luminosity AGNs, only in 8 ($\approx$\,20\%) of the 45 trios the situation is similar to that of Fig.~\ref{fig:l_luminosity}, in which one or the two controls present higher $\Sigma_{SFR}$ values in the center and decrease outwards, whereas in the AGN the star formation is only observed at the disk, with similar values to those quoted above. In most ($\approx$\,80\%) of the cases, the $\Sigma_{SFR}$ maps are similar between the AGN and controls, in the sense that in the controls, also the star formation is larger in the outer part of the galaxies than in their centers, while in the central part, instead of an AGN, there is a LIER.

\subsubsection{SFR histograms}

We have built histograms for the total SFR as listed -- in log$_{10}$ units of M$_\odot$\,yr$^{-1}$ -- in the last column of Table\,\ref{table1}. The total SFR values were obtained by adding the SFRs from all regions over the whole galaxy, where star formation is present, including the HII excitation regions and the transition regions. 

\begin{figure*}
    \includegraphics[width=2.3\columnwidth]{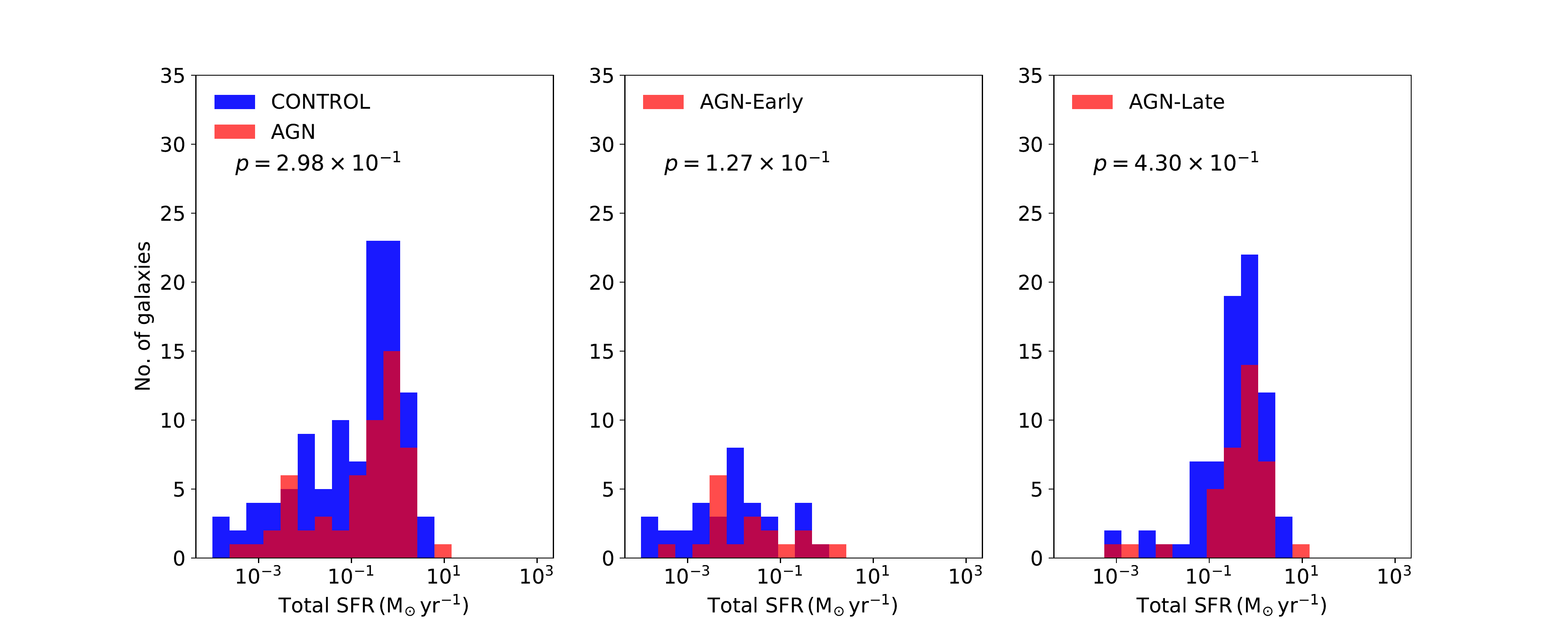}
    \caption{Histograms comparing the total SFR for the AGN and control galaxies. Left panel: whole sample; central panel: early-type hosts; right panel: late-type hosts. The $p$ value in each panel gives the probability that the two distributions are drawn from the same parent population.} 
    \label{fig:sfr}
\end{figure*}

Fig.~\ref{fig:sfr} shows the histograms of the total SFR for the AGN (in red), as compared to those of the control galaxies (in blue), separated in three panels: in the left panel we show the histograms for the whole sample, in the center panel only for the early-type galaxies and in the right panel for the late-type galaxies. The distributions for the AGN and controls in the three cases are very similar, what is confirmed by a 2-sample Anderson-Darling (A-D) test. The probability $p$ that the two distributions are drawn from the same parent population is given in the corresponding panels (the reference $p$ is 0.05, meaning that for smaller values, there is more than 95\% chance that the two distributions are distinct). 
According the A-D test for the above three cases, the probability that the distributions are statistically equal are, $\approx$\,30\%, $\approx$\,13\% and $\approx$\,43\% for the total, early-type and late-type samples, respectively.  

\begin{figure*}
    \includegraphics[width=2.3\columnwidth]{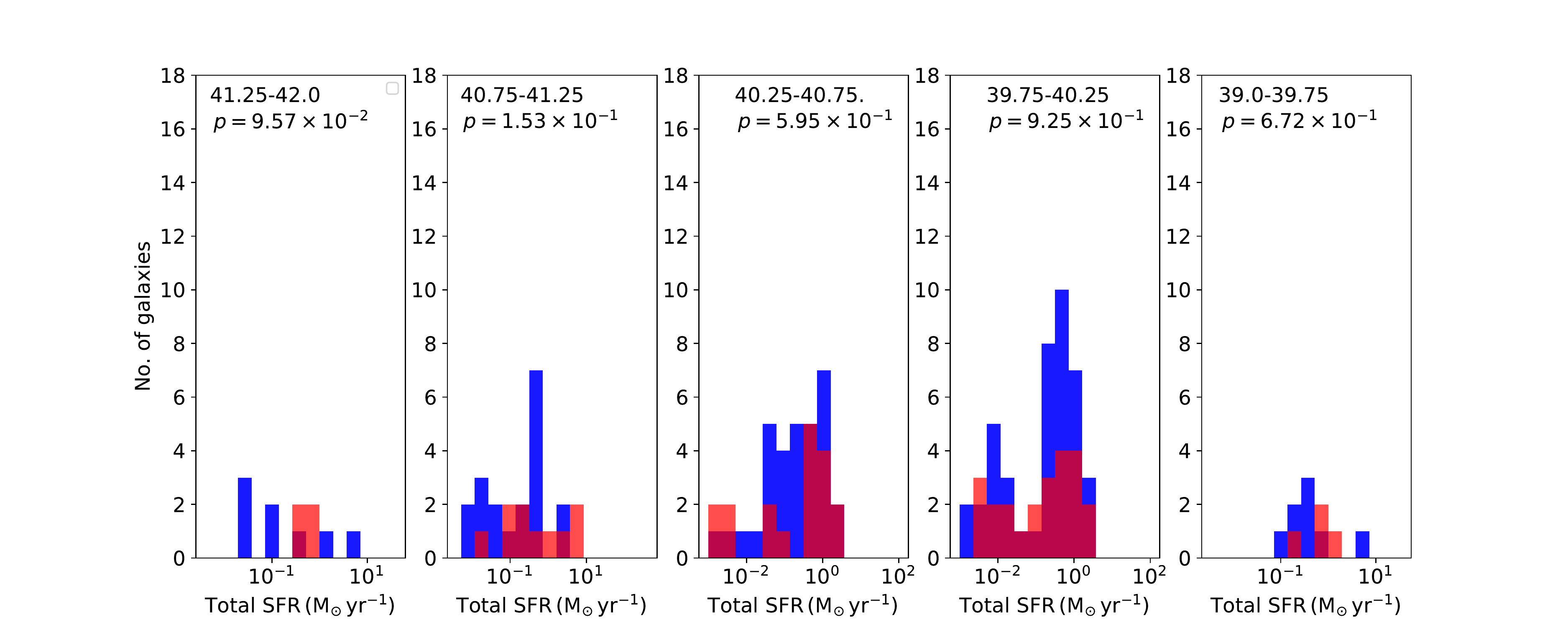}
    \caption{Histograms comparing the total SFR of AGN and control galaxies separated according to the AGN [OIII] luminosity log(L[OIII]). From left to right: decreasing L[OIII], with  values identified in the top of each panel in logarithm units of erg\,s$^{-1}$.
The $p$ value in each panel gives the probability that the two distributions are drawn from the same parent population.}
    \label{fig:sfr_l}
\end{figure*}

In Fig.~\ref{fig:sfr_l} we present similar histograms for the total  SFR now separated in different bins of AGN luminosity (log$_{10}$ L[OIII], in units of erg\,s$^{-1}$), from the most luminous (41.25 -- 42.0) to the left to the less luminous (39.0 -- 39.75) to the right. Again the distributions for AGN and controls are similar, with a probability that the distributions are statistically the same between 9\% and 92\% for the different luminosity bins, according to the A-D test.

It can thus be concluded, from the above two histograms, that the total SFR, over the whole galaxies is similar for AGN and control galaxies.

As we have carefully built the control sample to match the AGN host galaxy properties, we now compare the SFR of each AGN to those of its two control galaxies. Although the range of SFR of the two samples are similar, there may be a systematic difference between a galaxy in an AGN phase and one that is not (the corresponding control galaxies).

\begin{figure*}
    \includegraphics[width=2.1\columnwidth]{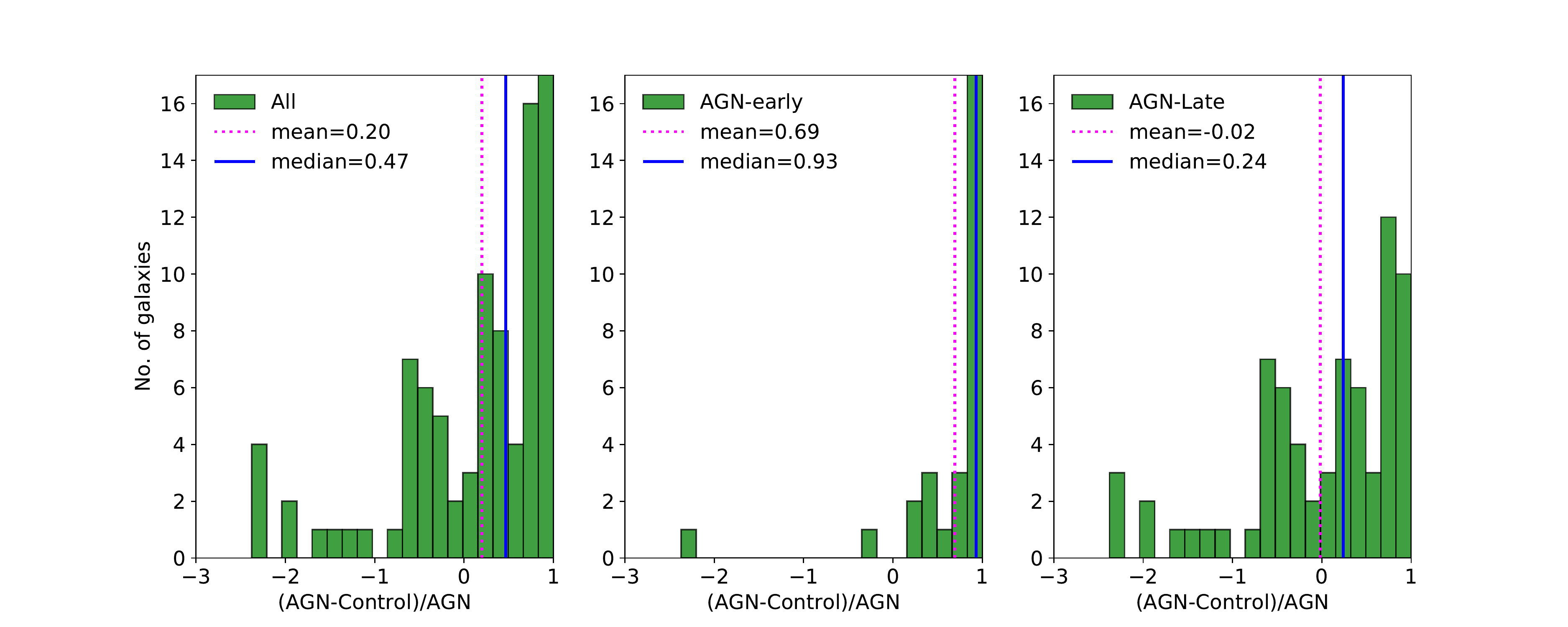}
    \caption{Histograms for the fractional difference in SFR between the AGN and each of its two control galaxies: (SFR$_{AGN}$-SFR$_{ctr}$)/SFR$_{AGN}$. Left panel: total sample; central panel: early-type hosts; right panel: late-type hosts.}
    \label{fig:sfr_diff}
\end{figure*}

We have thus built another histogram in which we present the distribution of differences between the SFR of the AGN and each of its control galaxies. In order to take into account the fact that the range of SFR values is very large, we have divided the difference in SFR by the AGN SFR. The histogram of these fractional differences is shown in Fig.~\ref{fig:sfr_diff}, for each pair AGN-control (112 combinations): (SFR$_{AGN}$--SFR$_{ctr}$)/SFR$_{AGN}$. The division by SFR$_{AGN}$ is necessary in order to bring the differences to a similar scale.

In these new histograms, when considering the whole sample, $\approx$61\% of the combinations AGN-control show positive fractional differences (SFR in the AGN larger than in the control), with a positive mean fractional difference of 0.2 and median of 0.47. When considering only the early-type sub-sample, this percentage increases to $\approx$\,76\%, mean of 0.69 and median of 0.93, while for the late-types this percentage decreases to $\approx$51\%. These results mean that, while for the early-type galaxies, most AGN have more star formation than its controls, for the late-type galaxies, in $\approx$\,50\% of the cases there is more star formation in AGN and in the other 50\% there is more star formation in the controls.

\subsection{L[OIII] maps}

The fourth row of panels (from top to bottom) of Figs. ~\ref{fig:elliptical} to~\ref{fig:h_luminosity} shows that, for these three AGN hosts -- early and late types and high-luminosity AGN -- L\rm{[OIII]} at the nucleus is higher than those of the control galaxies, as expected. But in the case of the low-luminosity AGN of Fig.~\ref{fig:l_luminosity} L\rm{[OIII]}
is lower in the AGN than in its controls. This can be due to the fact that $L\rm{[OIII]}$ in the control galaxies, due to star formation, can be higher than that of a low-luminosity AGN. This happens not only at the nucleus but also outside the nucleus, as seen also in Fig.~\ref{fig:l_luminosity}.

When inspecting Figs.\ref{fig:first_ap} -\ref{fig:last_ap} of the Appendix we note that the example of Fig~\ref{fig:l_luminosity} is more the exception than the rule, and most active galaxies do present the nuclear region more luminous in $L\rm{[OIII]}$ than their controls. There are only three more cases in which at least one control galaxy has the central region with higher $L\rm{[OIII]}$ than the AGN, but the profiles and the maps show that there is not a large difference between the AGN and its controls.

\subsection{Extinction}
The fifth row of Figs. ~\ref{fig:elliptical} to~\ref{fig:h_luminosity} show that the extinction maps and corresponding one-dimensional profiles present a very diverse behavior. For some AGN hosts and control galaxies, the extinction is larger at large radial distances; for others, the largest extinction values are found close to the center of the galaxy. There are also many cases of irregular extinction maps. We have not detected any significant difference between the extinction maps of the AGN and the control galaxies. As the estimate of extinction requires a valid flux measurement of both H$\alpha$ and H$\beta$, and in many cases we could not measure the H$\beta$ flux due to the low signal-to-noise ratio of this line, the resulting extinction maps often present ``holes'', what may have masked possible differences between AGN and control galaxies. 

\subsection{Ionized gas surface mass densities $\Sigma_{H^+}$}

The bottom panels of Figs.~\ref{fig:elliptical} to~\ref{fig:l_luminosity} show the $\Sigma_{H^+}$ maps and average one-dimensional profiles.

For the 20 early-type galaxies, including Fig.~\ref{fig:elliptical}, both the AGN and the non-active galaxies present ionized gas over the whole galaxy in 90\% of them, with the AGN -- as observed in the one-dimensional profiles -- showing higher $\Sigma_{H^+}$ than the controls at the center -- reaching several 10$^3$\,M$_\odot$\,kpc$^2$ within 0.2\,R$_{e}$ (the nuclear region) and decreasing outwards, while the control galaxies frequently have higher $\Sigma_{H^+}$ values than the AGN beyond 0.2\,R$_{e}$. In 2 (10\%) cases the $\Sigma_{H^+}$ is lower in the nuclear region of AGN than in the control galaxies.

The 38 late-type galaxies, as illustrated in Fig.~\ref{fig:spiral}, show that the ionized gas is spread all over the galaxies for both AGN and controls,  mostly due to star formation in the galaxy disks. According to the one-dimensional profiles, about half of the AGN have higher $\Sigma_{H^+}$ than the control galaxies within 0.2\,R$_{e}$ (the nuclear region). In the other half, the difference between AGN and controls is of the order of the typical azimuthal variation of the $\Sigma_{H^+}$ within each galaxy (shaded areas in the one-dimensional profiles). In general, the one-dimensional profiles do not show a large radial variation in $\Sigma_{H^+}$ values for both AGN and controls. 

In the $\Sigma_{H^+}$ maps of the high-luminosity AGN of  Fig.~\ref{fig:h_luminosity}, we observe much higher $\Sigma_{H^+}$ values for the AGN than in the controls within 0.2\,R$_{e}$, while outside 0.4\,R$_{e}$ the values are similar for the AGN and controls. 
Considering all high-luminosity AGN, approximately 64\% (11) (including the example of Fig.~\ref{fig:h_luminosity}), present higher values for the AGN than in the controls within 0.2\,R$_{e}$, while outside 0.4\,R$_{e}$, the values are similar for the AGN and controls. For the remaining 36\% (6 cases), in half the AGN show $\Sigma_{H^+}$ values smaller than their controls in the nucleus, and in the other half the AGN and controls have similar $\Sigma_{H^+}$ values. Some controls show an increase in the $\Sigma_{H^+}$ values outwards.

For the 45 low-luminosity AGN, including the example of Fig.~\ref{fig:l_luminosity}, approximately 45\% exhibit -- for both the AGN and control galaxies -- the ionized gas spread all over the galaxy. The one-dimensional profiles show that, while the AGN shows a higher $\Sigma_{H^+}$ inside 0.2\,R$_{e}$, the controls have higher values than the AGN beyond the nuclear region. For approximately 22\% of the low-luminosity AGN, the nuclei of the AGN and controls present similar $\Sigma_{H^+}$ values, whereas for 33\% the AGN have lower $\Sigma_{H^+}$ values than their controls. For regions beyond 0.2\,R$_{e}$, the profiles are mostly similar for the AGN and controls.

\subsection{Central ionized gas surface mass densities $\Sigma_{nuc}$}

\begin{figure*}
    \includegraphics[width=2.1\columnwidth]{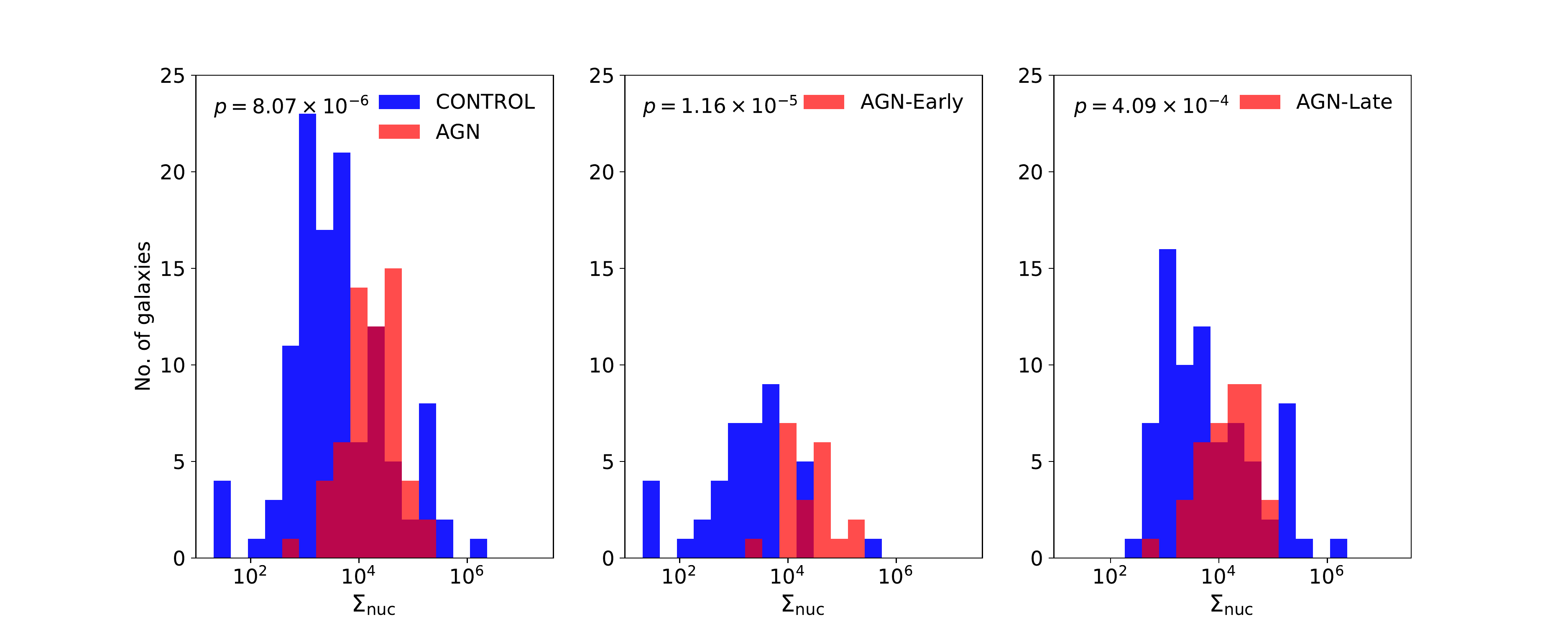}
    \caption{Histograms of central surface mass densities $\Sigma_{nuc}$ (in units of M$_{\odot}$\,kpc$^{-2}$) within the inner 0.2\,R$_{e}$ (effective radius); left: total sample; center: early-type sample; right: late-type sample.}
    \label{fig:CD}
\end{figure*}

\begin{figure*}
    \includegraphics[width=2.1\columnwidth]{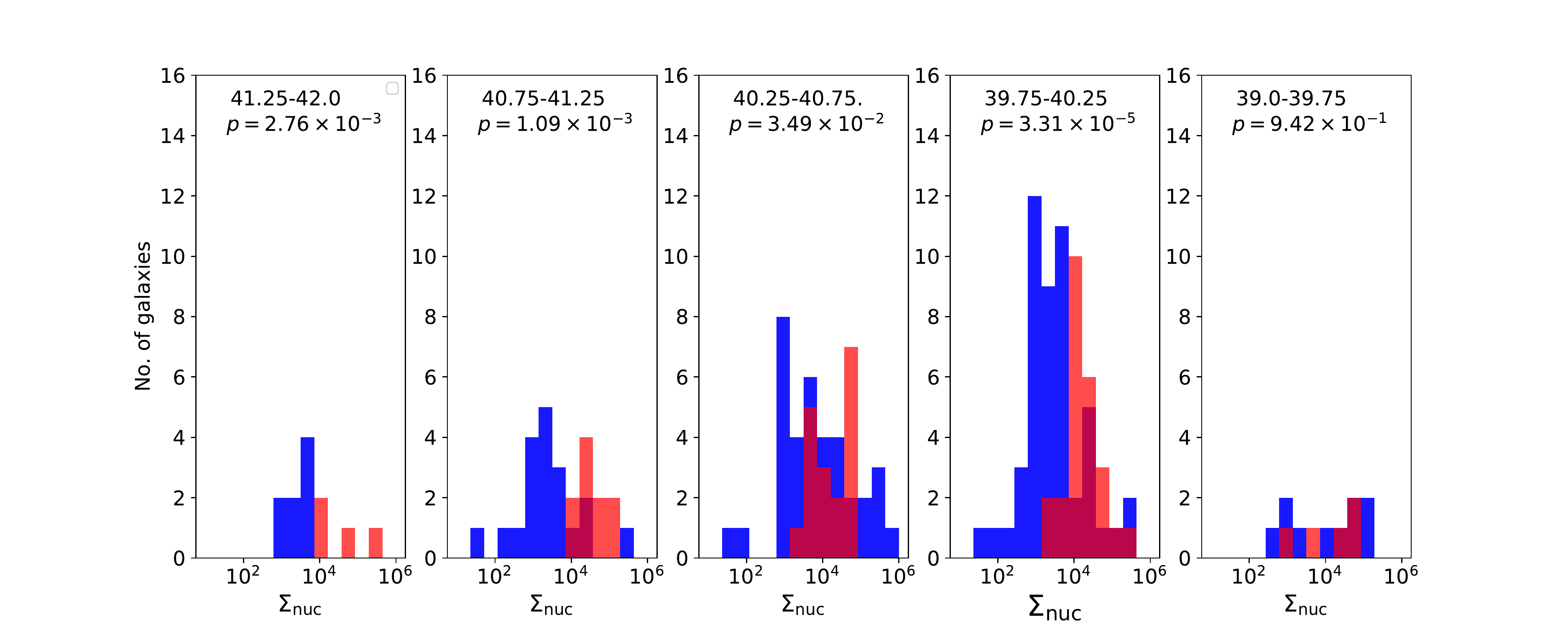}
    \caption{Histograms of $\Sigma_{nuc}$ separated in AGN luminosity bins -- shown in the top of each panel as log$_{10}$(L[OIII], in units of $10^{40}$\,erg\,s$^{-1}$, together with the A-D test $p$ value, from the most luminous (left) to the less luminous AGN (right). Colors as in Fig.~\ref{fig:CD}.}
    \label{fig:DL}
\end{figure*}

We have calculated the mean central surface mass density $\Sigma_{nuc}$ as the ionized gas surface mass density within a circular area of radius 0.2 R$_{e}$ around the nucleus. This value is listed in columns 4 and 8 of Table ~\ref{table1}, respectively for the AGN and controls. These values range from $7.35\,\times\,10^2$  to $2.61\,\times\,10^5$ for the AGN and $0.43\,\times\,10^2$  to $2.14\,\times\,10^6$ for the controls.

We show histograms of these values in Fig. ~\ref{fig:CD} and Fig. ~\ref{fig:DL}. In Fig.~\ref{fig:CD}, we show the distribution of $\Sigma_{nuc}$ values for the whole sample in the left panels, and separately for the early and late-type sub-samples in the central and right panels. 

According the A-D test, the probabilities $p$ that the AGN and control sample are drawn from the same population are smaller than 10$^{-5}$ for the total and early-type sample and $p = 4.09\times^{-4}$ for the late-type sample. The median values of $\Sigma_{nuc}$ are $1.74\times\,10^4$\,M$_{\odot}$\,kpc$^{-2}$ for all the AGN and $2.72\times\,10^3$\,M$_{\odot}$\,kpc$^{-2}$, for all the controls. 

In Fig.~\ref{fig:DL}, in which the AGN are separated in luminosity bins, even though the number of objects in each bin is small, 
the distributions are statistically different -- with the values of $p\le0.05$ for all bins except the lowest luminosity one.

\begin{figure*}
    \includegraphics[width=2.1\columnwidth]{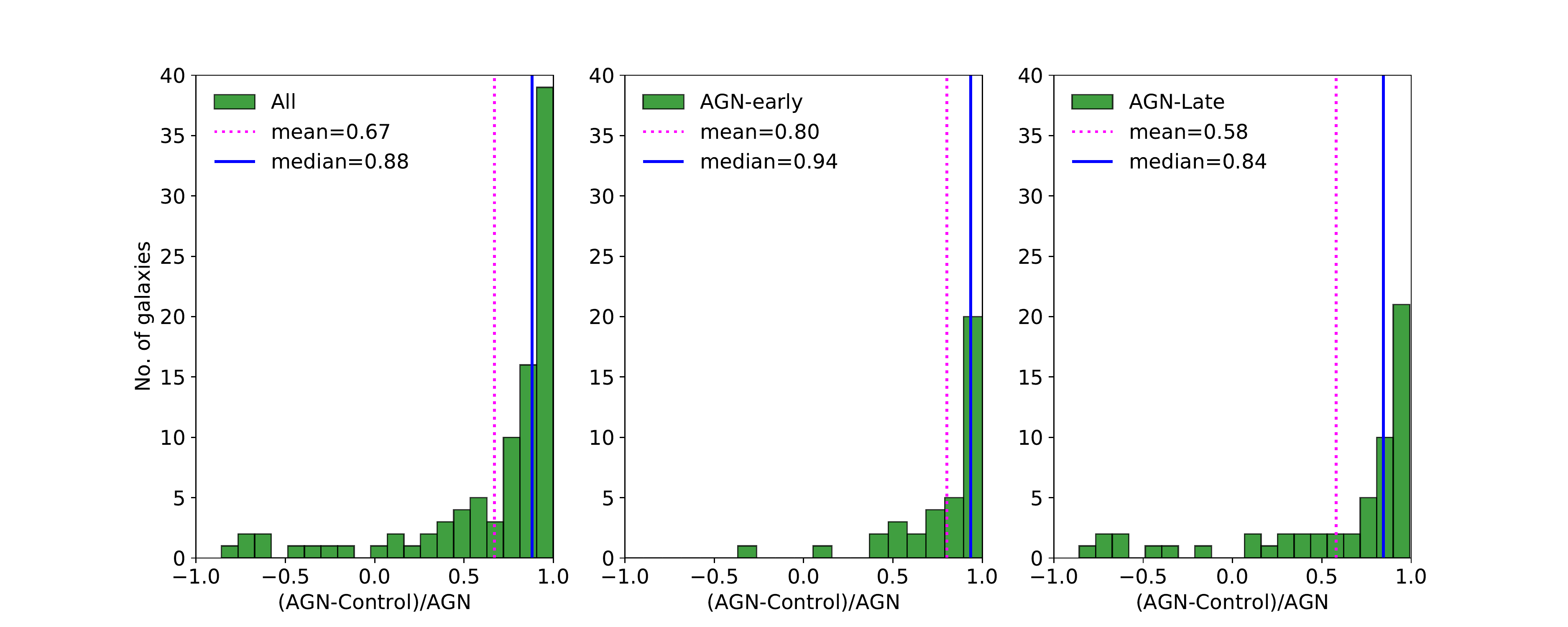}
    \caption{Histograms of the fractional differences $(\Sigma_{nuc,AGN}-\Sigma_{nuc,ctr})/\Sigma_{nuc,AGN}$; left: total sample; center: early-type sample; right: late-type sample.}
    \label{fig:D_diff}
\end{figure*}

We further explore the differences between the AGN and control galaxies via histograms of the fractional differences of $\Sigma_{nuc}$ for each pair AGN-control (112 combinations), as previously done for the SFR: ($\Sigma_{nuc,AGN}$--$\Sigma_{nuc,ctr}$)/$\Sigma_{nuc,AGN}$. The histograms for these values are shown in Fig.~\ref{fig:D_diff} for the total sample on the left and for the early and late-type sub-samples in the central and right panels. There were only a few cases for which the values are more negative than -1, which we have opted to exclude from the graph for visualization purposes. But even when taking them into account, we have that $\approx$74\% of the sample show positive fractional differences, indicating that $\Sigma_{nuc}$ is higher in the AGN than in the controls for most of the sample. When considering only the early-type sub-sample, this percentage increases to $\approx$\,93\%, while for the late-type sub-sample this percentage is $\approx$63\%. The median fractional differences of $\Sigma_{nuc}$ between the AGN and controls are: 0.88 for the total sample, 0.94 for the early-types and 0.84 for the late-types. 

In summary, $\Sigma_{nuc}$ is in most cases higher in AGN than in the control galaxies, with the largest fractional differences being observed for the early-type galaxies. 

\subsection{Total ionized gas mass}

\begin{figure*}
    \includegraphics[width=2.1\columnwidth]{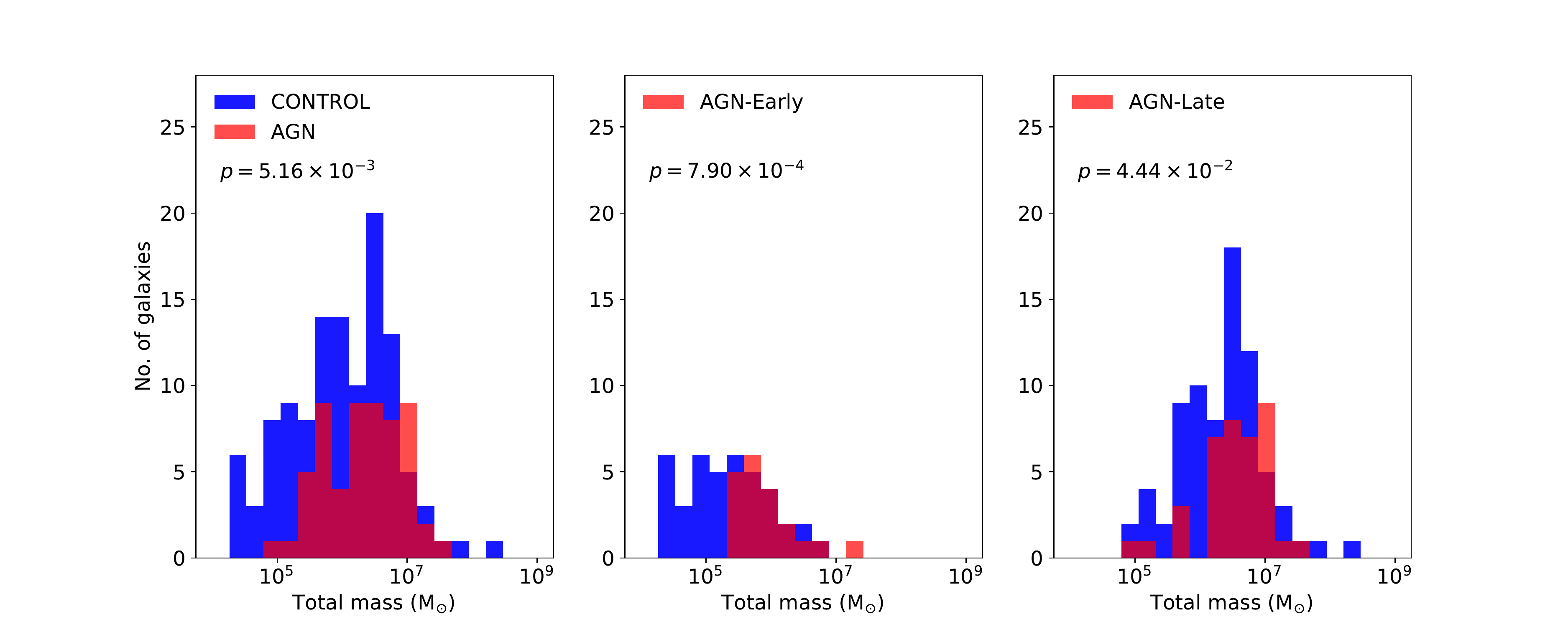}
    \caption{Histograms of the total mass of ionized gas; left: total sample; central: early-type sample; right: late-type sample.}
    \label{fig:TM}
\end{figure*}

\begin{figure*}
    \includegraphics[width=2.1\columnwidth]{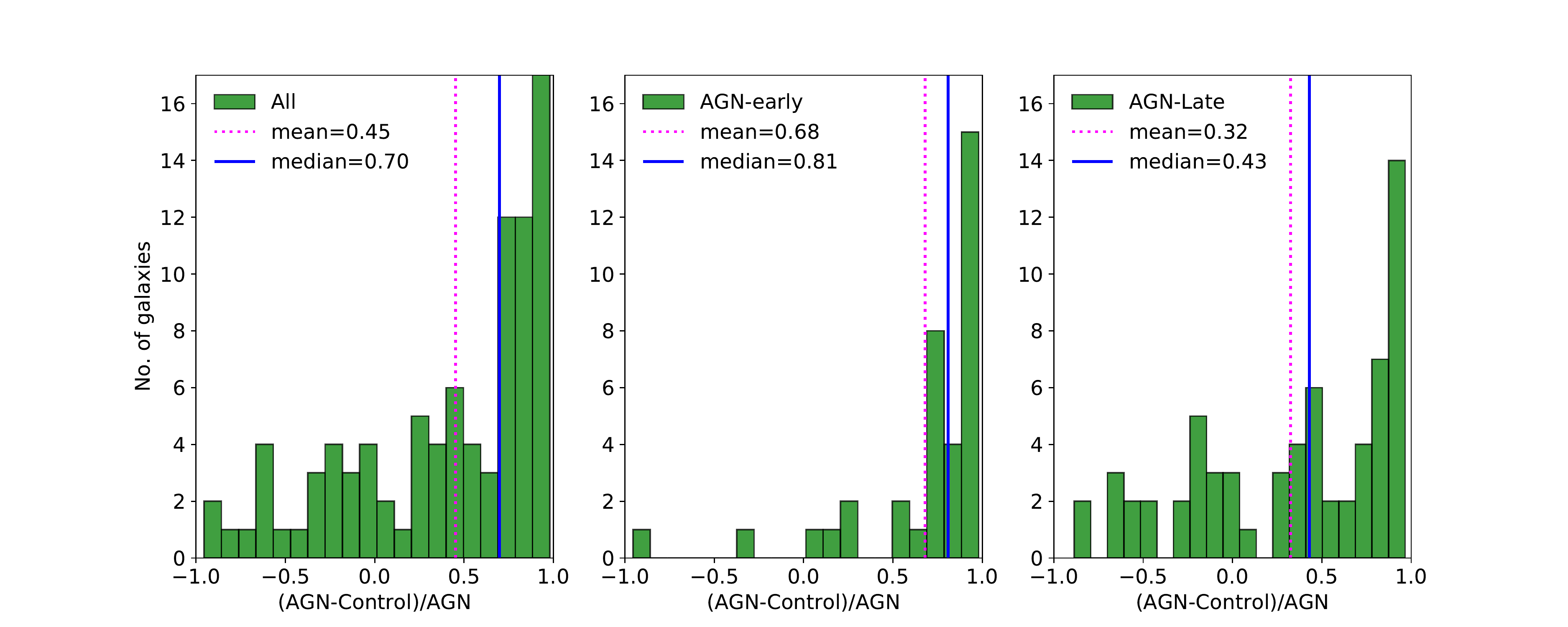}
    \caption{Histograms of the fractional differences $(M_{AGN}-M_{ctr})/M_{AGN}$; left: total sample; center: early-type sample; right: late-type sample.}
    \label{fig:TM_diff}
\end{figure*}

The total mass of ionized gas was obtained by integrating the whole H$\alpha$ emission from the galaxy. The corresponding values are shown in columns 3 and 7 of Table ~\ref{table1}, respectively, for the AGN and their controls.

The range of values for the total ionized gas mass are:
\noindent (1) For the total sample: $9.9\times10^{4} M_{\odot} - 2.9\times10^{7}\,M_{\odot}$ for the AGN, compared to $1.9\times10^{4} M_{\odot}$ - $1.7\times10^{8} M_{\odot}$ for the controls; 
\\
\noindent (2) For the early-type sub-sample: $2.2\times10^{5} M_{\odot} - 2.5\times10^{7} M_{\odot}$ for the AGN and $1.9\times10^{4} M_{\odot} - 6.3\times10^{6} M_{\odot}$ for the controls;
\\
(3) For the late-type sub-sample:  $9.9\times10^{4} M_{\odot} - 2.9\times10^{7} M_{\odot}$ for the AGN and  $9.3\times10^{4} M_{\odot} - 1.7\times10^{8} M_{\odot} $ for the controls.
\\ 

The ionized gas mass values obtained for our AGN sample are consistent with those obtained in previous studies of similar AGN hosts, such as \citet{H14}, where the total mass for a sample of 16 AGN, inside the inner 6--16\,kpc was found in the range (2-40)$\times10^{7}$ $M_{\odot}$ under the assumption that $n_{e}=100$ cm$^{-3}$ (and lower for $n_{e}=500$).
In \citet{G13} and \citet {G17} the total ionized gas mass obtained for a sample of radio galaxies ranges between 3.1$\times10^{5}$ $M_{\odot}$ and 4.1$\times10^{8}$ $M_{\odot}$, also consistent with the range of values that we obtained in this work.

The ranges of total ionized gas masses obtained for the AGN are similar to the ranges obtained for the controls, and this can be seen in the histograms in Fig.~\ref{fig:TM}. But the distribution of values is different for the AGN and control samples, as indicated by the A-D test that gives a probability larger than 95\%  that the distributions are not drawn from the same parent population. 

Calculating again the fractional differences of total ionized gas masses between each pair of AGN and control $(M_{AGN}-M_{ctr})/M_{AGN}$ (112 combinations), we obtain the results shown in the histograms of Fig.~\ref{fig:TM_diff} for fractional differences between $-$1 and 1. Again, as in the case of $\Sigma_{nuc}$, there are a few cases of values lower than $-$1, which we have excluded from the plot. But, even when considering these cases, $\approx$\,66\% of the total sample shows positive values, with $\approx$\,85\% for the early-type hosts and $\approx$\,57\% for the late-type hosts. The median values are: 0.70 for the total sample, 0.81 for the early-types and 0.43 for the late-types. Thus, even though the ranges of ionized gas masses of the AGN present large overlap with those of the control galaxies, when paired according to the host galaxy properties, most AGN show an excess of ionised gas mass (values larger than 0 in the histograms of Fig.~\ref{fig:TM_diff}) relative to their control pairs. 

In terms of AGN luminosity, the ranges of obtained total ionised gas masses are: (1) for the high-luminosity AGN, $5.2\times10^5 - 6.9\times10^{7} M_{\odot}$, which is skewed to larger values than the corresponding range for the controls of $6.2\times10^{4} - 2.2\times10^{7}$, while for the lower-luminosity AGN the range is $9.9\times10^{4} M_{\odot} - 2.9\times10^{7} M_{\odot}$ and $1.9\times10^{4} M_{\odot} - 1.7\times10^{8} M_{\odot}$ for the controls, thus showing complete overlap in the values.

\subsection{Average gradients of $\Sigma_{H^+}$}

\begin{figure*}
    \centering
    \includegraphics[width=2.1\columnwidth]{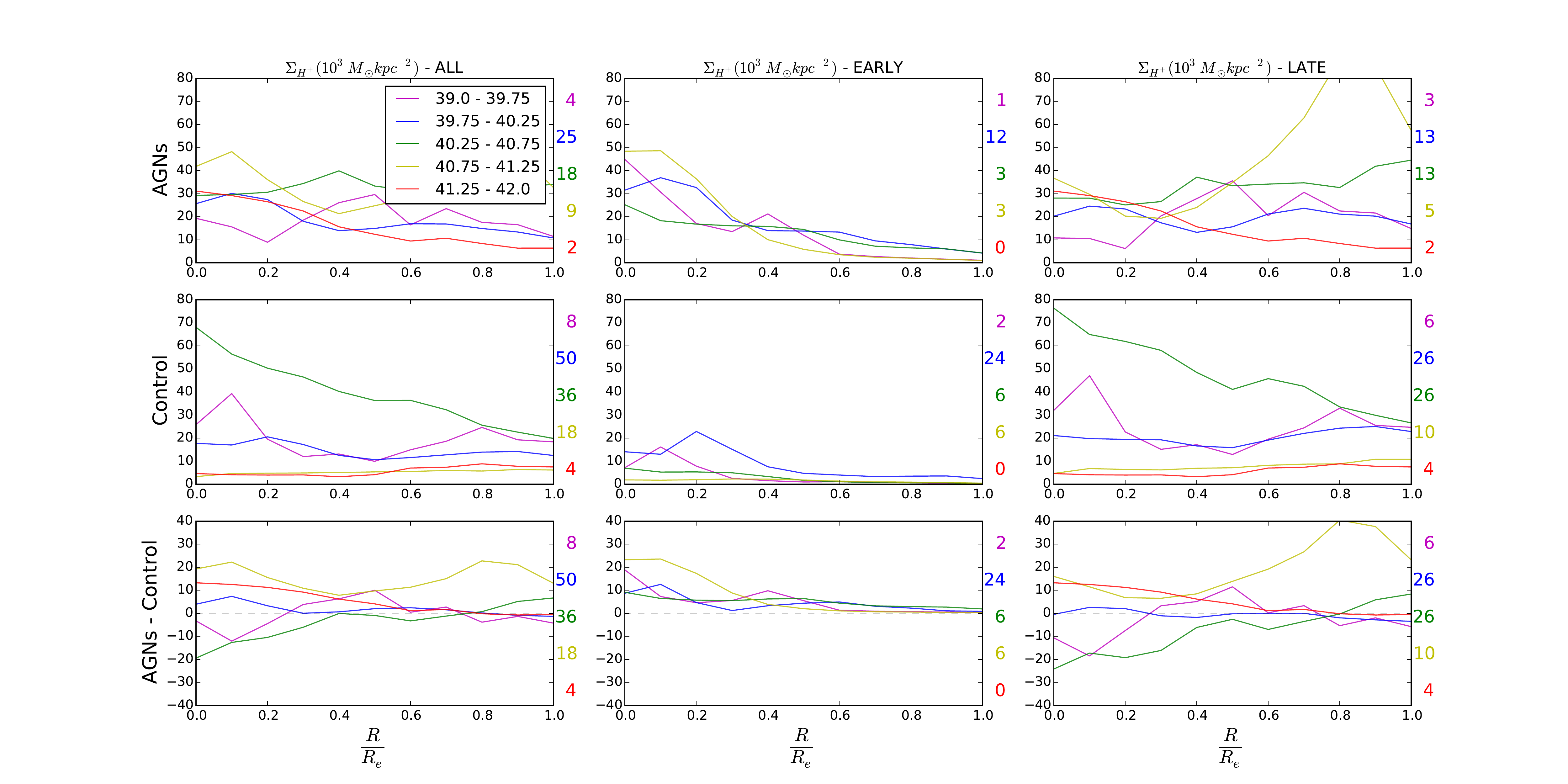}
    \caption{Average profiles of the ionized gas surface mass densities (in units of $10^3 M_\odot kpc^{-2}$) of groups of AGN separated according to the AGN luminosity (according to the insert in the top left panel). Top panels, from left to rigth: total AGN sample, early-type AGN, late-type AGN; middle panels: control sample corresponding to each top panel; bottom panels: difference between each AGN and control sample group.}
    \label{fig:prof}
\end{figure*}

We have used the azimuthaly-averaged profiles of $\Sigma_{H^+}$ of each galaxy to calculate the average profile of groups of AGN separated according to their luminosity, as well as of their control galaxies, following the methodology described in \citet{Mallmann18}. 
In Fig.~\ref{fig:prof} we show these for each of the following AGN luminosity bins of log(L[OIII]), with L[OIII] in erg\,s$^{-1}$: 41.25 -- 42.0 (shown in red), 40.75 -- 41.25 (light green), 40.25 -- 40.75 (dark green), 39.75 -- 40.25 (blue), 39.0 -- 39.75 (purple).

In this figure, the left column of panels shows the average profiles for the total sample, the central column shows the results for the the early-type sub-sample and in the right column, for the late-type sub-sample.
The first line of panels show the AGN profiles, the middle shows the corresponding control galaxies and the bottom shows the difference between AGN and controls.

Considering the whole sample, there is a tendency for the higher luminosity AGN to show larger values of gas surface mass density than the control galaxies from the nucleus up to about 0.4\,R$_{e}$; beyond this radius there is no difference, what can be understood as due to the fact that the presence of a luminous AGN requires more gas than in non-AGN, combined with the presence of a nuclear source of radiation, what is restricted to the inner regions of the galaxy. As the AGN luminosity decreases,
the difference becomes smaller, and in the case of the lowest luminosity bin, the control galaxies have even more ionized gas than the AGN.

The central and right panels of Fig.~\ref{fig:prof} show that the tendency discussed above seems to hold both for the early and late-type galaxies, even though in our sample we do not have an early-type AGN in the highest luminosity bin.

\section{Summary and conclusions}

We have mapped the gas excitation, star-formation rate SFR and ionized gas mass distributions of 62 AGN host galaxies as compared with a control sample of 112 galaxies using MaNGA-SDSS-IV datacubes. The sample comprises 20 early-type, 38 late-type galaxies and 4 galaxies in interaction with a companion. In terms of the AGN luminosity, 17 have $L(\rm{[OIII]})\lambda 5007 > 3.8\times 10^{40}\,\mbox{erg}\ \mbox {s}^{-1}$, that we have called high-luminosity, and 45 have lower luminosity than this (called low-luminosity AGN). The results of our measurements and analysis are summarized below.

\begin{itemize}

\item Nuclear excitation:
We have mapped the gas excitation using the BPT and WHAN diagrams. Defining the nuclear region as having a radius $\le$0.2\,R$_e$ (effective radius), 2/3 of the high-luminosity AGN have nuclear Seyfert (Sy) excitation, and the rest have either LINER or mixed Sy/LINER excitation. Of the 45 low-luminosity AGN, 30\% have Sy, 40\% LINER and 30\% mixed nuclear excitation. Regarding the controls, using only the BPT diagram, many nuclear regions result classified as AGN -- especially in the early-type hosts, but the WHAN diagram reveals that most of these cases are LIERs (excited by evolved hot stars). In the late-type galaxies, besides LIERs there are also a number of controls with starburst nuclei. 

\item Extra-nuclear excitation:
We have measured the extent $R$ of the NLR defined as the region surrounding the nucleus with AGN excitation, obtaining values ranging from 1.5 to 27\,kpc, and found the relation $R\propto\,(L(OIII])^{0.5}$, similar to that valid for the Broad-Line Region and recently obtained also for the NLR of luminous AGN by \citet{sb18} and previously by \citet{bennert02}. 

Both in AGN and controls there are some extra-nuclear regions (detached from the nucleus) that are classified as AGN according to the BPT diagram; using also the WHAN diagram, most of these regions become LIERs, although the emission of some of them could be signatures of warm ionized gas (seen mostly in highly inclined galaxies) and some could also be signatures of ``relic AGN".

\item Star formation rate:
The SFR surface density $\Sigma_{SFR}$ was obtained in the region of the galaxies dominated by star formation, mostly in the galaxy disks. The $\Sigma_{SFR}$ maps are similar for the AGN and controls. Some control galaxies show star formation also in the nuclear region, reaching the highest values there in the case of late-type galaxies. 

The total SFR of both the AGN and control galaxies range from 10$^{-3}$ to 10 \,M$_\odot$\,yr$^{-1}$, with a difference between AGN and controls seen only for the early-type galaxies: 75\% of the early-type AGN have higher SFR than their controls. This result suggests that the gas that is feeding the AGN also triggers star-formation in the outer parts of the galaxy, and that the AGN is not quenching star-formation there. As we do not see differences in SFR between AGN and controls for the late-type galaxies, we do not see signature of quenching in the disks of late-type galaxies as well.

\item Surface mass densities of ionized gas $\Sigma_{H^+}$:
Within 0.2\,R$_{e}$, $\Sigma_{H^+}=\Sigma_{nuc}$ is larger for the AGN than for the control galaxies in 75\% of the sample, with median values of $1.74\times10^4$\,M$_\odot$\,kpc$^{-2}$ for the AGN and $2.72\times10^3$\,M$_\odot$\,kpc$^{-2}$ for the controls.
The con-trol galaxies of the early types usually have higher
$\Sigma_{H^+}$ values in the outer regions than in their nuclei. In the case of late-type hosts, the regions of highest $\Sigma_{H^+}$ are more extended (throughout the galaxy disks) than in the early-types, with similar values in the galaxy disks for the AGN and controls, while in the nucleus, it is higher in the AGN for half of the sample and lower in the other half.

\item Gradients of $\Sigma_{H^+}$:
Average gradients constructed from the above $\Sigma_{H^+}$ maps are usually steeper for the AGN than for the controls from the center to about 0.4\,R$_{e}$, with the difference decreasing as the AGN luminosity decreases.

\item Total ionized gas masses:
We have obtained lower limits for the integrated ionized gas masses, with the total values estimated to be 10\%--40\% higher, depending on the galaxy. The lower limits 
range from $9.9\times10^{4}\,M_{\odot}$ to $2.9\times10^{7}\,M_{\odot}$ for the AGN, compared to the range of $1.9\,\times10^{4}\,M_{\odot}$ to $1.7\times10^{8}\,M_{\odot}$ for the controls. When pairing each AGN with its controls, for the early-type hosts, 85\% of the AGN have more ionized gas than the controls, while this percentage decreases to 57\% for the late-type hosts.

\end{itemize}

In summary, the main differences between AGN and controls is observed for early-type galaxies, that have: (1) higher SFR than the controls in 75\% of the cases; (2) larger ionized gas masses than the controls in 85\% of the cases. This suggests a connection between the star-formation and nuclear activity in the early-type galaxies: the same gas that feeds the AGN is also triggering star formation in its vicinity. And these results also do not support that AGN feedback is quenching star formation for the luminosities probed by this nearby sample.

Another important result is the increase of the extent of the region excited by the AGN with its luminosity, reaching the galaxy limits for the highest luminosities probed here, of log(L[OIII])=41.8, and suggesting that this radiation will escape the galaxy for higher luminosities.

\section*{Acknowledgements}
JCN thanks to CNPq for financial support. TSB, RAR, RR and SR acknowledge the support of the Brazilian funding agencies FAPERGS and CNPq. We would like to thank the support of the Instituto Nacional de Ci\^encia e Tecnologia (INCT) e-Universe project (CNPq grant 465376/2014-2).

Funding for the Sloan Digital Sky Survey IV has been provided by the Alfred P. Sloan Foundation, the U.S. Department of Energy Office of Science, and the Participating Institutions. SDSS acknowledges support and resources from the Center for High-Performance Computing at the University of Utah. The SDSS web site is www.sdss.org.

SDSS is managed by the Astrophysical Research Consortium for the Participating Institutions of the SDSS Collaboration including the Brazilian Participation Group, the Carnegie Institution for Science, Carnegie Mellon University, the Chilean Participation Group, the French Participation Group, Harvard-Smithsonian Center for Astrophysics, Instituto de Astrof\'isica de Canarias, The Johns Hopkins University, Kavli Institute for the Physics and Mathematics of the Universe (IPMU) / University of Tokyo, Lawrence Berkeley National Laboratory, Leibniz Institut f\"ur Astrophysik Potsdam (AIP), Max-Planck-Institut f\"ur Astronomie (MPIA Heidelberg), Max-Planck-Institut f\"ur Astrophysik (MPA Garching), Max-Planck-Institut f\"ur Extraterrestrische Physik (MPE), National Astronomical Observatories of China, New Mexico State University, New York University, University of Notre Dame, Observat\'orio Nacional / MCTI, The Ohio State University, Pennsylvania State University, Shanghai Astronomical Observatory, United Kingdom Participation Group, Universidad Nacional Aut\'onoma de M\'exico, University of Arizona, University of Colorado Boulder, University of Oxford, University of Portsmouth, University of Utah, University of Virginia, University of Washington, University of Wisconsin, Vanderbilt University, and Yale University.





\begin{thebibliography}{99}
\bibitem[\protect\citeauthoryear{Abolfathi et al.}{2018}]{dr14} Abolfathi, B., 2018, ApJS, 235, 42.
\bibitem[\protect\citeauthoryear{Baldwin, Phillips \& Terlevich }{1981}]{bpt81}
Baldwin, J.A., Phillips, M.M., Terlevich, R., 1981, PASP,93, 5
\bibitem[\protect\citeauthoryear{Belfiore et al.}{2015}]{B15}
Belfiore, F., Maiolino, R., Bundy, K., et al.\ 2015, \mnras, 449, 867 
\bibitem[\protect\citeauthoryear{Bennert et al.}{2002}]{bennert02} Bennert, N., Falcke, H., Schulz, H., Wilson, A. S. \& Wills, B. J., 2002,
\apj, 574, L105
\bibitem[\protect\citeauthoryear{Blanton et al.}{2017}]{Blanton+17} 
Blanton, M.~R., Bershady, M.~A., Abolfathi, B., et al., 2017, \aj, 154, 28 
\bibitem[\protect\citeauthoryear{Brinchmann, Kunth \& Durret}{2008}]{Br+08}
Brinchmann J., Kunth D., Durret F., 2008, A{\&A}, 485, 657
\bibitem[\protect\citeauthoryear{Brusa, et al.}{2016}]{Brusa+16}
Brusa M., et al., 2016, A{\&}A, 588, A58
\bibitem[\protect\citeauthoryear{Bundy et al.}{2015}]{bundy15}
Bundy, K., et al., 2015, ApJ 798, id. 7
\bibitem[\protect\citeauthoryear{Cardelli, Clayton \& Mathis}{1989}]{cardelli98}  
Cardelli J.~A., Clayton G.~C., Mathis J.~S., 1989, 
\bibitem[Carniani et al.(2016)]{Carniani+16}
Carniani, S., Marconi, A., Maiolino, R., et al.\ 2016, \aap, 591, A28 
\bibitem[\protect\citeauthoryear{Cid Fernandes et al.}{2010}]{Cid+10}
Cid Fernandes, R., Stasi{\'n}ska, G., Schlickmann, M. S., Mateus, A., Vale Asari, N., Schoenell, W., Sodrhoenell, W., Sodr\'e 2010, L.2010,\mnras, 403, 1036 
\bibitem[\protect\citeauthoryear{Couto et al.}{2013}]{G13}
Couto, G.~S., Storchi-Bergmann, T., Axon, D.~J., et al.\ 2013, \mnras, 435, 2982 
\bibitem[\protect\citeauthoryear{Couto et al.}{2017}]{G17}
Couto, G.~S., Storchi-Bergmann, T., \& Schnorr-M{\"u}ller, A.\ 2017, \mnras, 469, 1573
\bibitem[\protect\citeauthoryear{Diamond-Stanic \& Rieke} {2012}]{ds12} Diamond-Stanic, A. \& Rieke, G. H., 2012, \apj, 746, 168
\bibitem[\protect\citeauthoryear{Drory et al.}{2015}]{Drory+15}
Drory, N., MacDonald, N., Bershady, M.~A., et al.\ 2015, \aj, 149, 77 
\bibitem[\protect\citeauthoryear{Esquej et al.}{2014}]{Esquej+14} Esquej, P. et al. 2014, \apj, 780, 86
\bibitem[\protect\citeauthoryear{Greene et al.}{2011}]{greene11},
Greene, J. E., Zakamska, N. L., Ho, L. C., \& Barth, A. J. 2011, \apj, 732, 9
\bibitem[\protect\citeauthoryear{Gunn et al.}{2006}]{Gunn+06}
Gunn, J.~E., Siegmund, W.~A., Mannery, E.~J., et al.\ 2006, \aj, 131, 2332 
\bibitem[\protect\citeauthoryear{Ilha et al.}{2018}]{Ilha18}
Ilha, G. S., et al., 2018, \mnras, 484, 252
\bibitem[\protect\citeauthoryear{Hainline et al.}{2013}]{hainline+13} Hainline, K. N, Hickox, R., Greene, J., Myers, A. D. \& Zakamska4, N. L. 2013, \apj, 774, 145
\bibitem[\protect\citeauthoryear{Harrison et al.}{2014}]{H14}
Harrison, C.~M., Alexander, D.~M., Mullaney, J.~R., \& Swinbank, A.~M.\ 2014, \mnras, 441, 3306 
\bibitem[\protect\citeauthoryear{Hicks et al.}{2013}]{Hicks13}
Hicks, E.~K.~S., Davies, R.~I., Maciejewski, W., et al.\ 2013, \apj, 768, 107
\bibitem [\protect\citeauthoryear{Kaspi et al.}{2005}]{kaspi05} Kaspi, S., Maoz, D., Netzer, H., Peterson, B. M., Vestergaard, M. \& Jannuzi, B. T., 2005, \apj, 629, 61
\bibitem [\protect\citeauthoryear{Kennicutt}{1998}]{Kennicutt}
Kennicutt, Jr., 1998,ARAA,36,189
\bibitem [\protect\citeauthoryear{Kauffmann et al.}{2003}]{kauf03}
Kauffmann G., Heckman T. M., Tremonti C., et al., 2003, MNRAS, 346, 1055
\bibitem [\protect\citeauthoryear{Kewley et al.}{2001}]{Kewley01}
Kewley L. J., Dopita M. A., Sutherland R. S., Heisler C. A.,Trevena J., 2001, ApJ, 556, 121
\bibitem [\protect\citeauthoryear{Kewley et al.}{2006}]{kewley06}
Kewley,  L.J.,  Groves,  B.,  Kauffmann,  G.,  Heckman,  T.,2006, MNRAS, 372, 961
\bibitem[\protect\citeauthoryear{Law et al.}{2015}]{Law+15}
Law, D.~R., Yan, R., Bershady, M.~A., et al.\ 2015, \aj, 150, 19 
\bibitem[\protect\citeauthoryear{Law et al.}{2016}]{Law+16}
Law, D.~R., Cherinka, B., Yan, R., et al.\ 2016, \aj, 152, 83 
\bibitem [\protect\citeauthoryear{Lintott et al.}{2011}]{L11}
Lintott, et al.,2011, MNRAS, 410,166.
\bibitem [\protect\citeauthoryear{Mallmann et al.}{2018}]{Mallmann18} Mallmann, N. et al., 2018, MNRAS, 478, 5491.
\bibitem[\protect\citeauthoryear{Liu et al.}{2013}]{liu+13}
Liu, G., Zakamska, N. L., Greene, J. E., Nesvadba,
N. P. H., \& Liu, X. 2013,\mnras, 436, 2576
\bibitem[\protect\citeauthoryear{Martini et al.}{2003}]{Martini03}
Martini, P., Regan, M.~W., Mulchaey, J.~S., \& Pogge, R.~W.\ 2003, \apj, 589, 774 
\bibitem[\protect\citeauthoryear{Mushotzky et al.}{2014}]{mushotzky+14} Mushotzky, R. F., Taro Shimizu, T., Mel\'endez, M. \& and Koss, M. 2014, \apj, L34
\bibitem[\protect\citeauthoryear{Osterbrock \& Ferland}{2006}]{of06}
Osterbrock,  D.E.,  Ferland,  G.J.,  1989,Astrophysics  of Gaseous  Nebulae and Active Galactic Nuclei,  2nd.  ed., University Science Books, California
\bibitem[\protect\citeauthoryear{Peterson}{1997}]{p97}
Peterson, B.M., 1997, An Introduction to Active Galactic Nuclei, Cambridge University Press, Cambridge
\bibitem[\protect\citeauthoryear{Peterson}{2014}]{peterson14}
Peterson B.~M., 2014, SSRv, 183, 253
\bibitem [\protect\citeauthoryear{Rembold et al.}{2017}]{Rembold17}
Rembold, S.~B., Shimoia, J.~S., Storchi-Bergmann, T., et al.\ 2017, \mnras, 472, 4382 
\bibitem[\protect\citeauthoryear{Rosario et al.}{2018}]{R18}
Rosario, D.~J., Burtscher, L., Davies, R.~I., et al.\ 2018, \mnras, 473, 5658 
\bibitem [\protect\citeauthoryear{Riffel et al.}{2017}]{Riffel18}
Riffel, R.~A., Storchi-Bergmann, T., Riffel, R., et al.\ 2018, \mnras, 474, 1373 
\bibitem [\protect\citeauthoryear{Sarzi et al.}{2006}]{Marc06}
Sarzi, M., Falc{\'o}n-Barroso, J., Davies, R.~L., et al.\ 2006, \mnras, 366, 1151 
\bibitem [\protect\citeauthoryear{Schmitt et al.}{2003a}]{Schmitt03a}
Schmitt, H. R., Donley, J. L., Antonucci, R. R. J.,Hutchings, J. B., \& Kinney, A. L. 2003, \apjs, 148, 327 
\bibitem [\protect\citeauthoryear{Schmitt et al.}{2003b}]{Schmitt03b}
Schmitt, H. R., Donley, J. L., Antonucci, R. R. J., et al., 2003, \apj, 597, 768
\bibitem[\protect\citeauthoryear {Somerville et al.}{2008}]{S+08}
Somerville, R.~S., Hopkins, P.~F., Cox, T.~J., Robertson, B.~E., \& Hernquist, L.\ 2008, \mnras, 391, 481 
\bibitem[\protect\citeauthoryear{Sim{\~o}es Lopes et al.}{2007}]{SL07}
Sim{\~o}es Lopes, R.~D., Storchi-Bergmann, T., de F{\'a}tima Saraiva, M., \& Martini, P.\ 2007, \apj, 655, 718
\bibitem[\protect\citeauthoryear{Storchi-Bergmann et al.}{1996}]{sb96}
Storchi-Bergmann, T., Rodriguez-Ardila, A., Schmitt, H. R., Wilson, A. S. \& Baldwin, J. A. 1996, \apj, 472, 83
\bibitem[\protect\citeauthoryear{Storchi-Bergmann et al.}{2018}]{sb18}
Storchi-Bergmann T., et al., 2018, \apj, 868, 14
\bibitem[\protect\citeauthoryear{Storchi-Bergmann \& Schnorr-M\"uller}{2019}]{SB19}
Storchi-Bergmann, T. \& Schnorr-M\"uller, A., 2019, Nature Astronomy, 3, 48
\bibitem[\protect\citeauthoryear{Veilleux \& Osterbrock}{1987}]{VB87}
Veilleux, S., \& Osterbrock, D.~E.\ 1987, \apjs, 63, 295
\bibitem[\protect\citeauthoryear{Wylezalek et al.}{2018}]{W18}
Wylezalek, D., Zakamska, N.~L., Greene, J.~E., et al.\ 2018, \mnras, 474, 1499 
\bibitem[\protect\citeauthoryear{Yan et al.}{2016}]{Yan+16}
Yan, R., Tremonti, C., Bershady, M.~A., et al.\ 2016, \aj, 151, 8 
\end{thebibliography}




\appendix

\section{AGN - Control Galaxies Images}

\begin{figure*}
   \includegraphics[width=2.1\columnwidth]{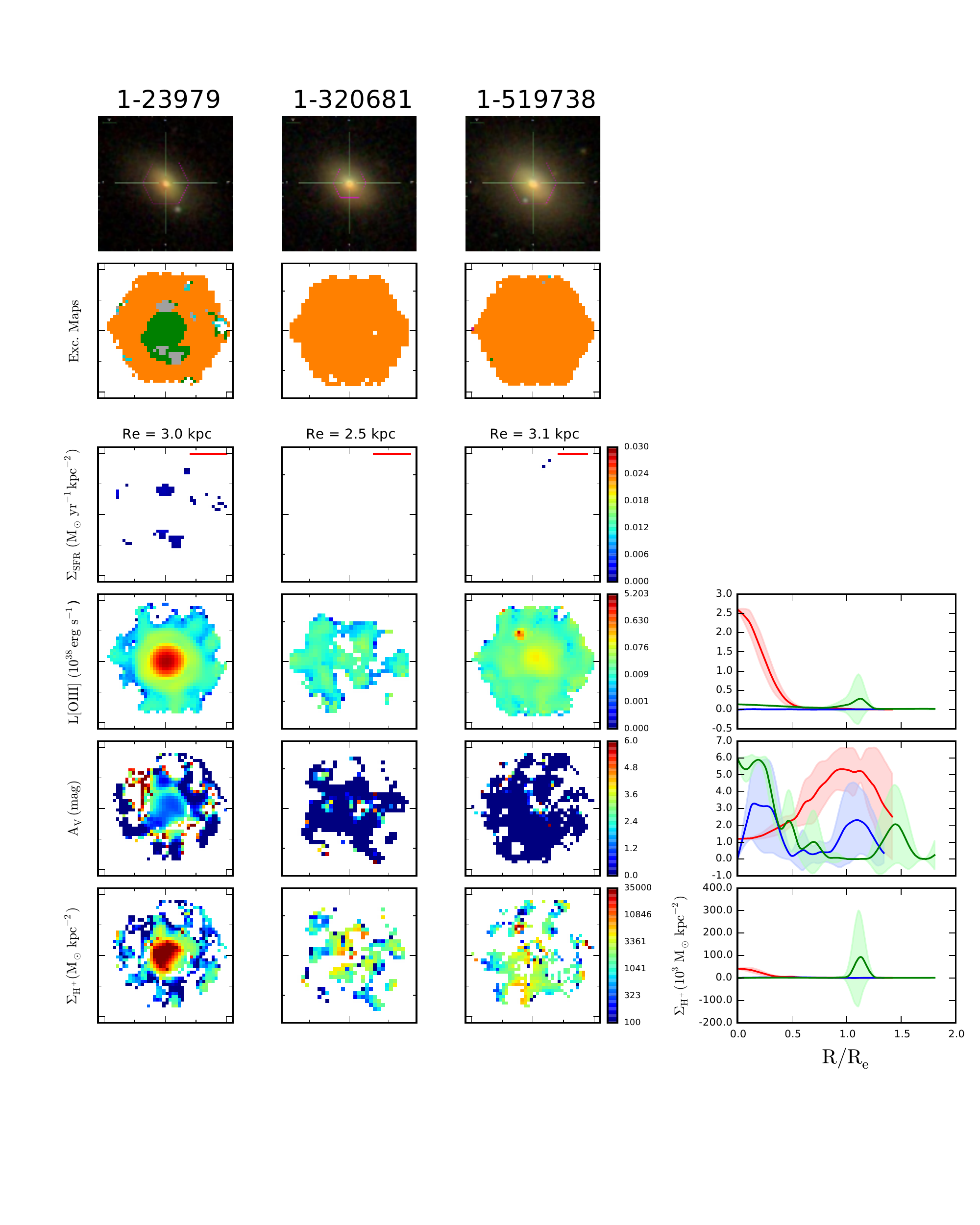}
   \vspace*{-30mm}
   \caption{ early-type and  weak-luminosity AGN
       }
   \label{fig:first_ap}
\end{figure*}

\begin{figure*}
   \includegraphics[width=2.1\columnwidth]{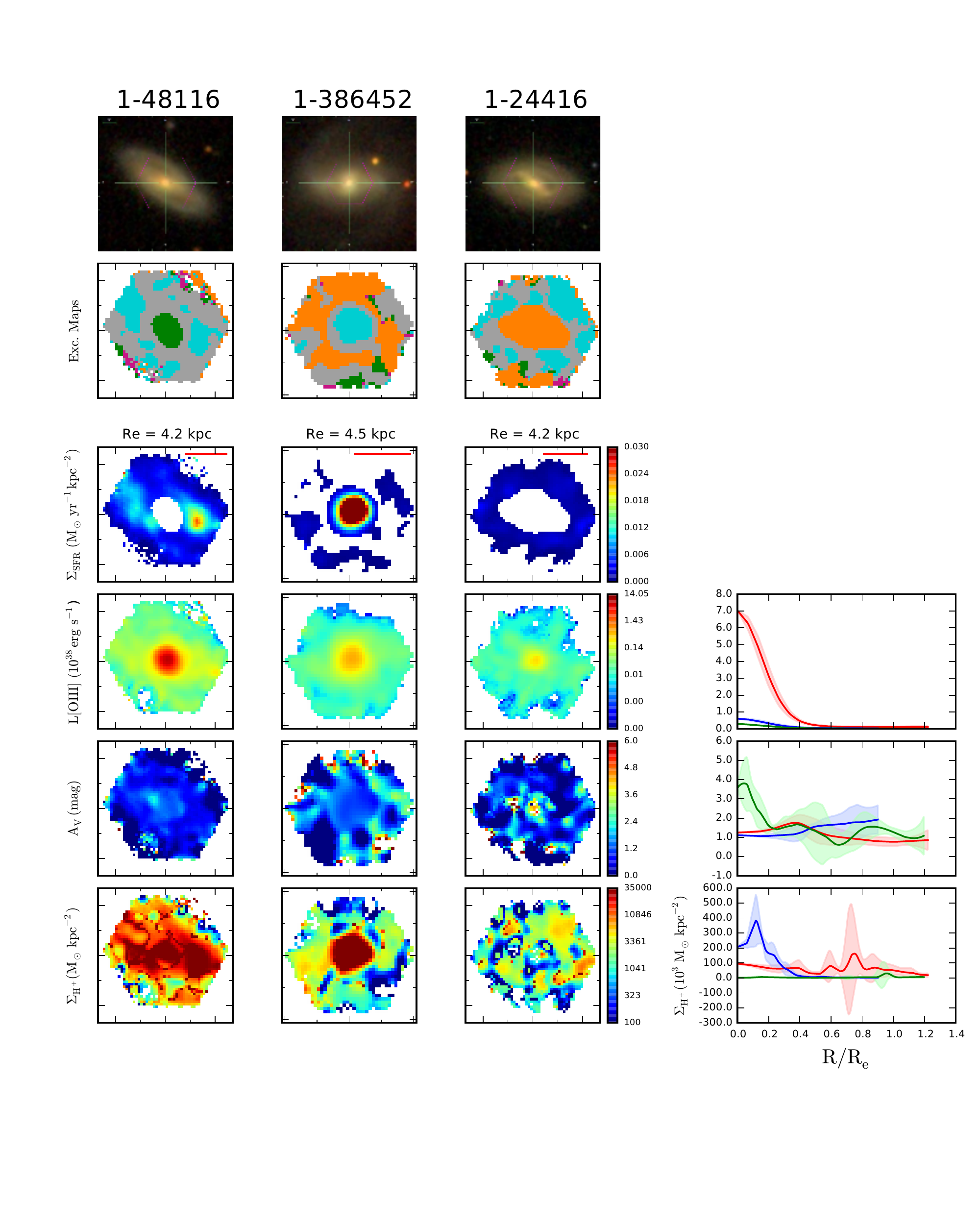}
   \vspace*{-30mm}
   \caption{late type and  weak-luminosity AGN
       }
 \label{fig:last_ap}
   \end{figure*}

We present all the comparisons between AGN and its control galaxies in Figs. \ref{fig:first_ap} -\ref{fig:last_ap}

\begin{figure*}
   \includegraphics[width=2.1\columnwidth]{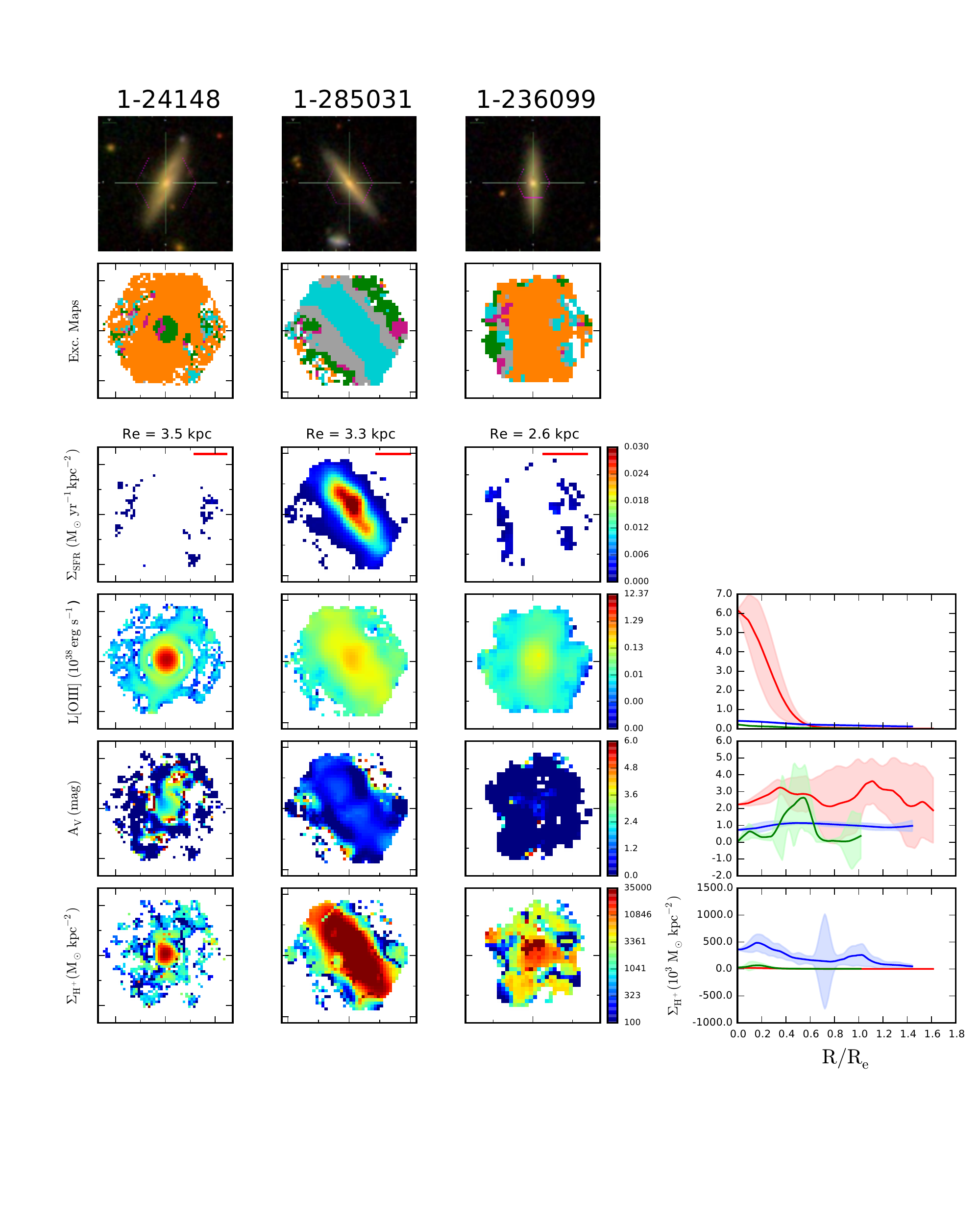}
   \vspace*{-30mm}
   \caption{ late type and  weak-luminosity AGN
         }
\end{figure*}

\begin{figure*}
   \includegraphics[width=2.1\columnwidth]{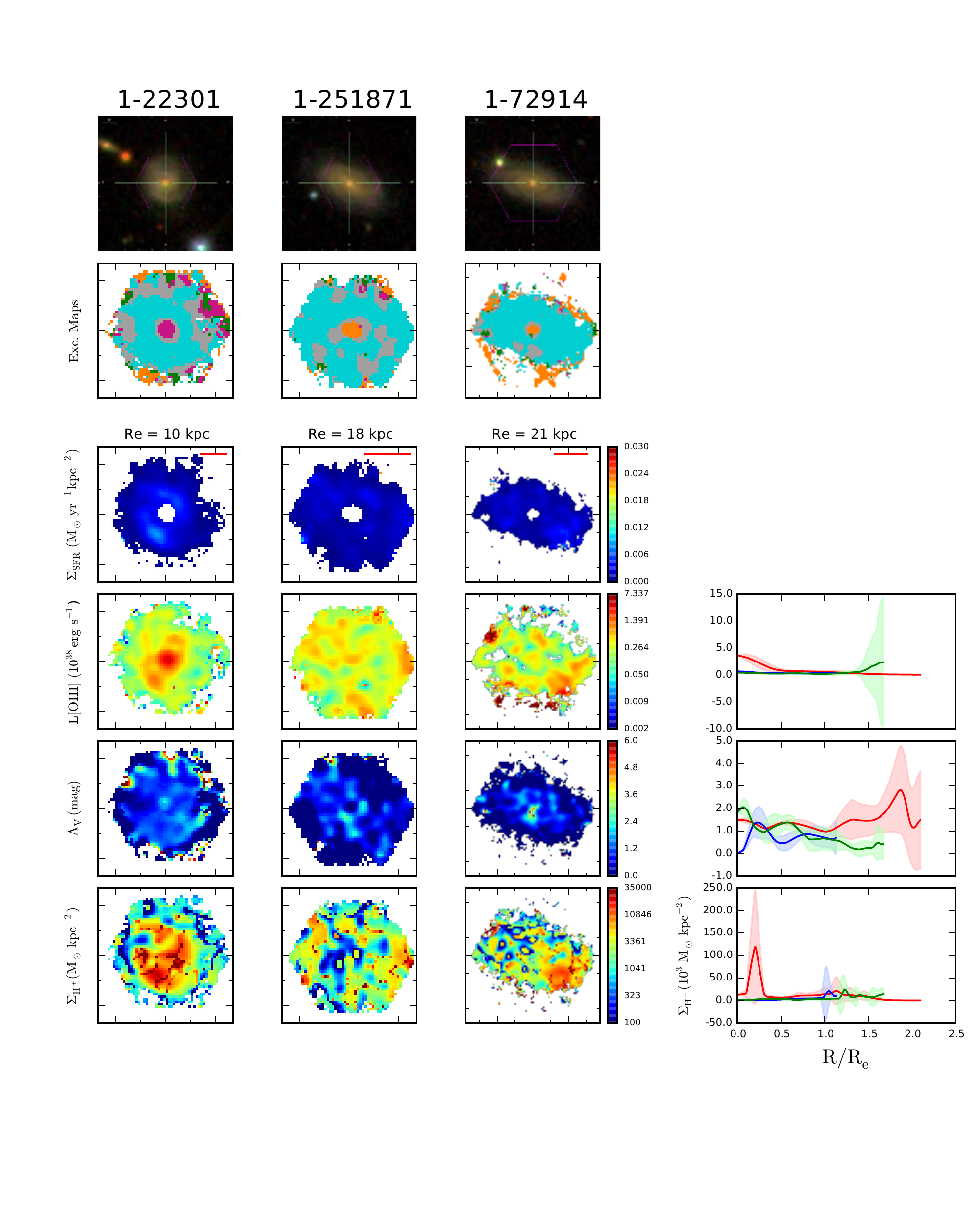}
   \vspace*{-30mm}
   \caption{ late type and  weak-luminosity AGN
        }
\end{figure*}

\begin{figure*}
   \includegraphics[width=2.1\columnwidth]{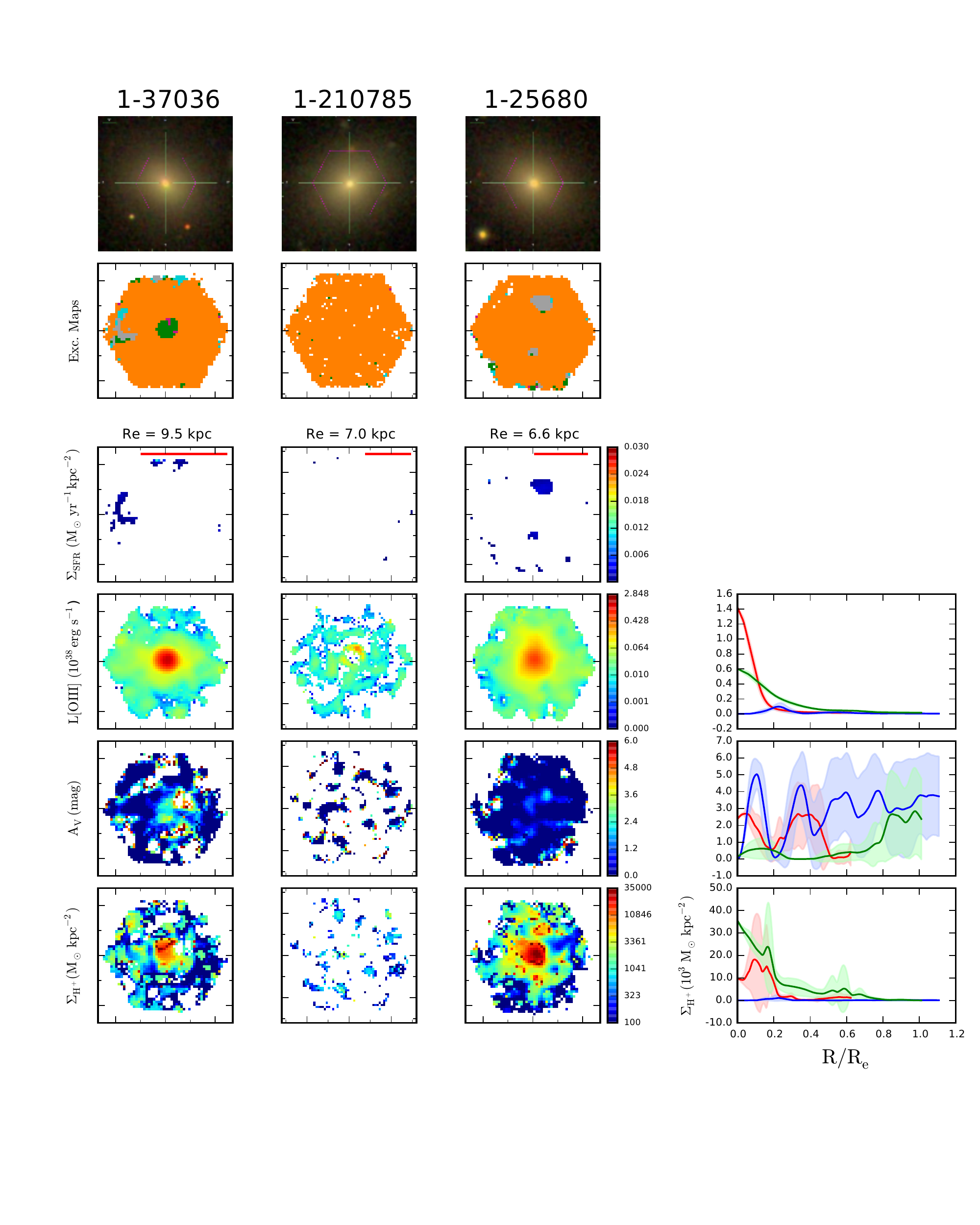}
   \vspace*{-30mm}
   \caption{ eatly-type and  weak-luminosity AGN
       }
\end{figure*}

\begin{figure*}
   \includegraphics[width=2.1\columnwidth]{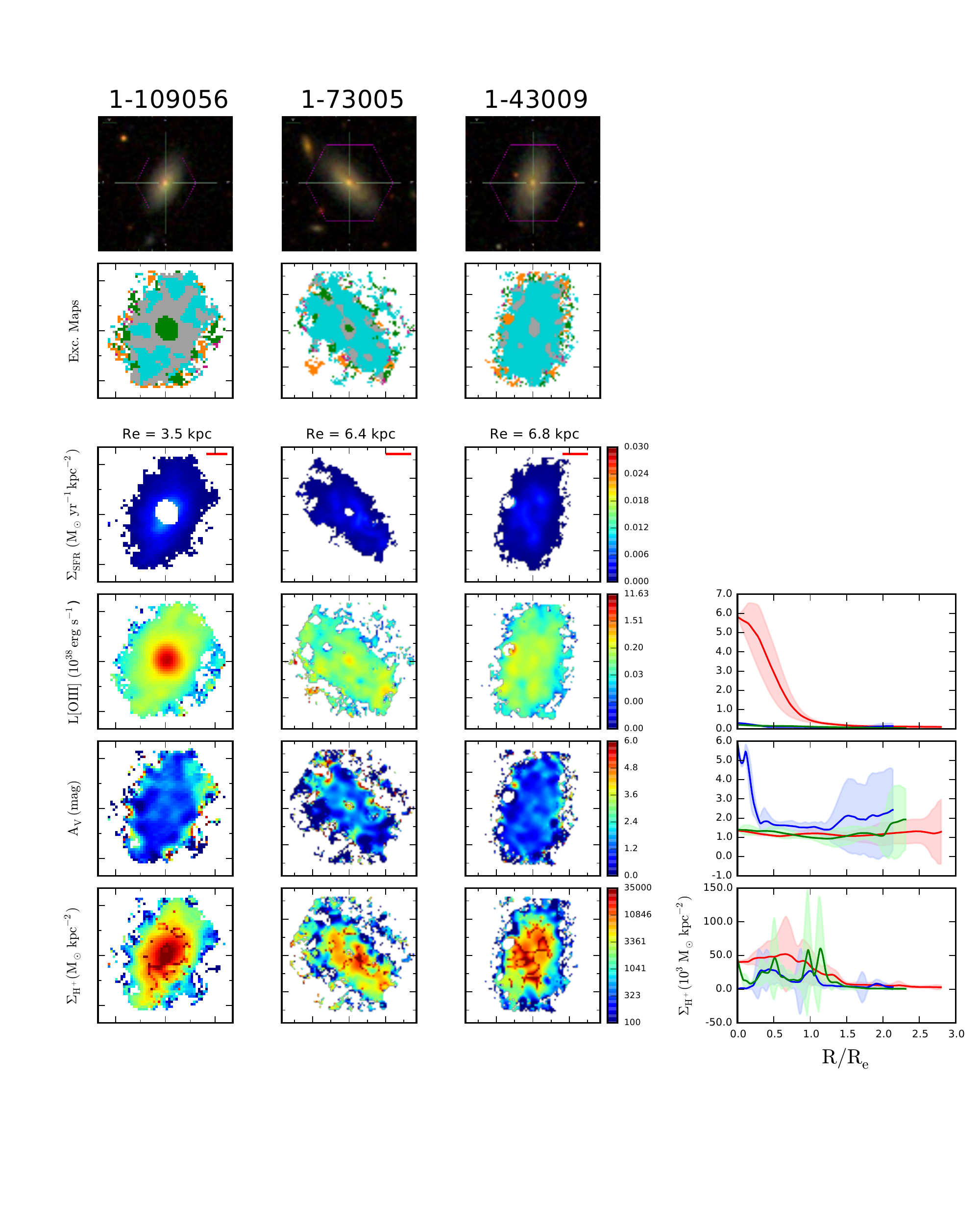}
   \vspace*{-30mm}
   \caption{ late type and  weak-luminosity AGN
       }
\end{figure*}

\begin{figure*}
   \includegraphics[width=2.1\columnwidth]{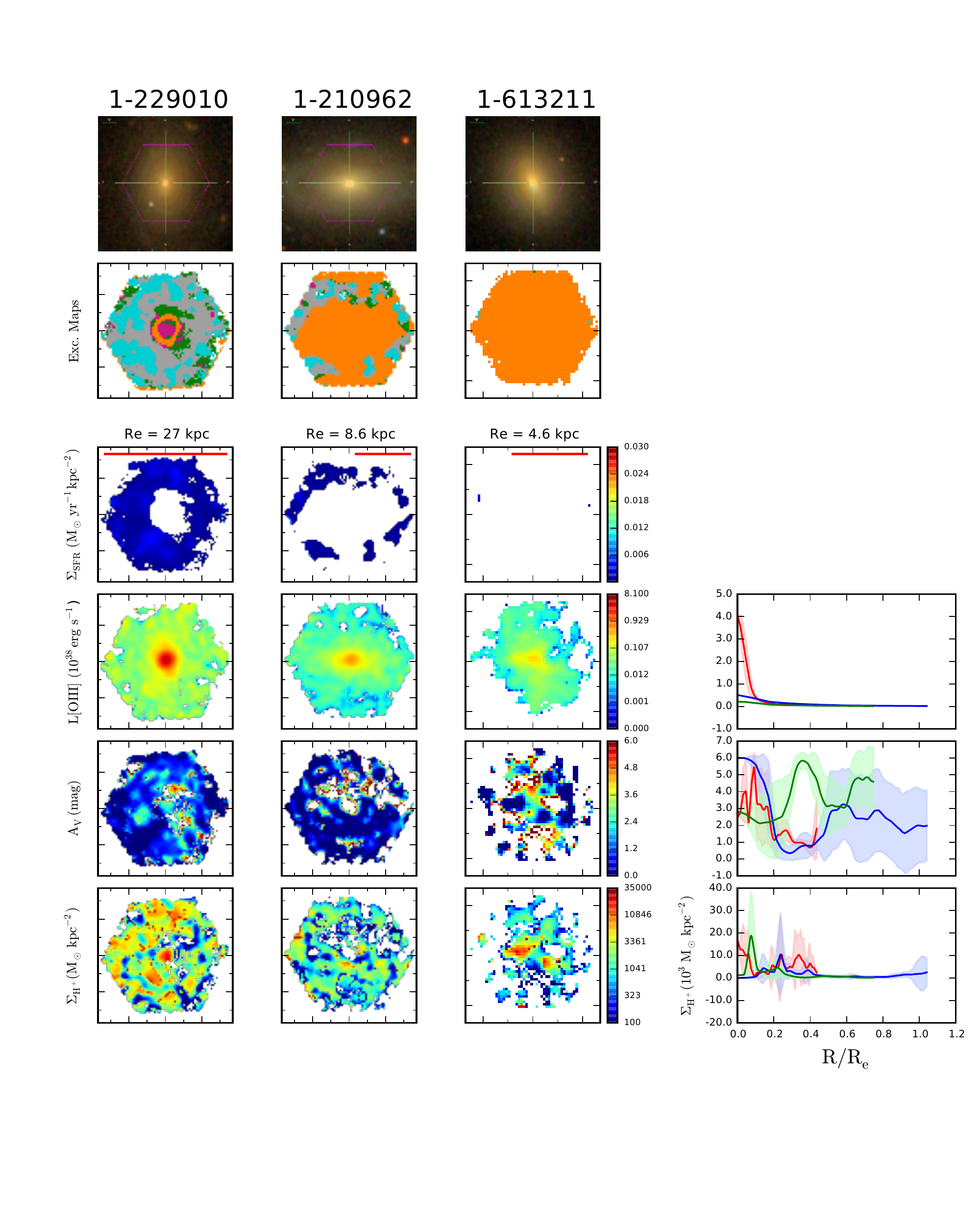}
   \vspace*{-30mm}
   \caption{ late type and  weak-luminosity AGN
      }
\end{figure*}

\begin{figure*}
   \includegraphics[width=2.1\columnwidth]{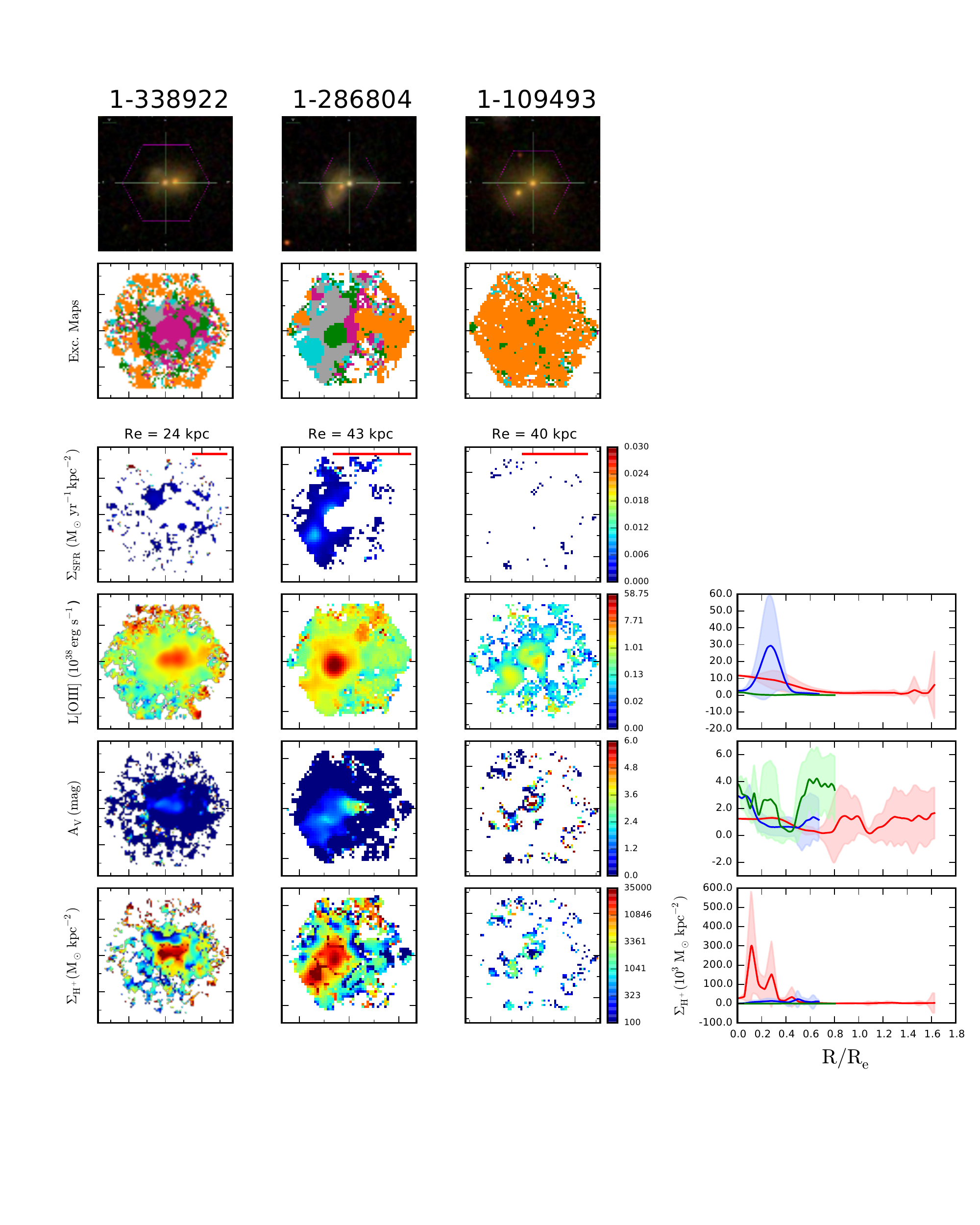}
   \vspace*{-30mm}
   \caption{ merger and strong-luminosity AGN     
   }
\end{figure*}

\begin{figure*}
   \includegraphics[width=2.1\columnwidth]{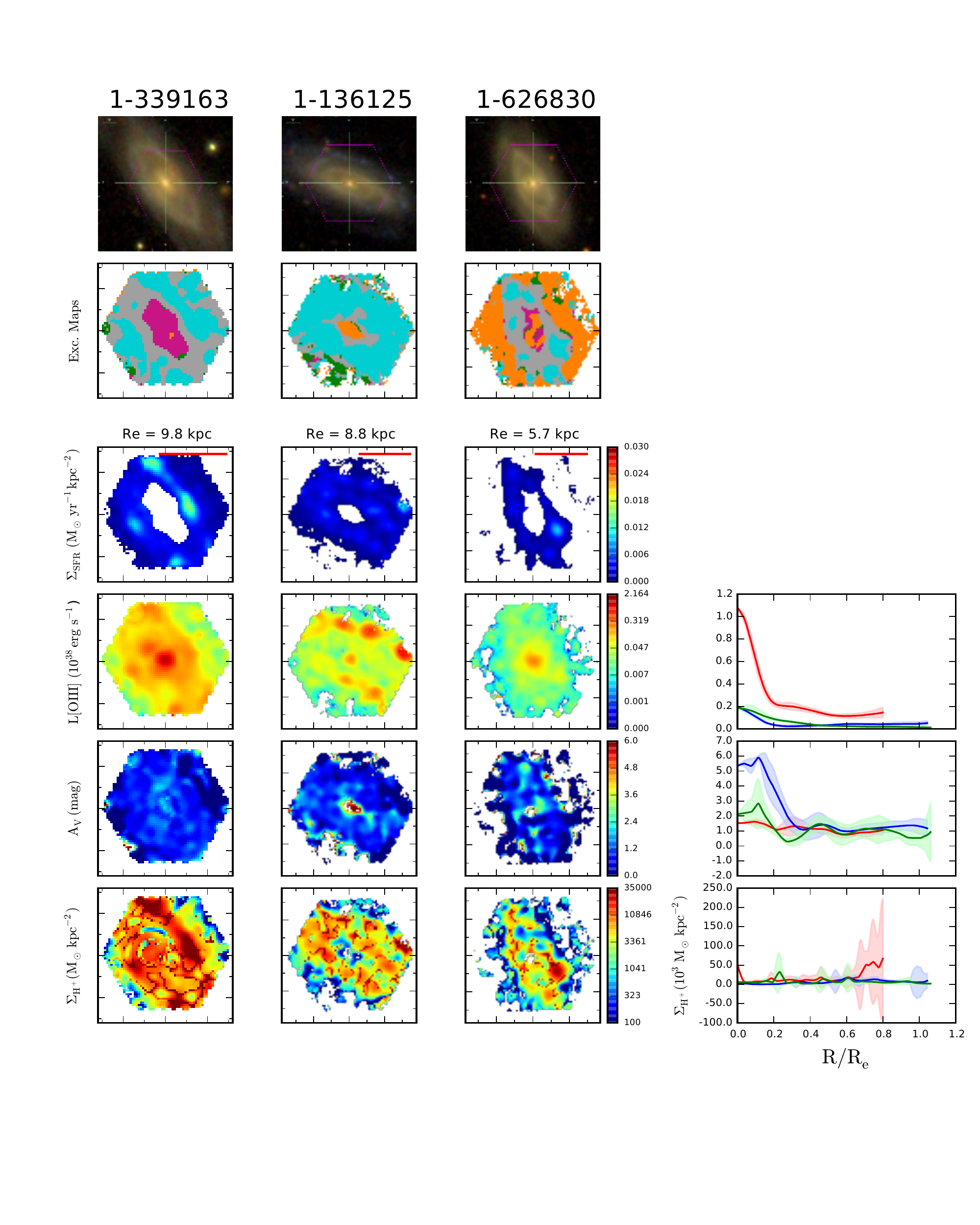}
   \vspace*{-30mm}
   \caption{ late type and  weak-luminosity AGN     
   }
\end{figure*}

\begin{figure*}
   \includegraphics[width=2.1\columnwidth]{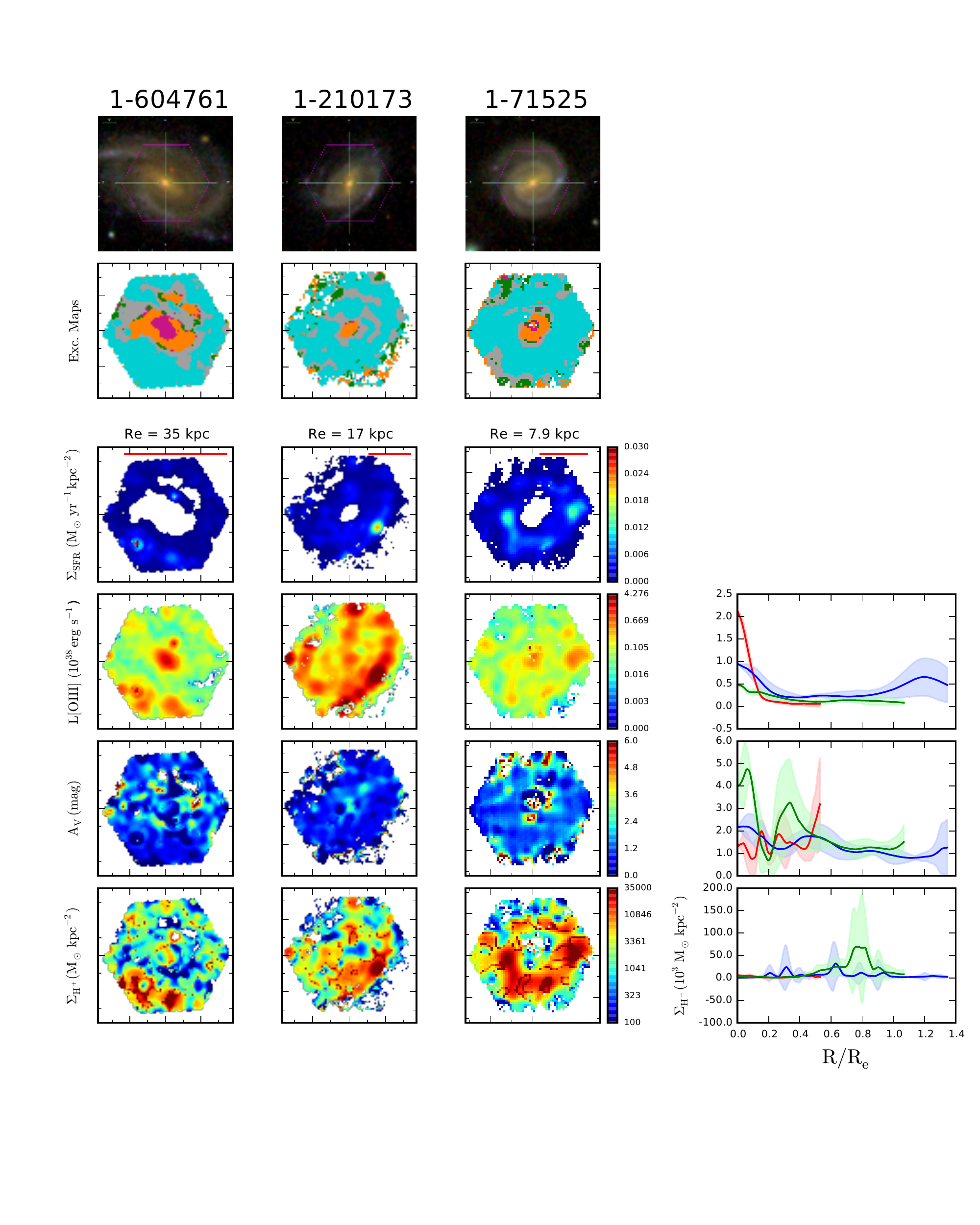}
   \vspace*{-30mm}
   \caption{ late type and  weak-luminosity AGN    
   }
\end{figure*}

\begin{figure*}
   \includegraphics[width=2.1\columnwidth]{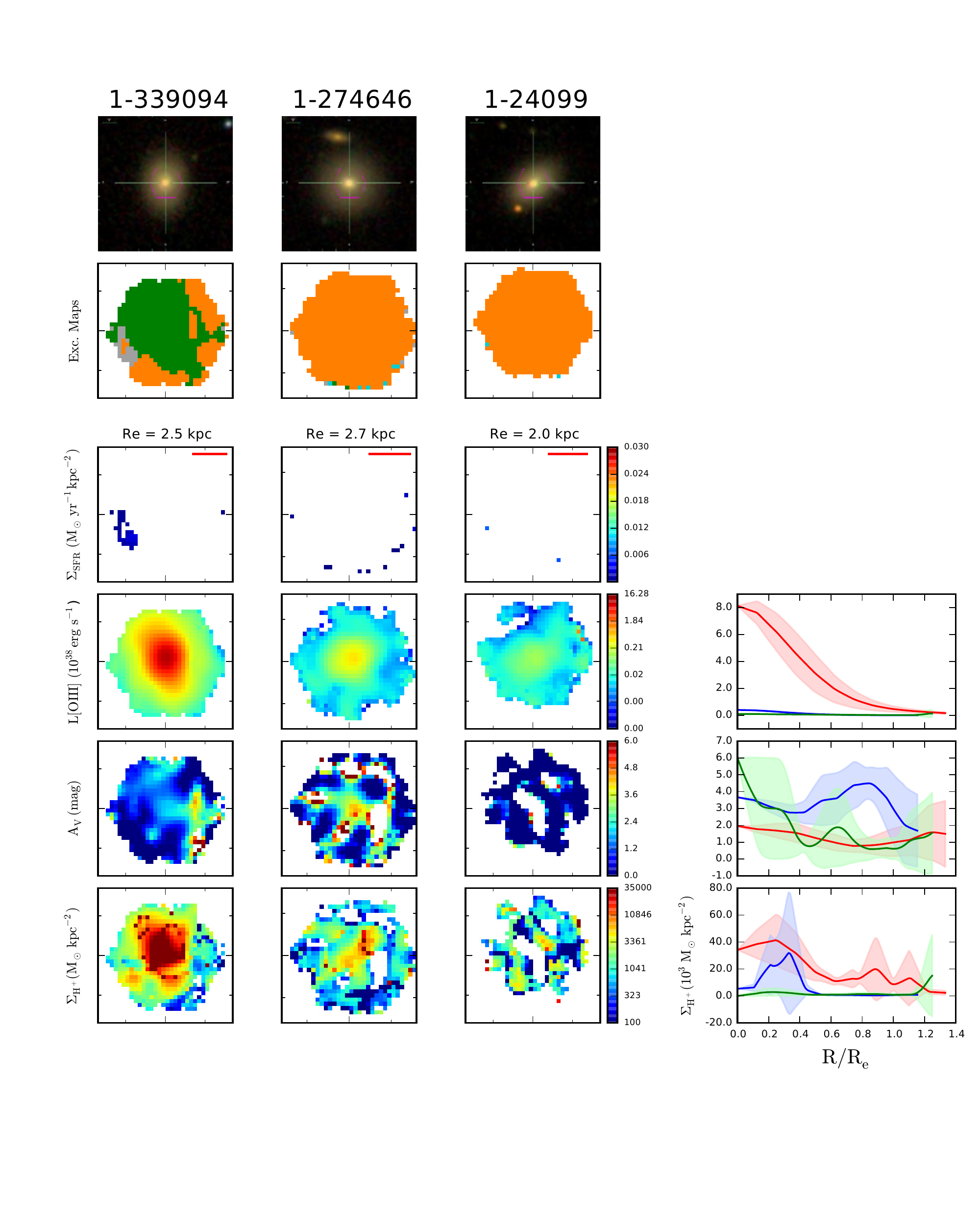}
   \vspace*{-30mm}
   \caption{ early-type and strong-luminosity AGN     
   }
\end{figure*}

\begin{figure*}
   \includegraphics[width=2.1\columnwidth]{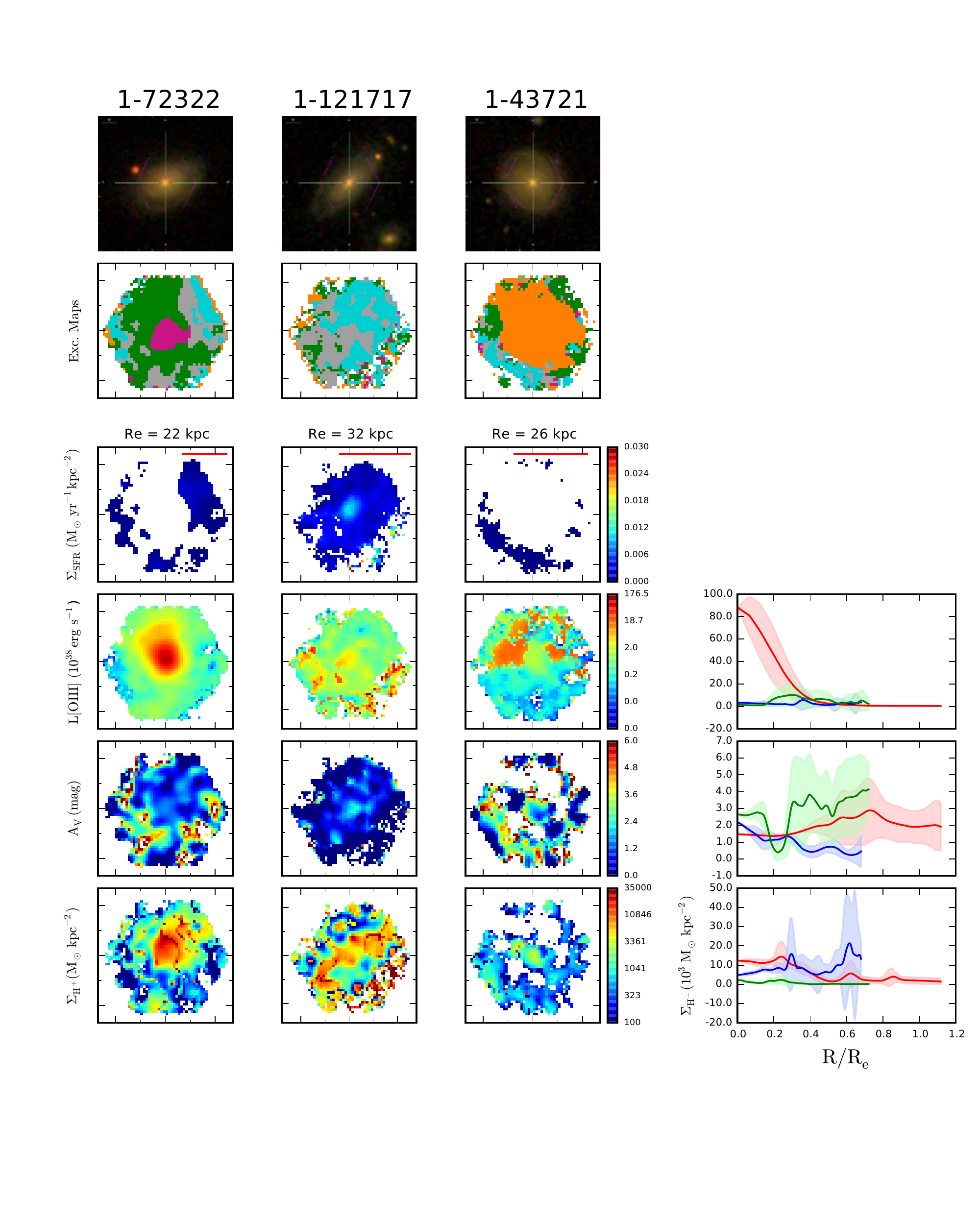}
   \vspace*{-30mm}
   \caption{ late-type and strong-luminosity AGN
   }
\end{figure*}

\begin{figure*}
   \includegraphics[width=2.1\columnwidth]{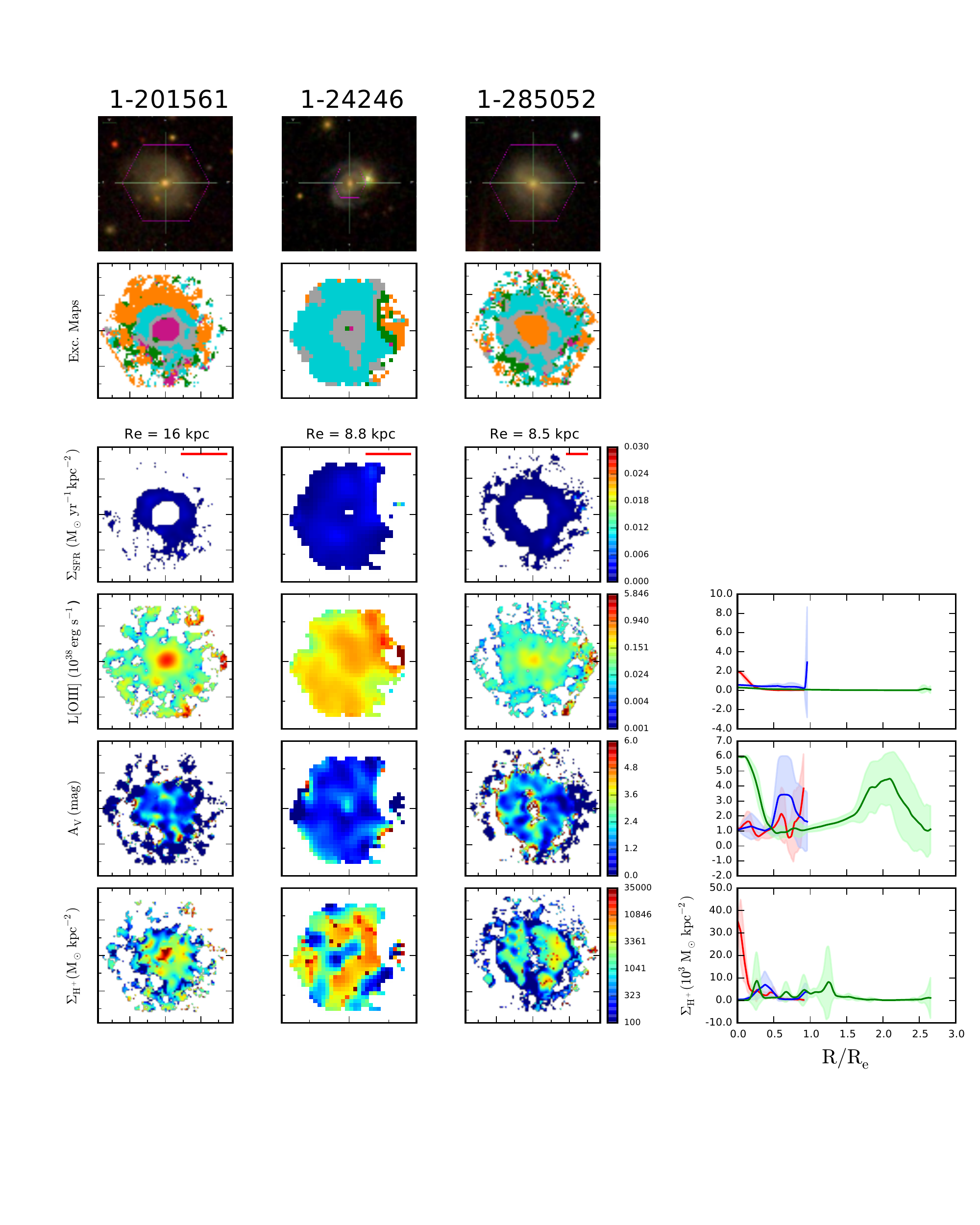}
   \vspace*{-30mm}
   \caption{late type and  weak-luminosity AGN 
   }
\end{figure*}

\begin{figure*}
   \includegraphics[width=2.1\columnwidth]{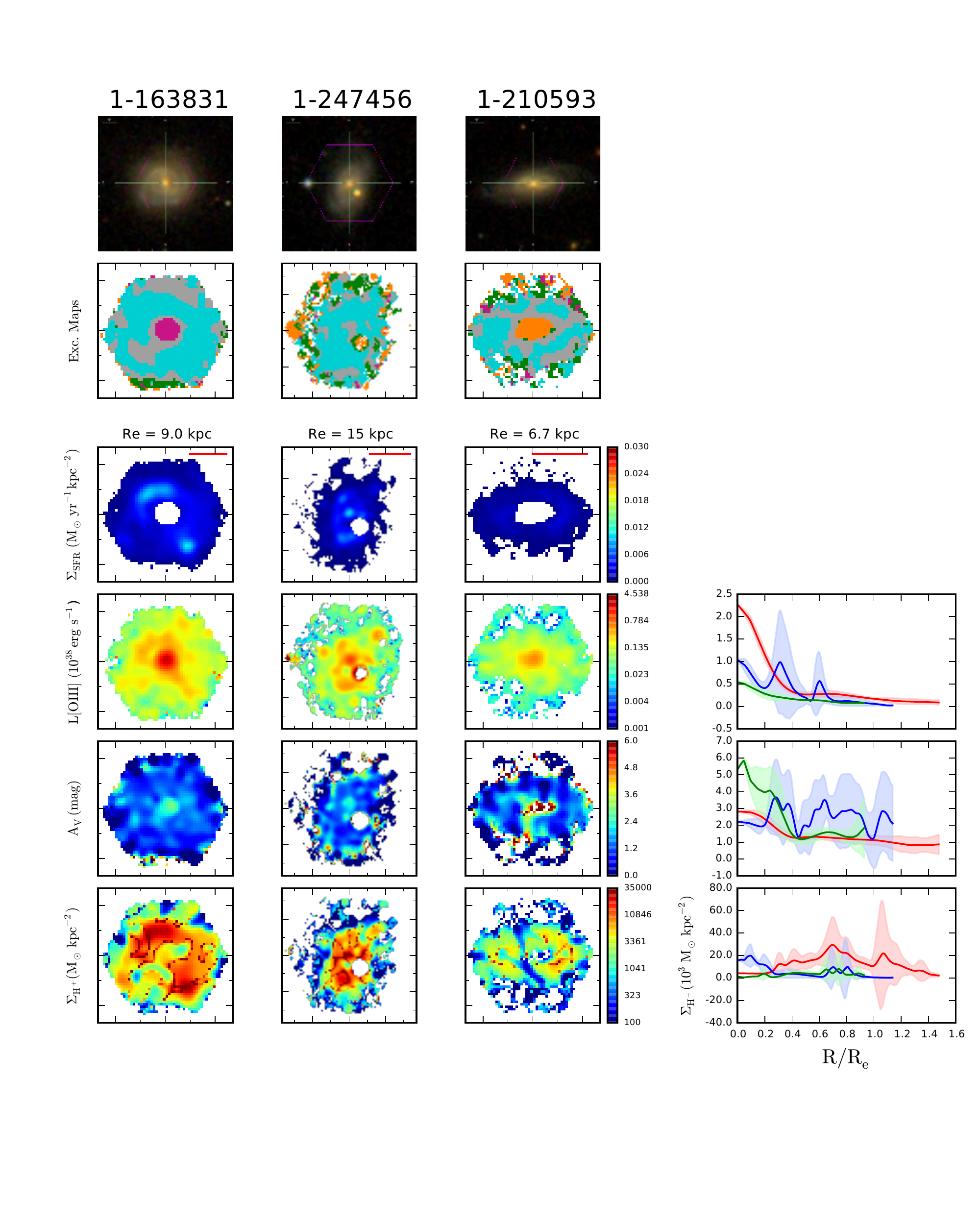}
   \vspace*{-30mm}
   \caption{late type and  weak-luminosity AGN     
   }
\end{figure*}

\begin{figure*}
   \includegraphics[width=2.1\columnwidth]{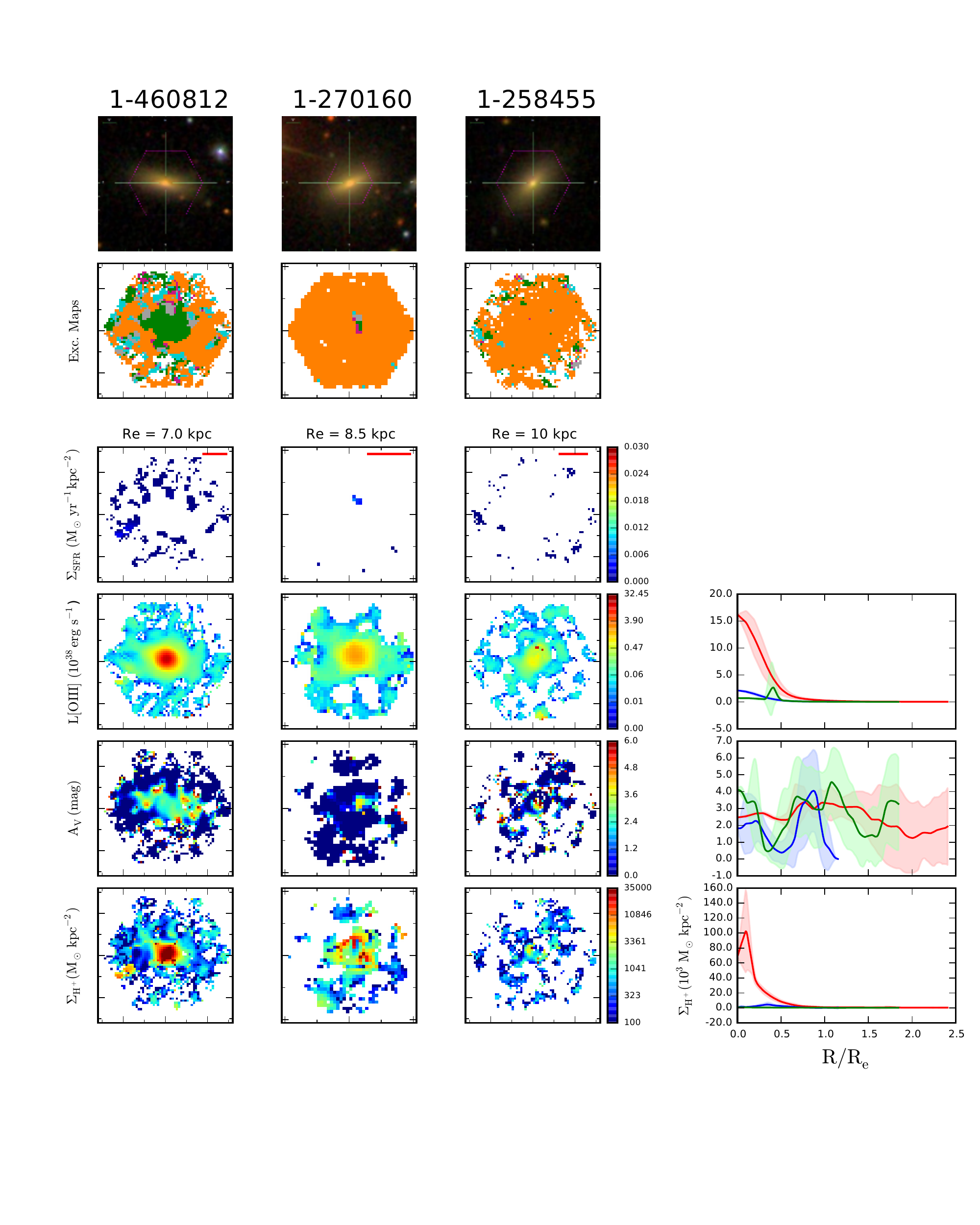}
   \vspace*{-30mm}
   \caption{early-type and strong-luminosity AGN     
   }
\end{figure*}

\begin{figure*}
   \includegraphics[width=2.1\columnwidth]{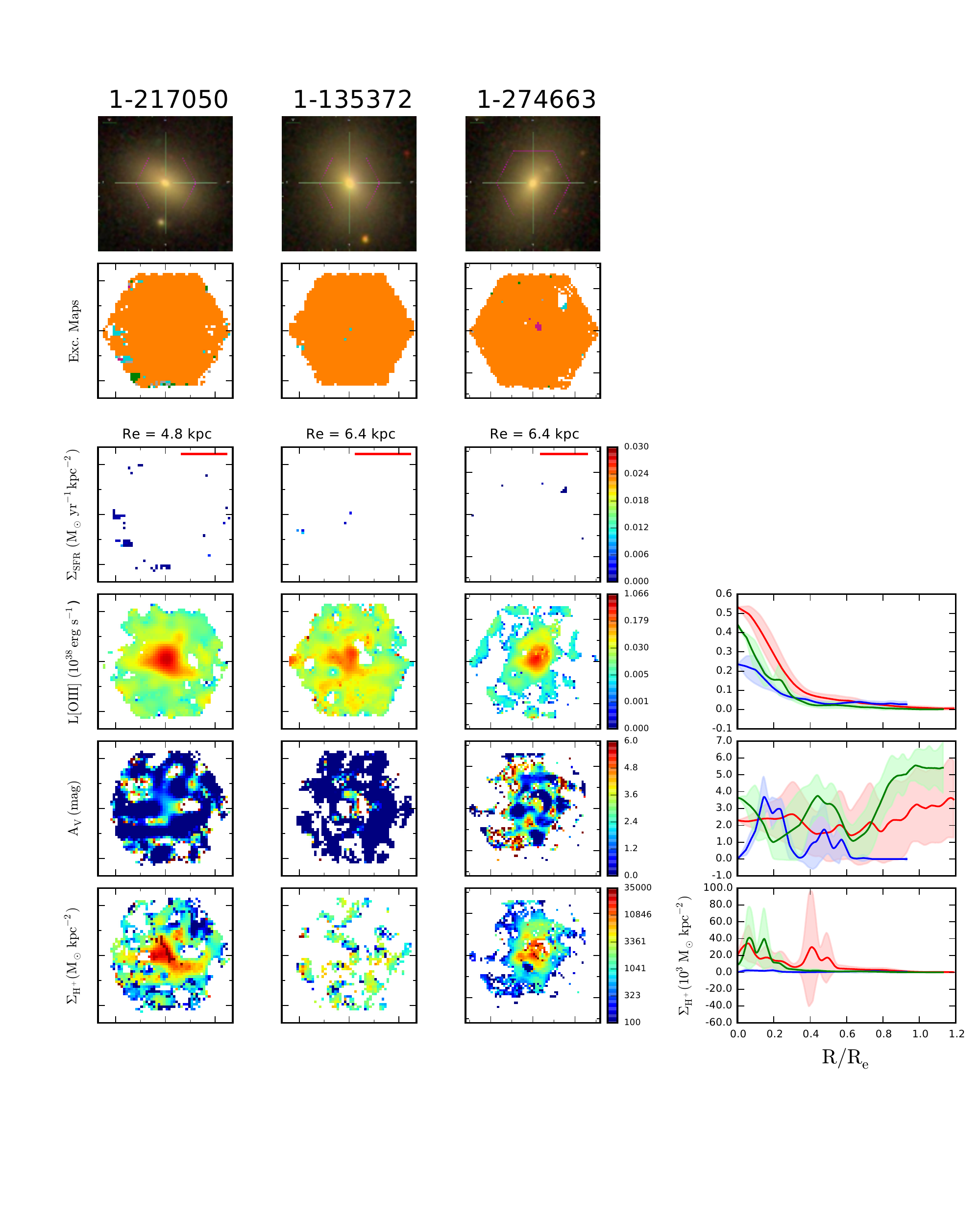}
   \vspace*{-30mm}
   \caption{early-type and weak-luminosity AGN
   }
\end{figure*}

\begin{figure*}
   \includegraphics[width=2.1\columnwidth]{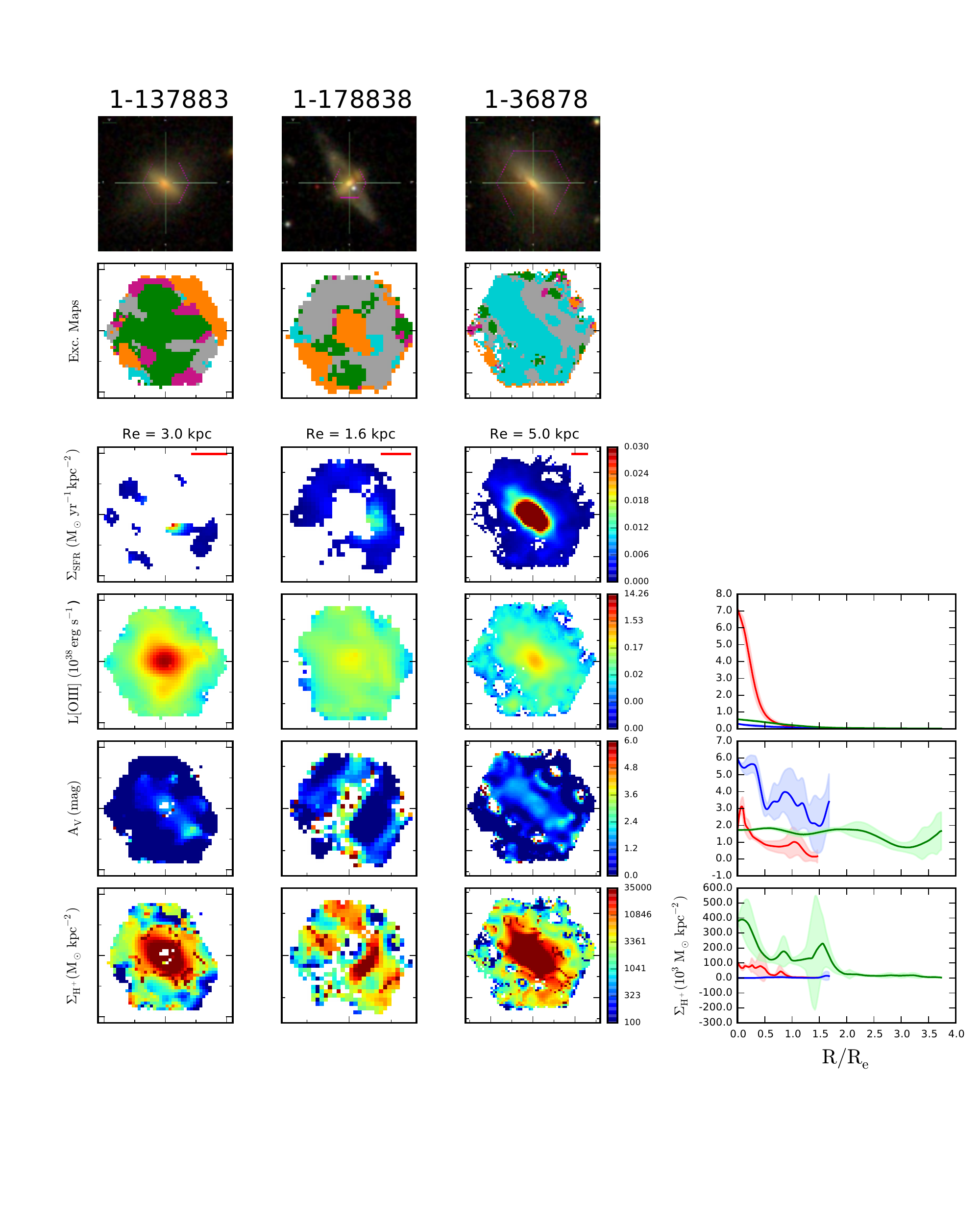}
   \vspace*{-30mm}
   \caption{ E/S type and strong-luminosity AGN
   }
\end{figure*}

\begin{figure*}
   \includegraphics[width=2.1\columnwidth]{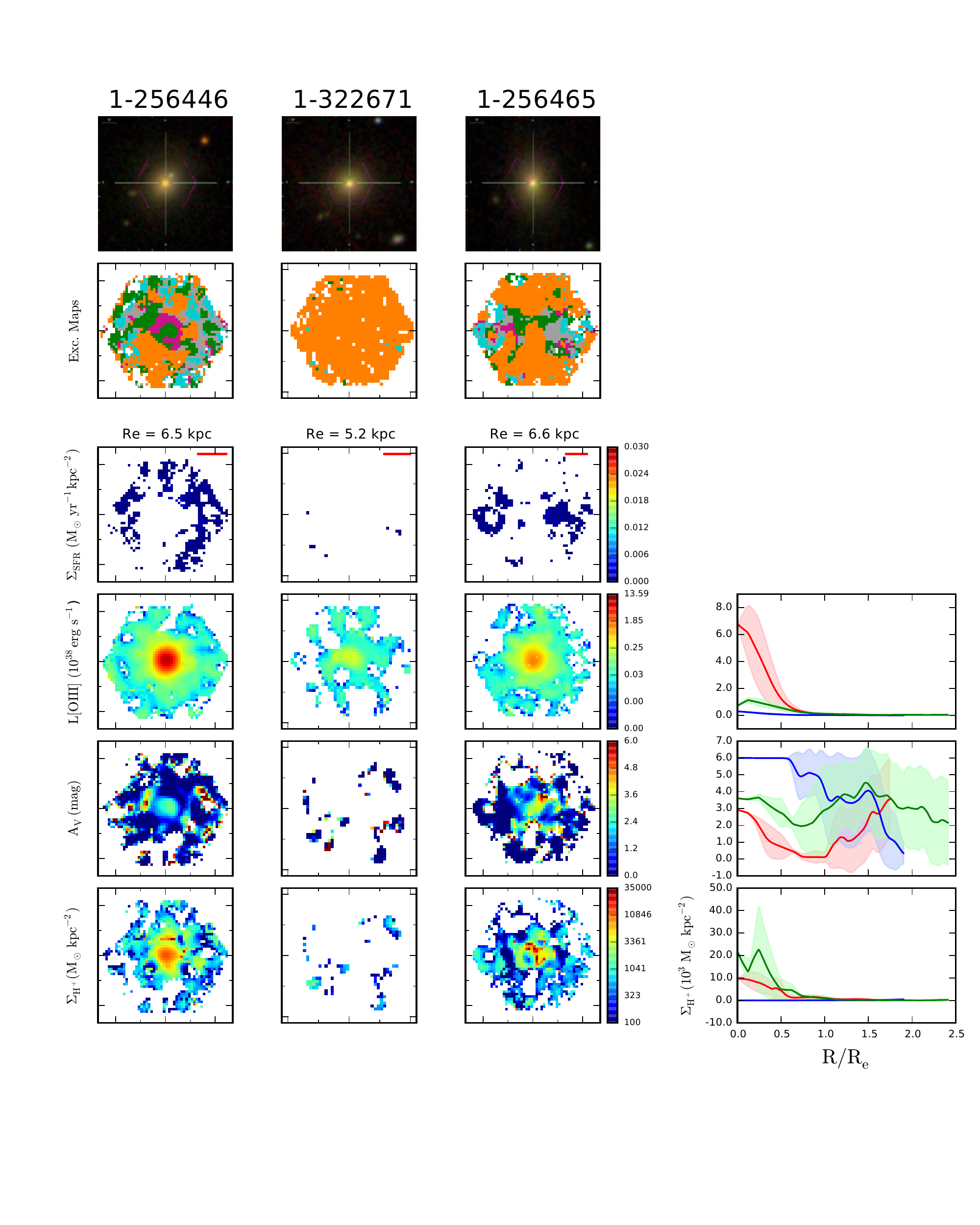}
   \vspace*{-30mm}
   \caption{early-type and weak-luminosity AGN  
   }
\end{figure*}

\begin{figure*}
   \includegraphics[width=2.1\columnwidth]{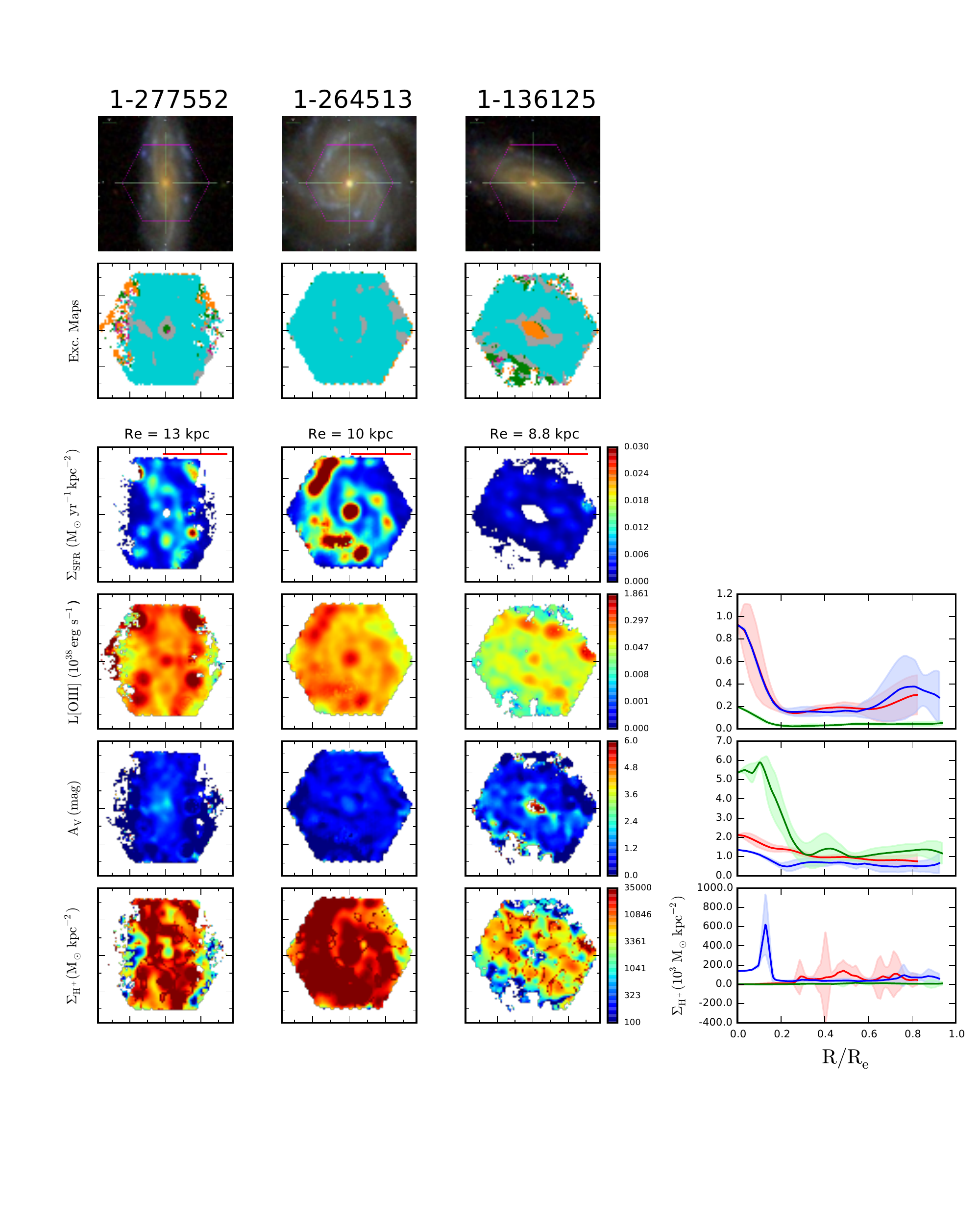}
   \vspace*{-30mm}
   \caption{late type and  weak-luminosity AGN     
   }
\end{figure*}

\begin{figure*}
   \includegraphics[width=2.1\columnwidth]{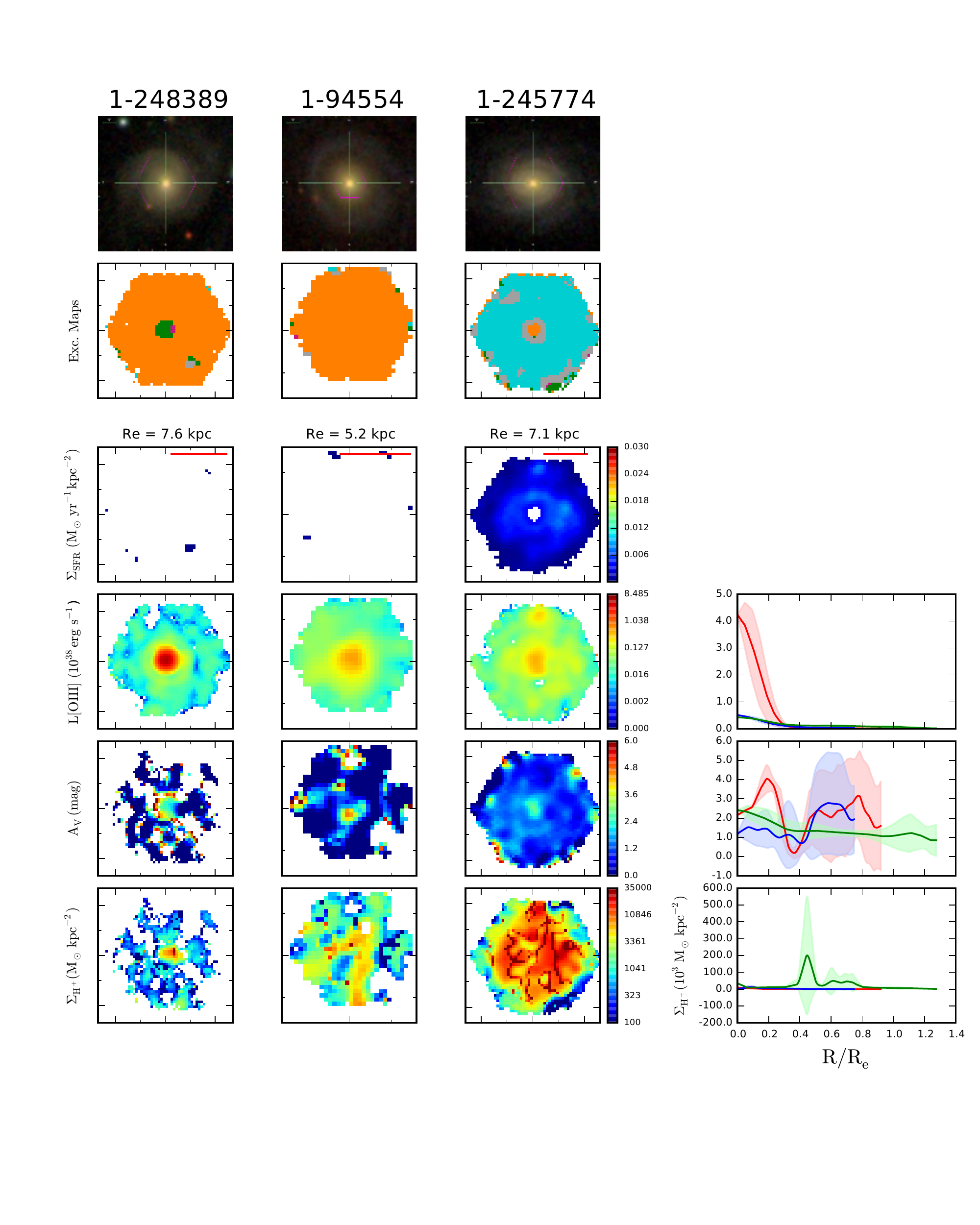}
   \vspace*{-30mm}
   \caption{late type and  weak-luminosity AGN      
   }
\end{figure*}

\begin{figure*}
   \includegraphics[width=2.1\columnwidth]{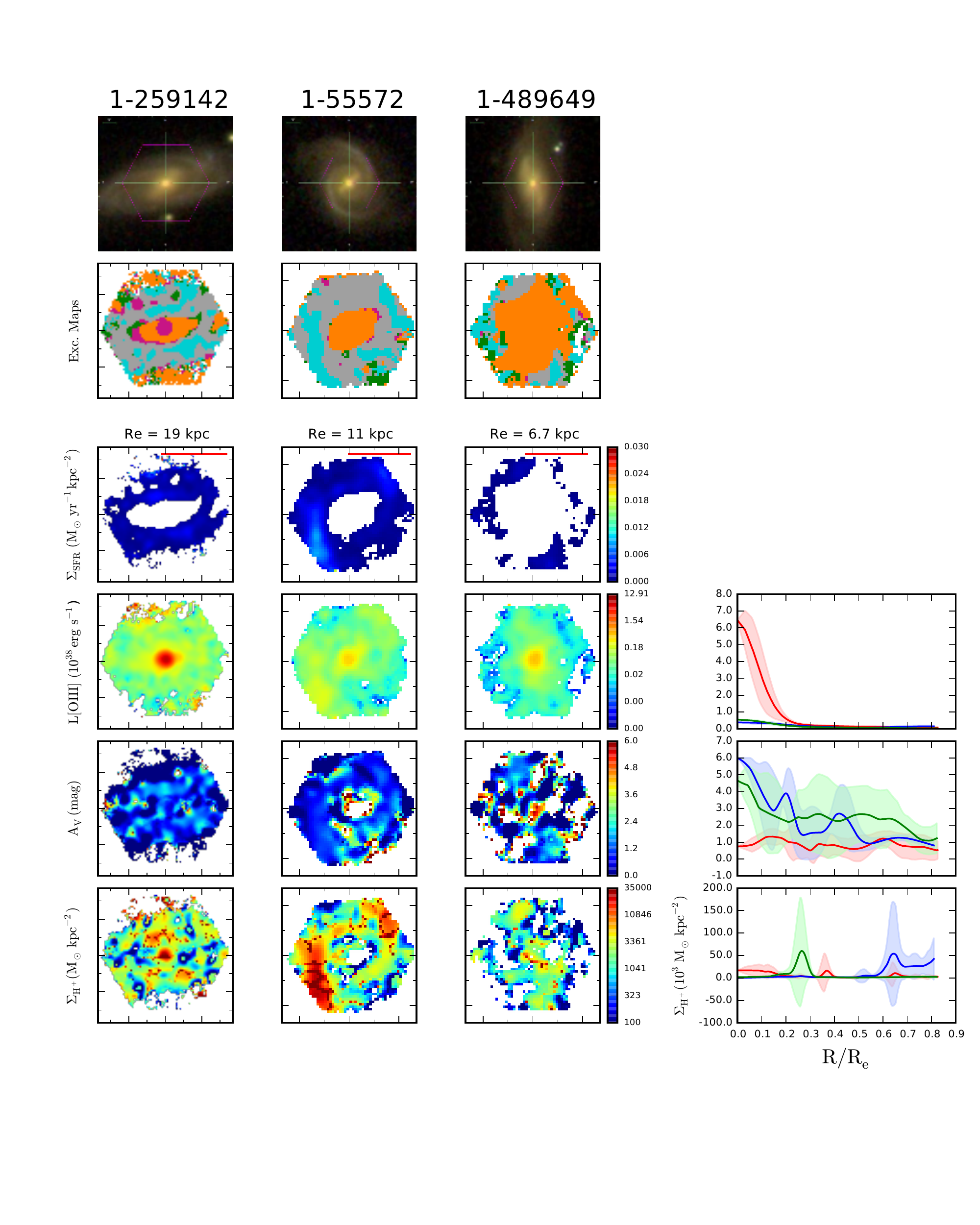}
   \vspace*{-30mm}
   \caption{late type and  weak-luminosity AGN
   }
\end{figure*}

\begin{figure*}
   \includegraphics[width=2.1\columnwidth]{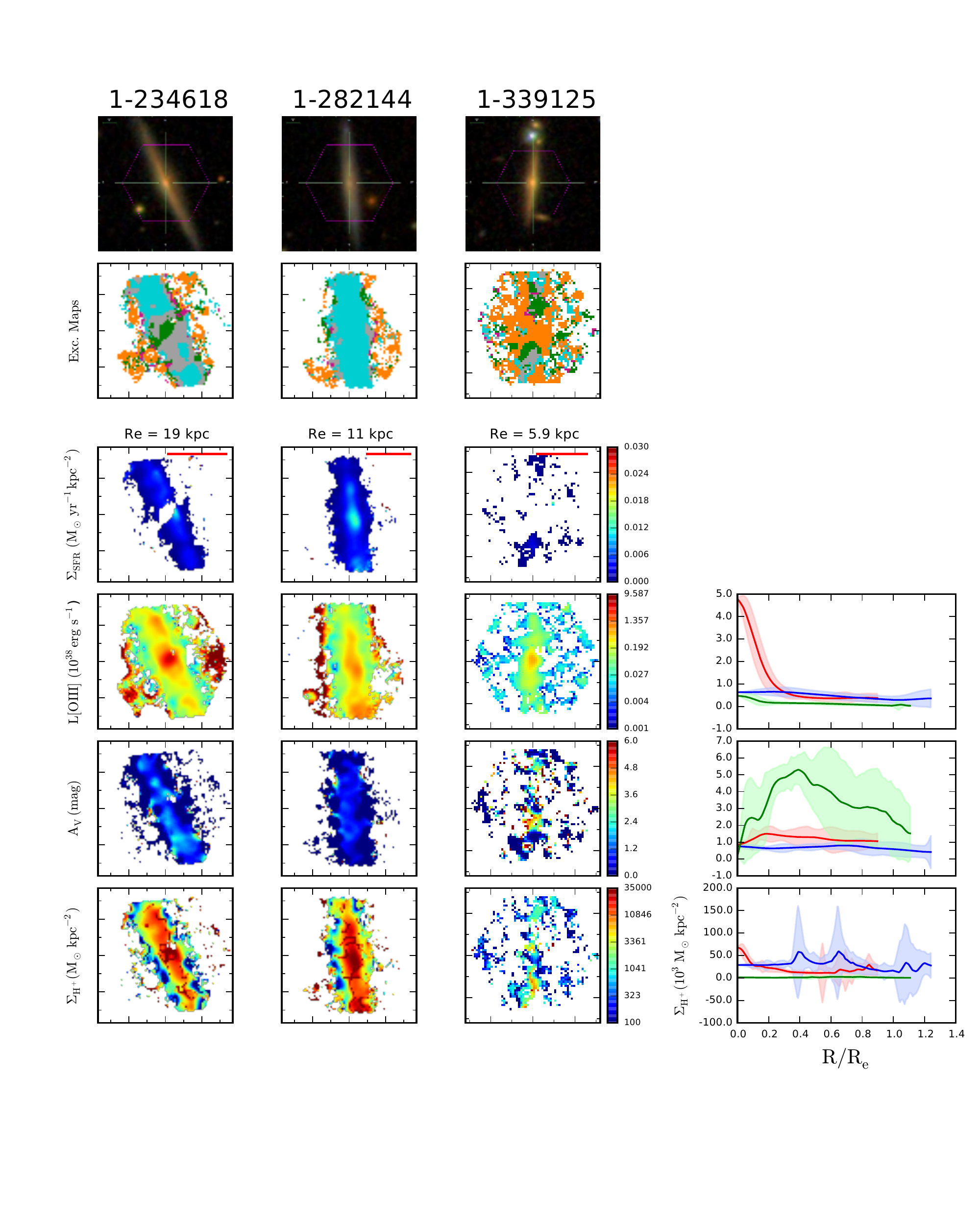}
   \vspace*{-30mm}
   \caption{late type and  weak-luminosity AGN
   }
\end{figure*}

\begin{figure*}
   \includegraphics[width=2.1\columnwidth]{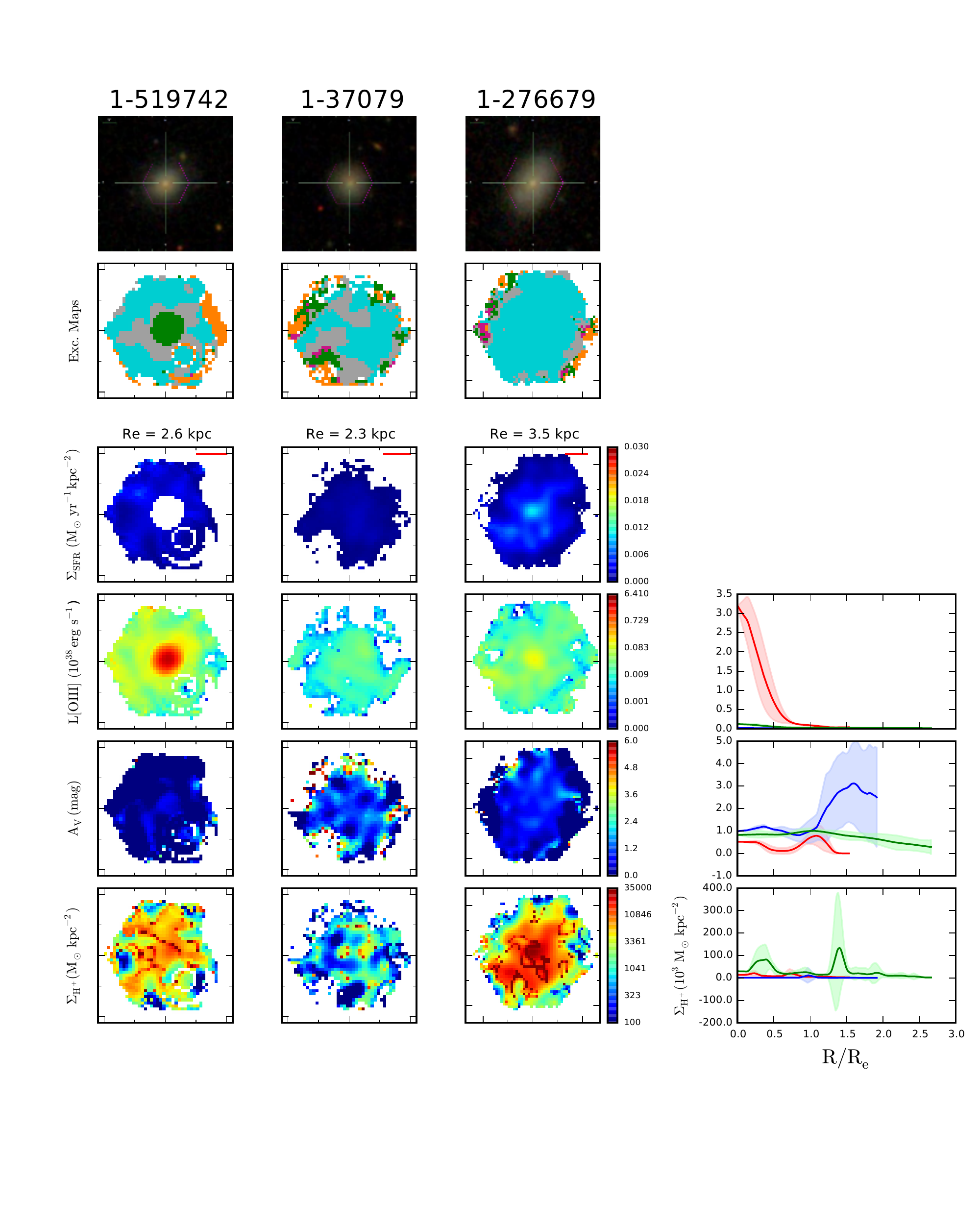}
   \vspace*{-30mm}
   \caption{late type and  weak-luminosity AGN     
   }
\end{figure*}

\begin{figure*}
   \includegraphics[width=2.1\columnwidth]{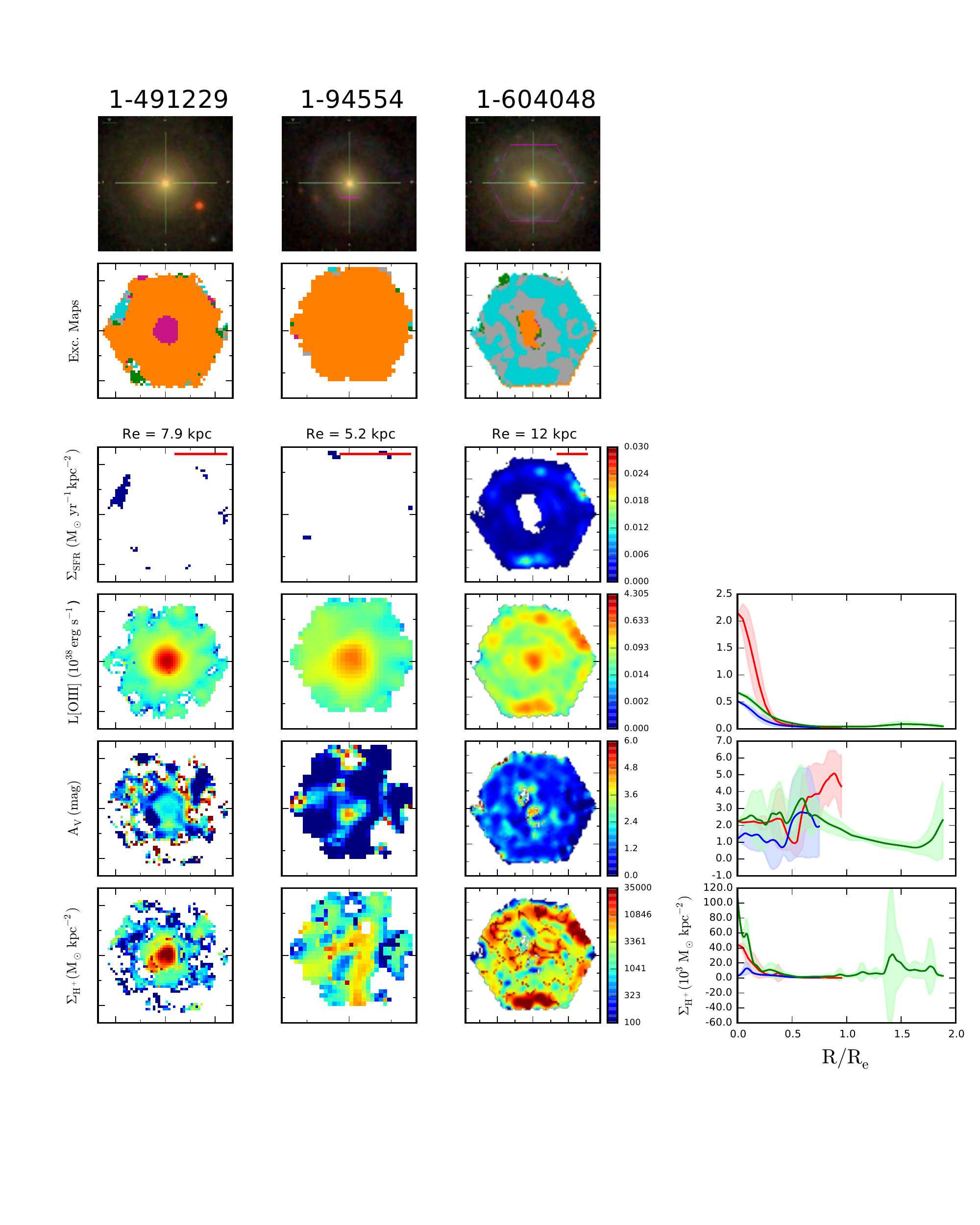}
   \vspace*{-30mm}
   \caption{early-type and weak-luminosity AGN     
   }
\end{figure*}

\begin{figure*}
   \includegraphics[width=2.1\columnwidth]{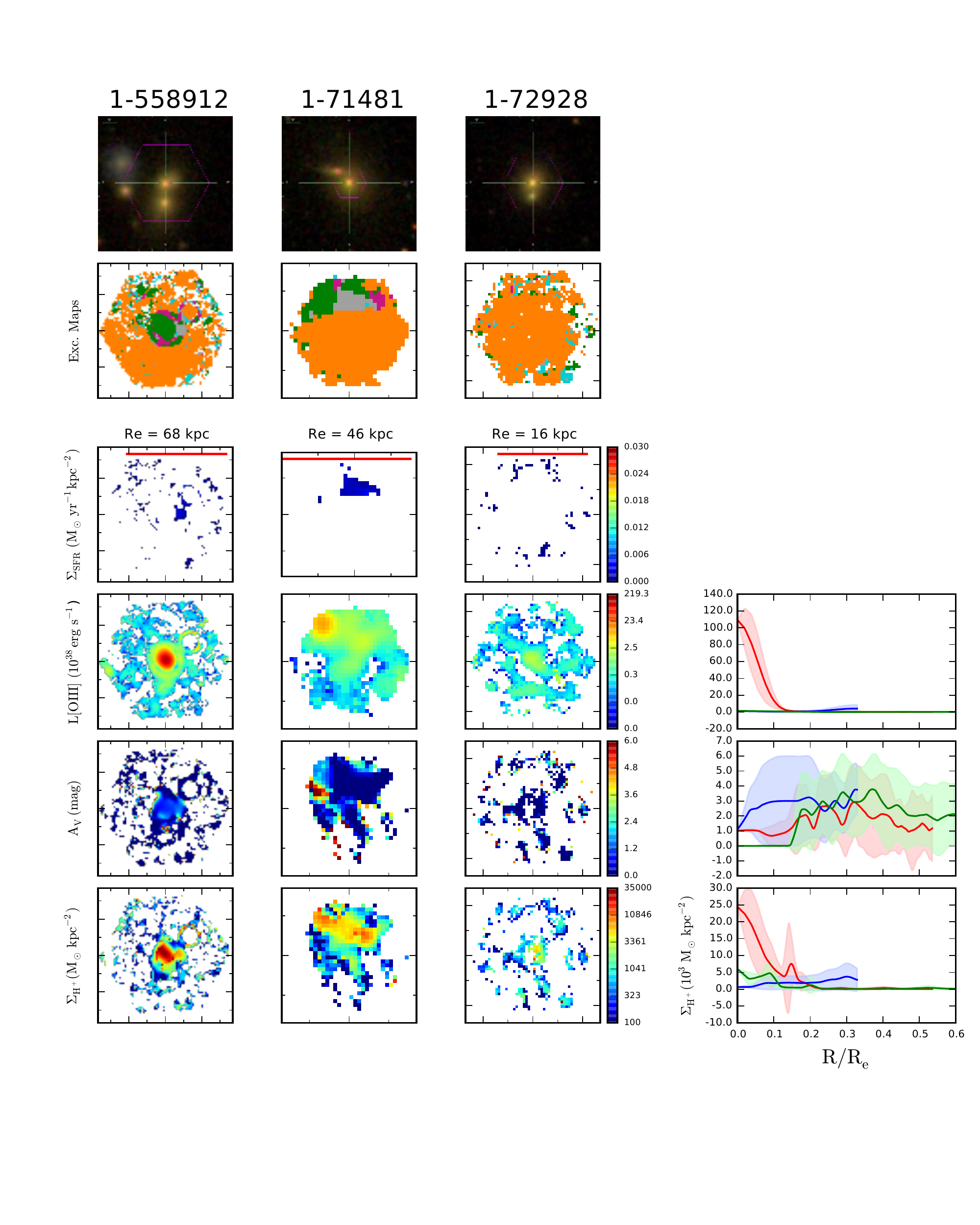}
   \vspace*{-30mm}
   \caption{merger and strong-luminosity AGN
   }
\end{figure*}

\begin{figure*}
   \includegraphics[width=2.1\columnwidth]{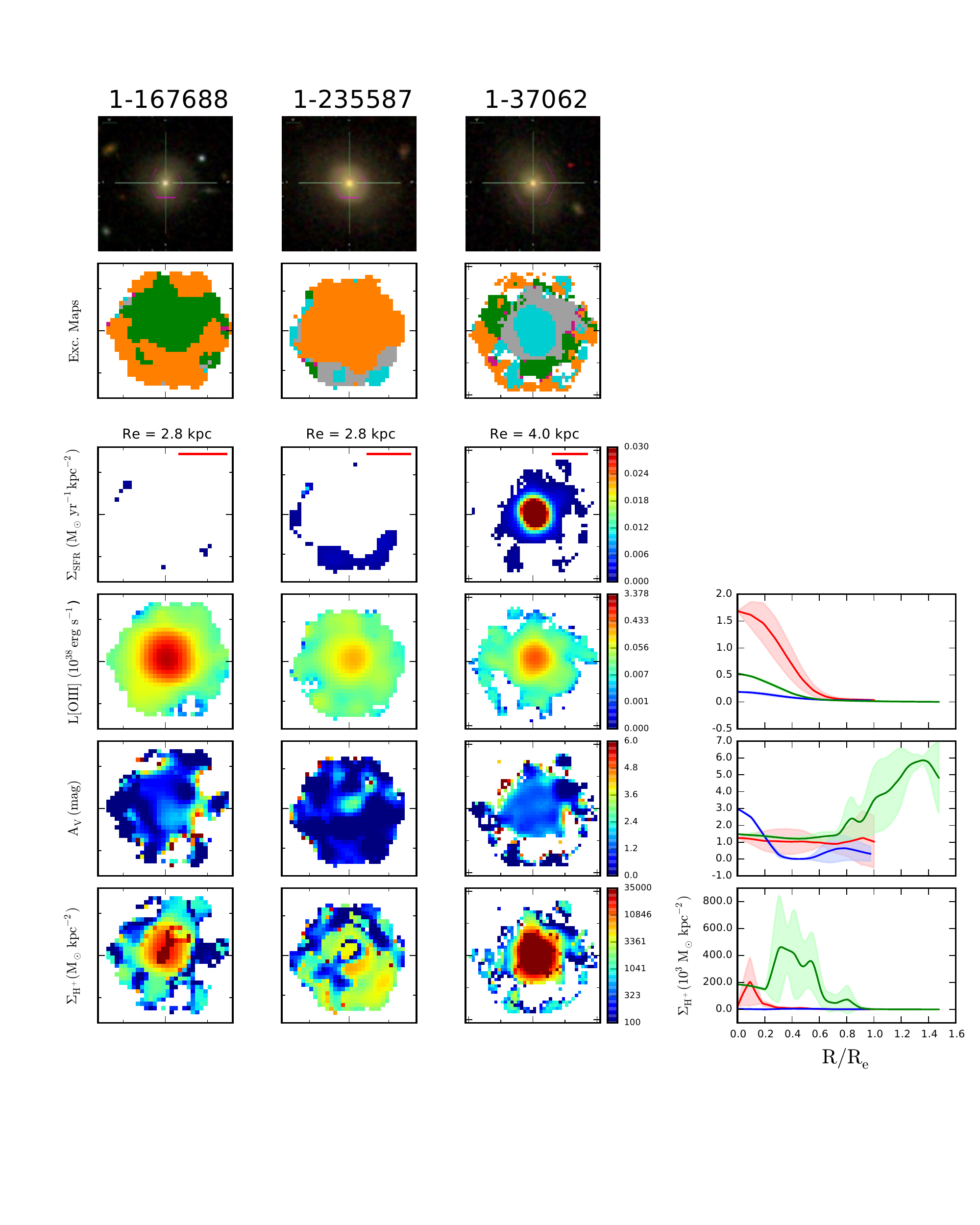}
   \vspace*{-30mm}
   \caption{early-type and weak-luminosity AGN
   }
\end{figure*}

\begin{figure*}
   \includegraphics[width=2.1\columnwidth]{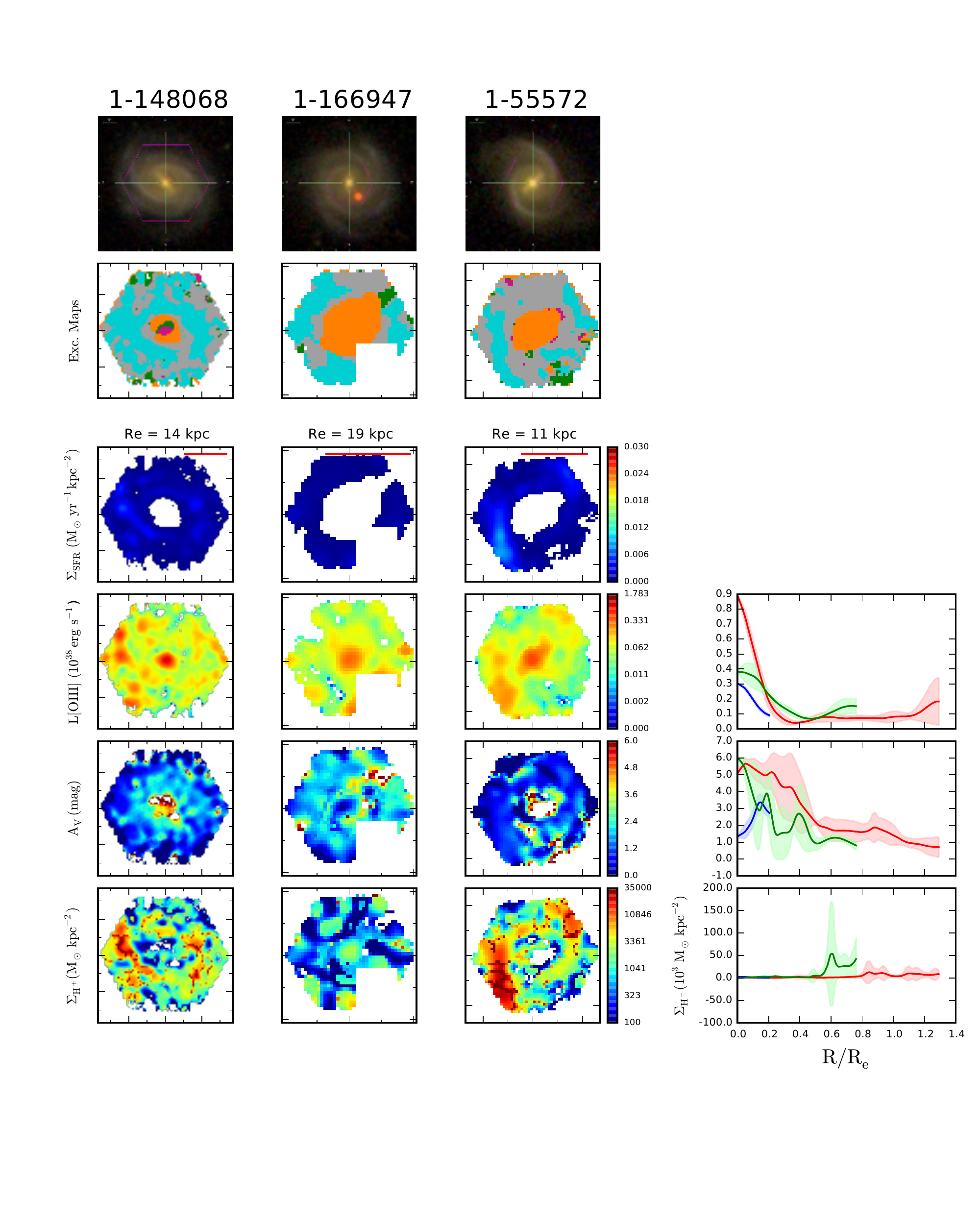}
   \vspace*{-30mm}
   \caption{late type and  weak-luminosity AGN
   }
\end{figure*}

\begin{figure*}
   \includegraphics[width=2.1\columnwidth]{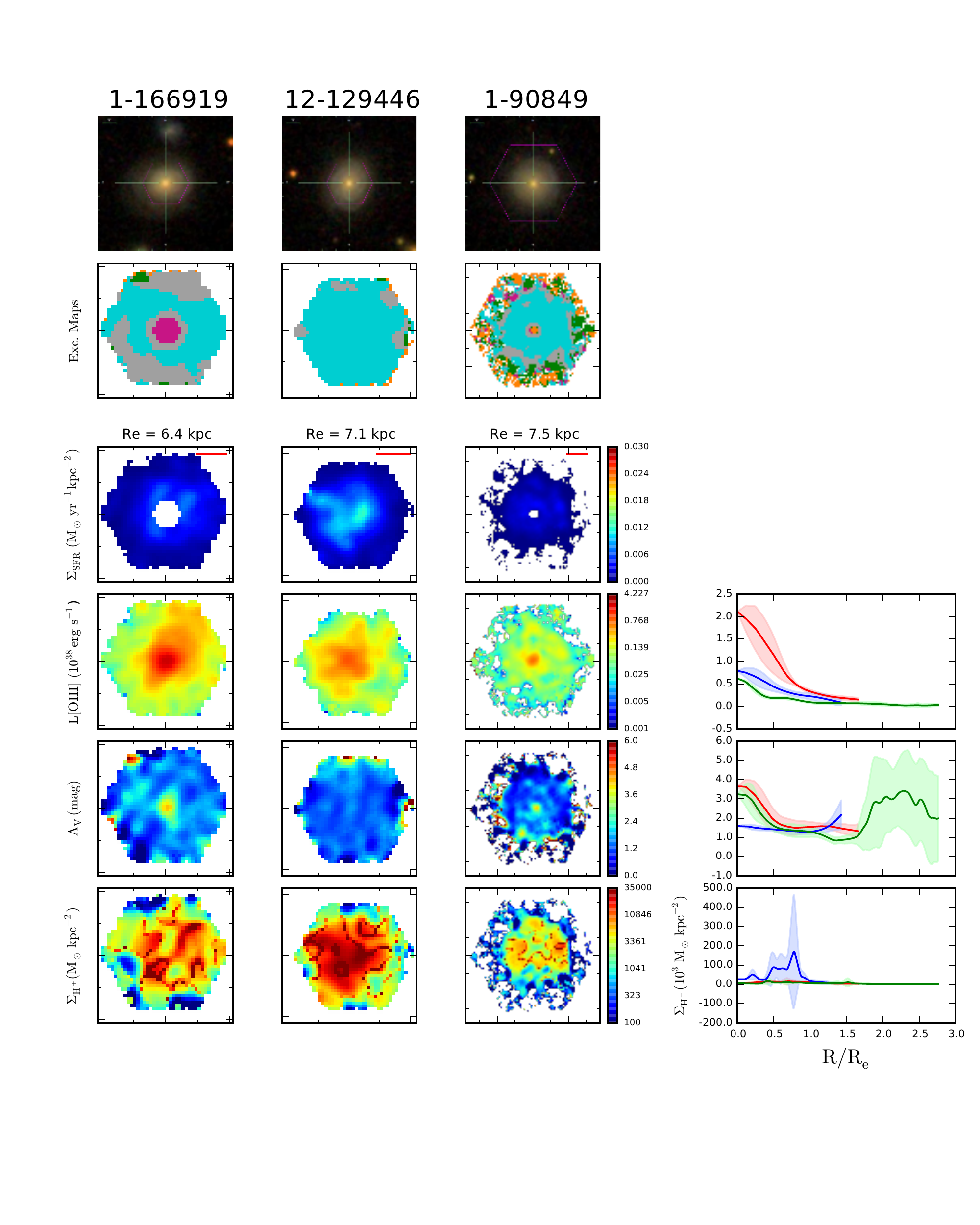}
   \vspace*{-30mm}
   \caption{late type and  weak-luminosity AGN
   }
\end{figure*}

\begin{figure*}
   \includegraphics[width=2.1\columnwidth]{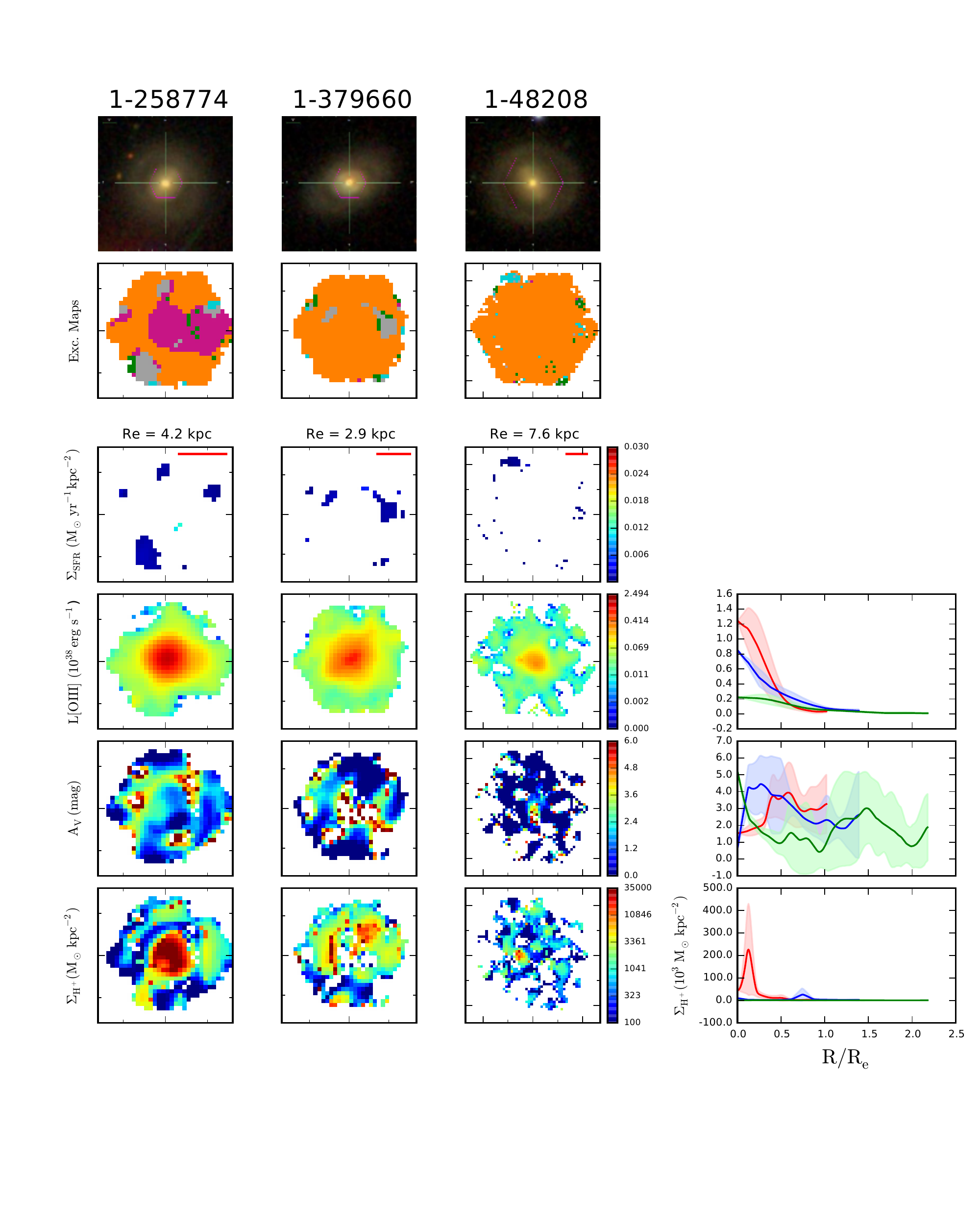}
   \vspace*{-30mm}
   \caption{late type and  weak-luminosity AGN      
   }
\end{figure*}

\begin{figure*}
   \includegraphics[width=2.1\columnwidth]{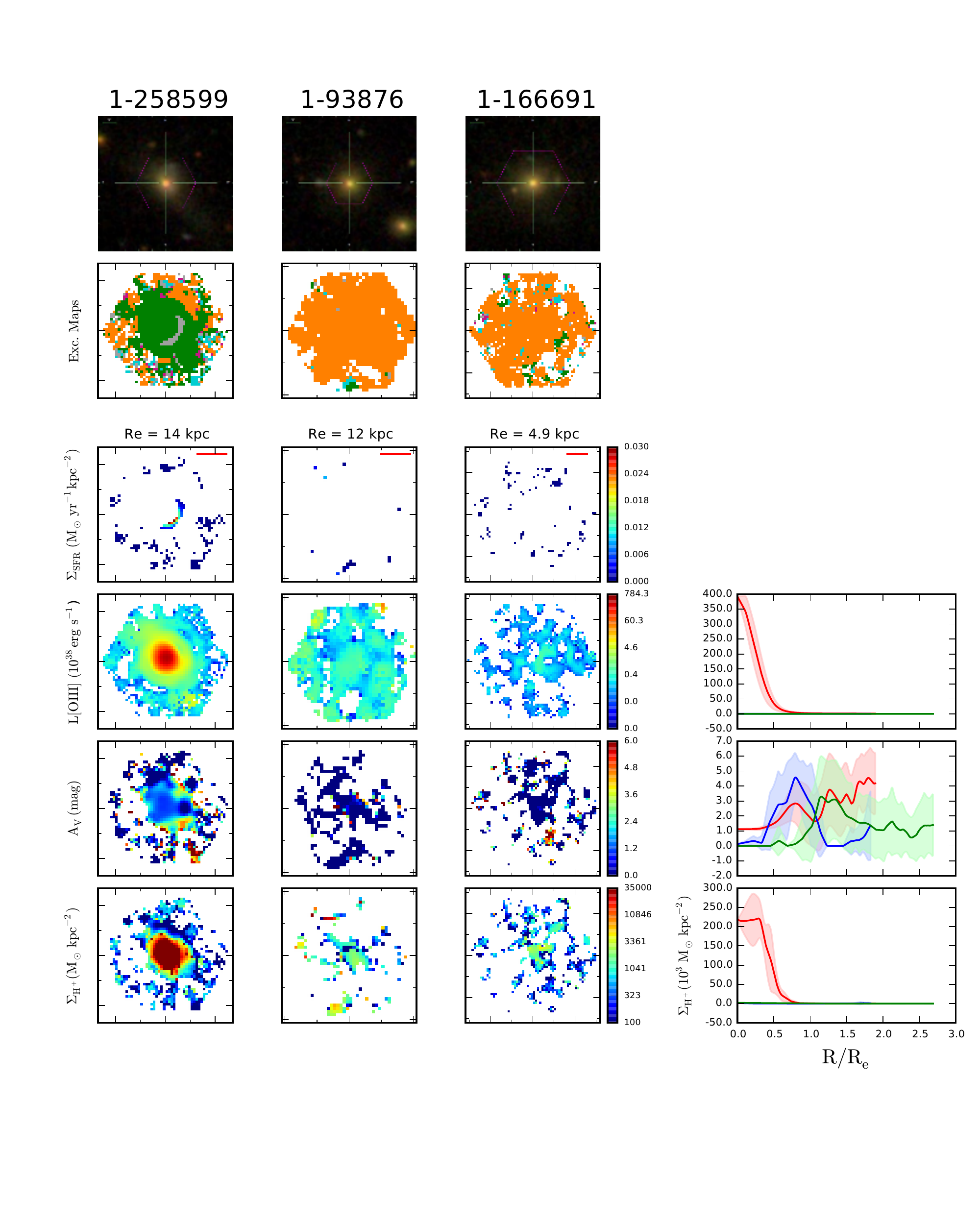}
   \vspace*{-30mm}
   \caption{early-type and strong-luminosity AGN
   }
\end{figure*}

\begin{figure*}
   \includegraphics[width=2.1\columnwidth]{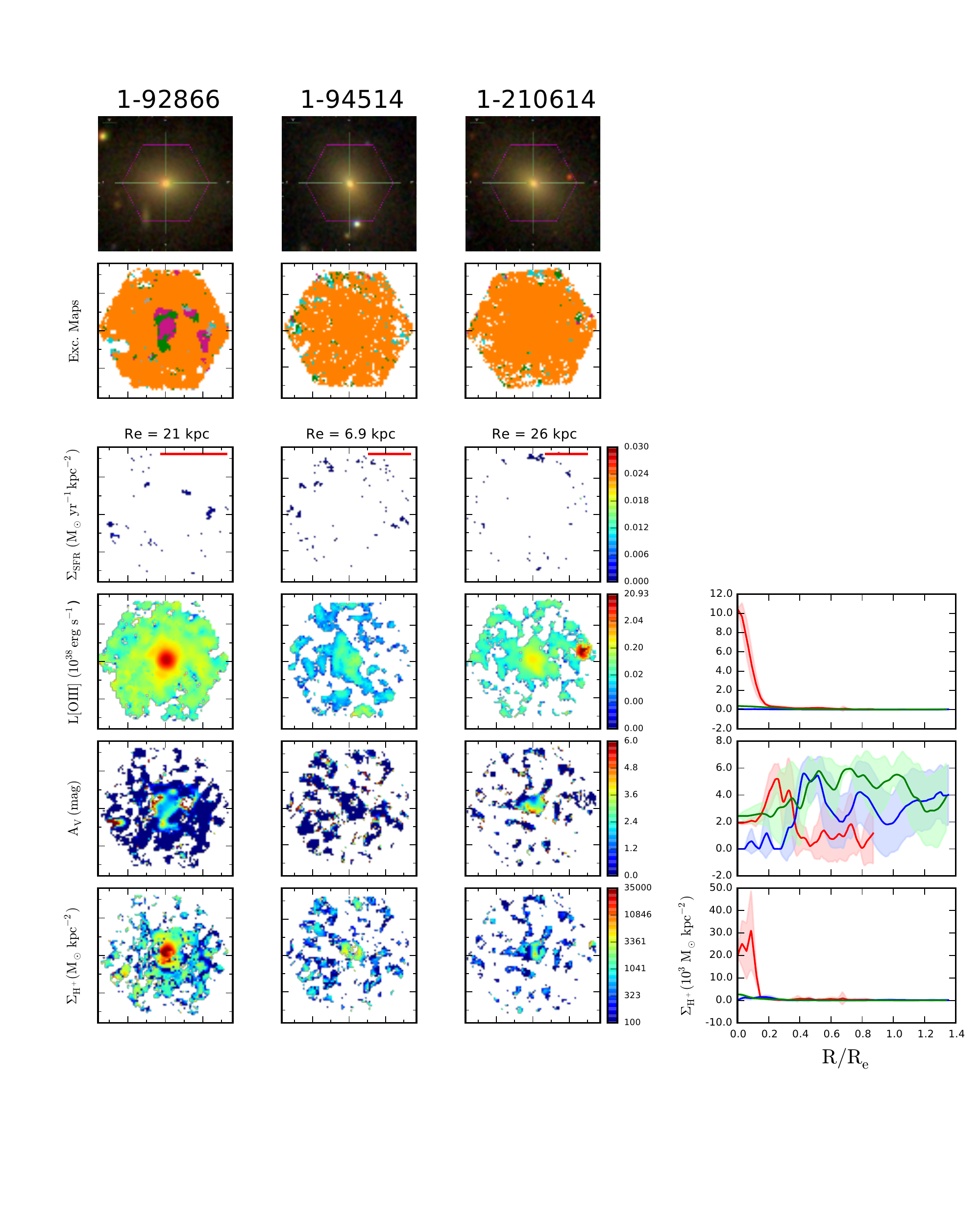}
   \vspace*{-30mm}
   \caption{early-type and strong-luminosity AGN    
   }
\end{figure*}

\begin{figure*}
   \includegraphics[width=2.1\columnwidth]{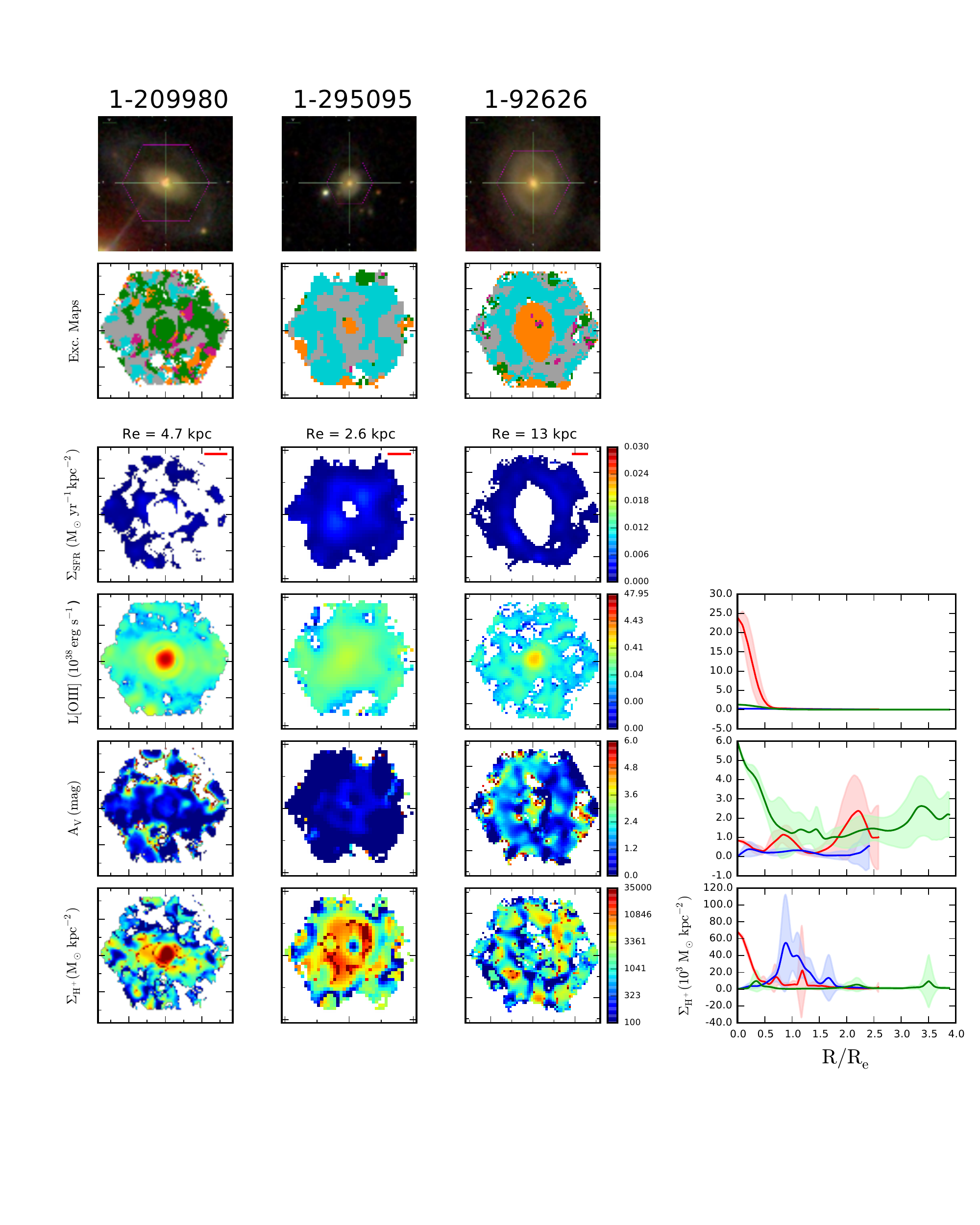}
   \vspace*{-30mm}
   \caption{late-type and strong-luminosity AGN      
   }
\end{figure*}

\begin{figure*}
   \includegraphics[width=2.1\columnwidth]{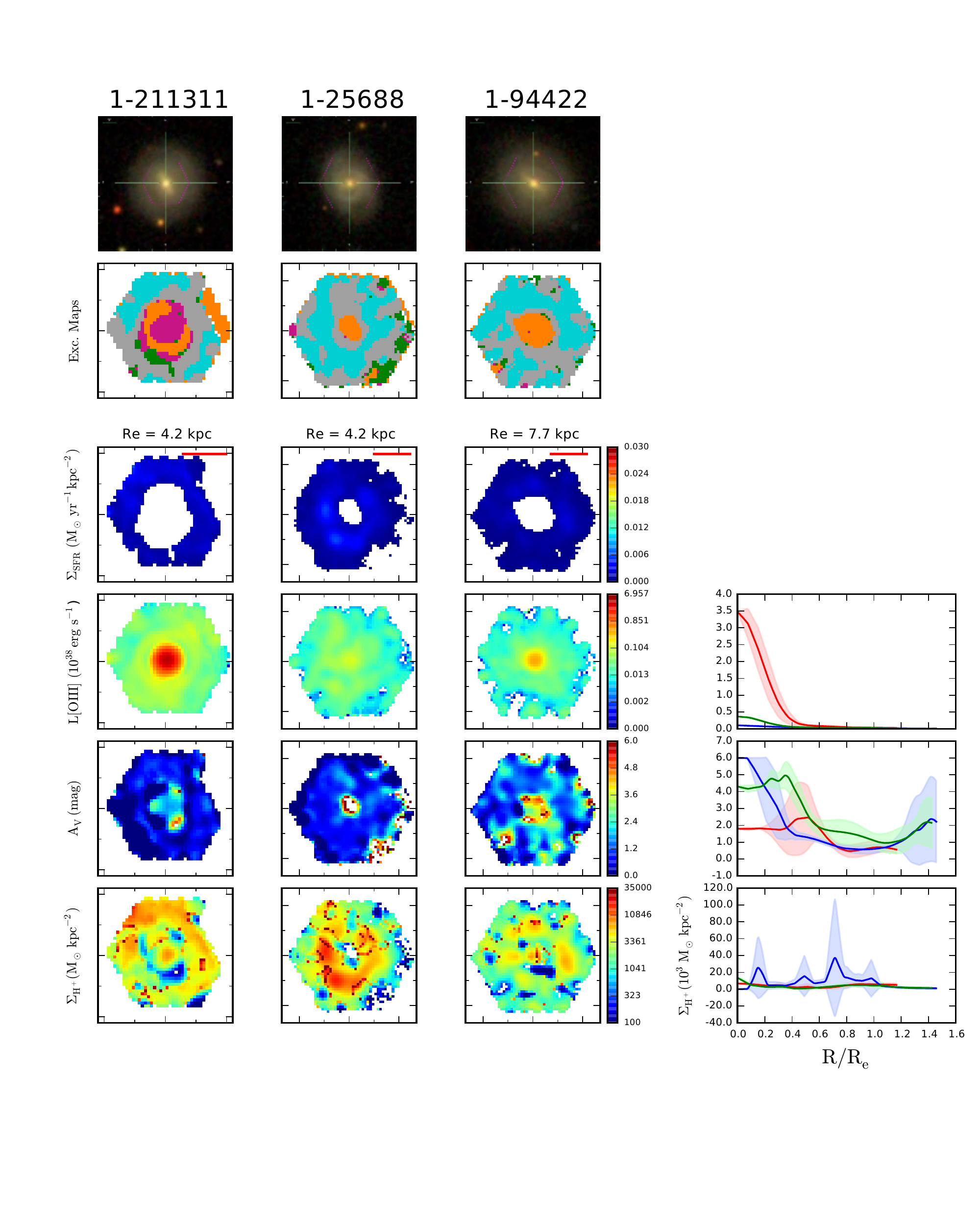}
   \vspace*{-30mm}
   \caption{E/S type and weak-luminosity AGN     
   }
\end{figure*}

\begin{figure*}
   \includegraphics[width=2.1\columnwidth]{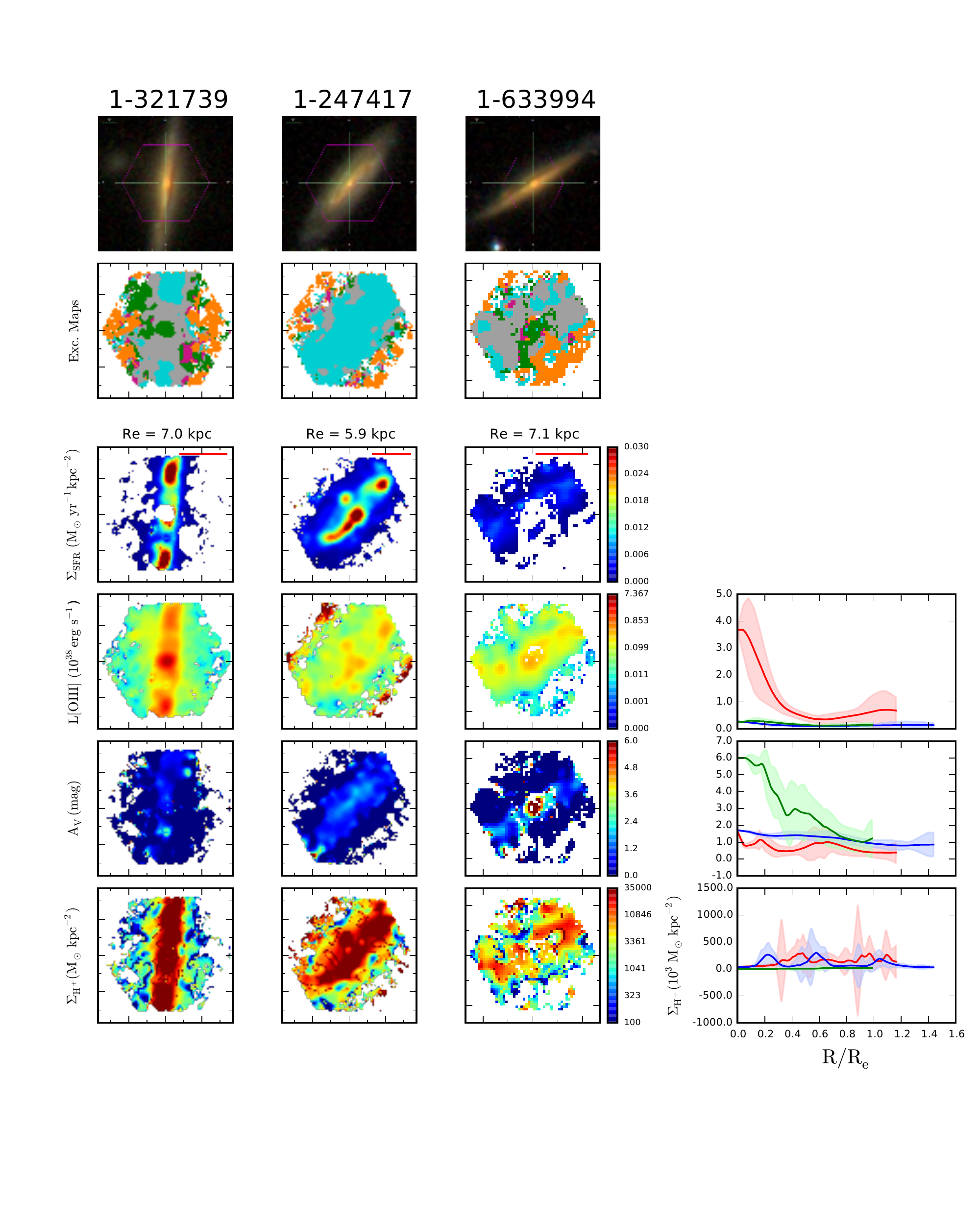}
   \vspace*{-30mm}
   \caption{late type and  weak-luminosity AGN      
   }
\end{figure*}

\begin{figure*}
   \includegraphics[width=2.1\columnwidth]{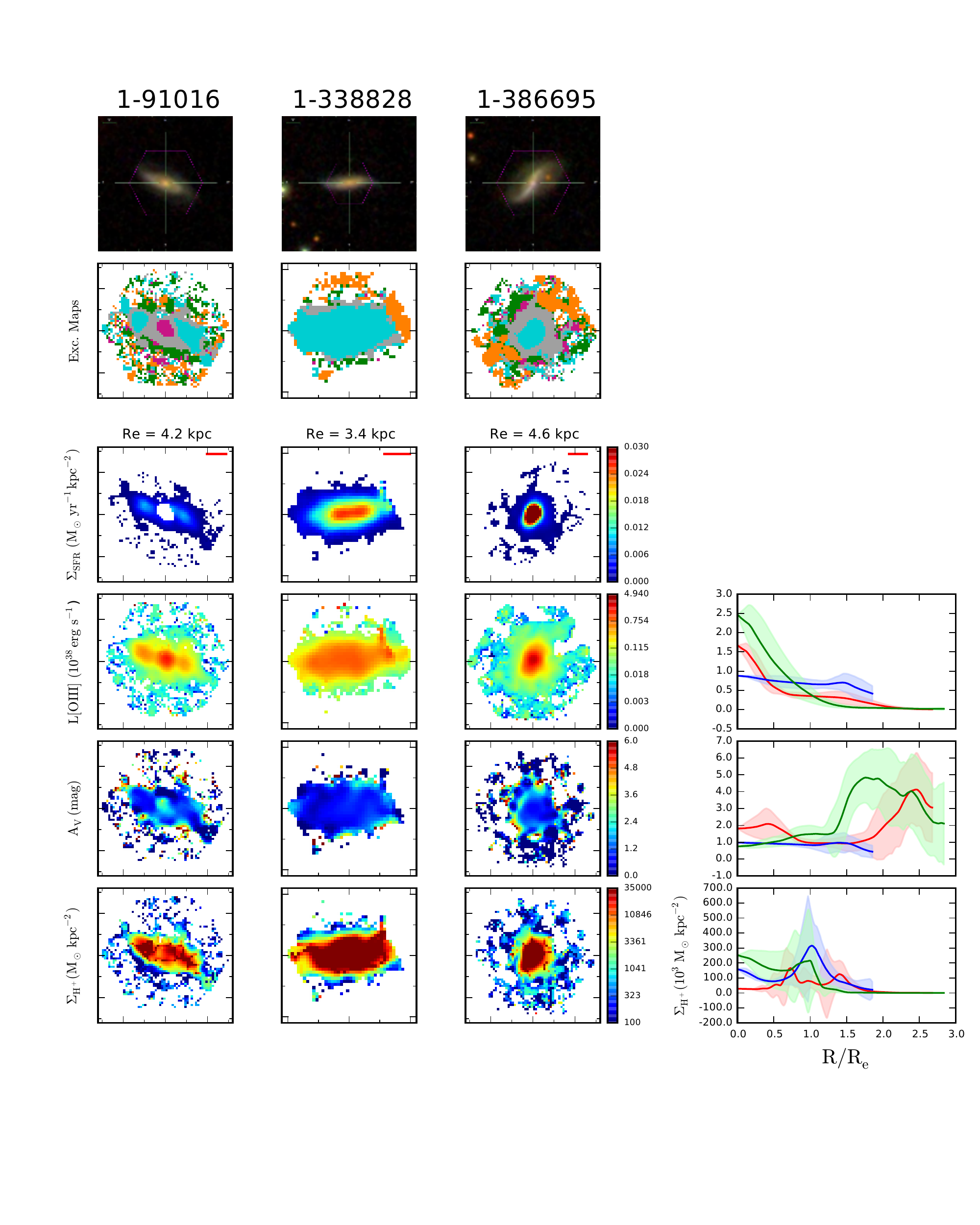}
   \vspace*{-30mm}
   \caption{late type and  weak-luminosity AGN
   }
\end{figure*}

\begin{figure*}
   \includegraphics[width=2.1\columnwidth]{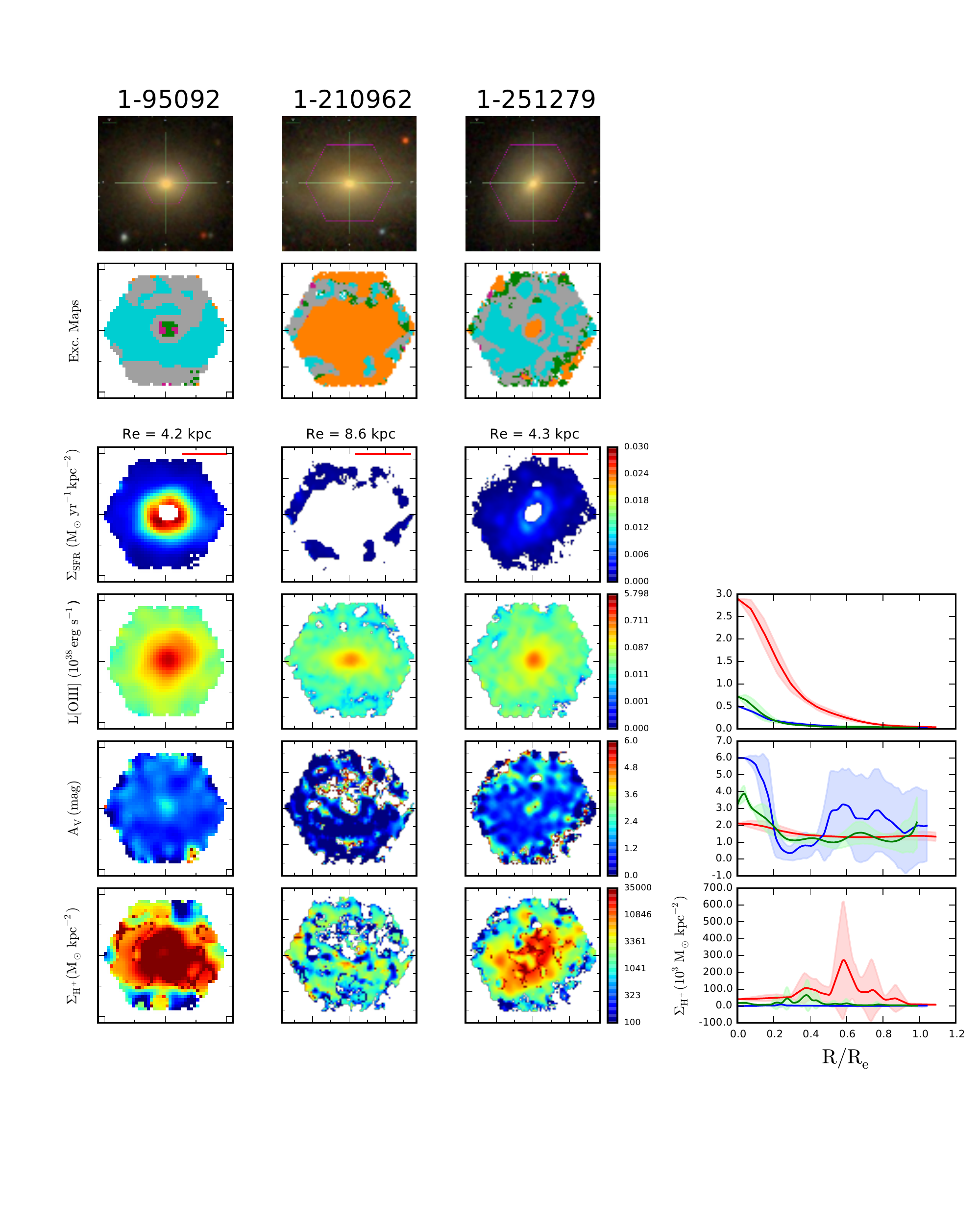}
   \vspace*{-30mm}
   \caption{early-type and weak-luminosity AGN    
   }
\end{figure*}

\begin{figure*}
   \includegraphics[width=2.1\columnwidth]{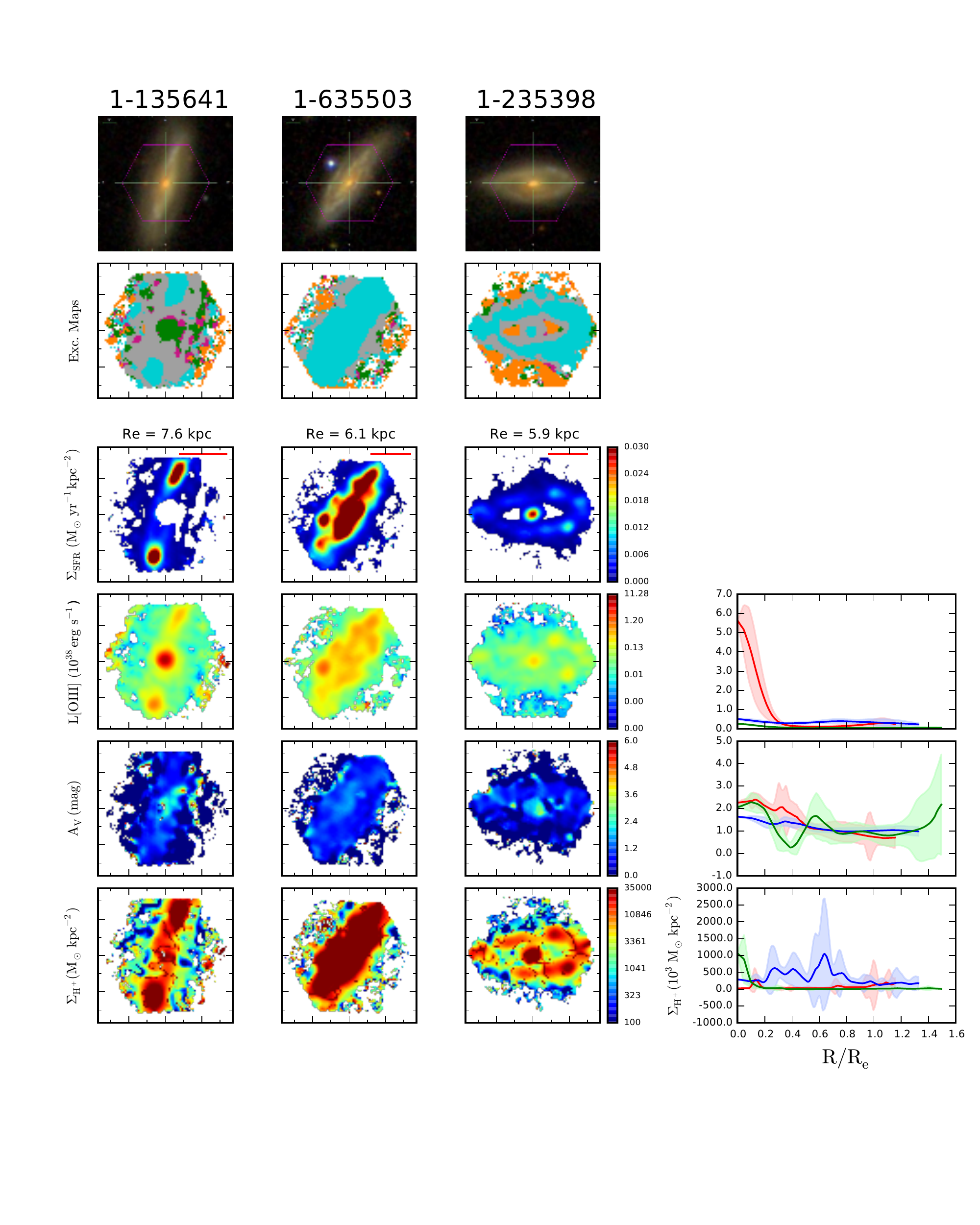}
   \vspace*{-30mm}
   \caption{late type and  weak-luminosity AGN    
   }
\end{figure*}

\begin{figure*}
   \includegraphics[width=2.1\columnwidth]{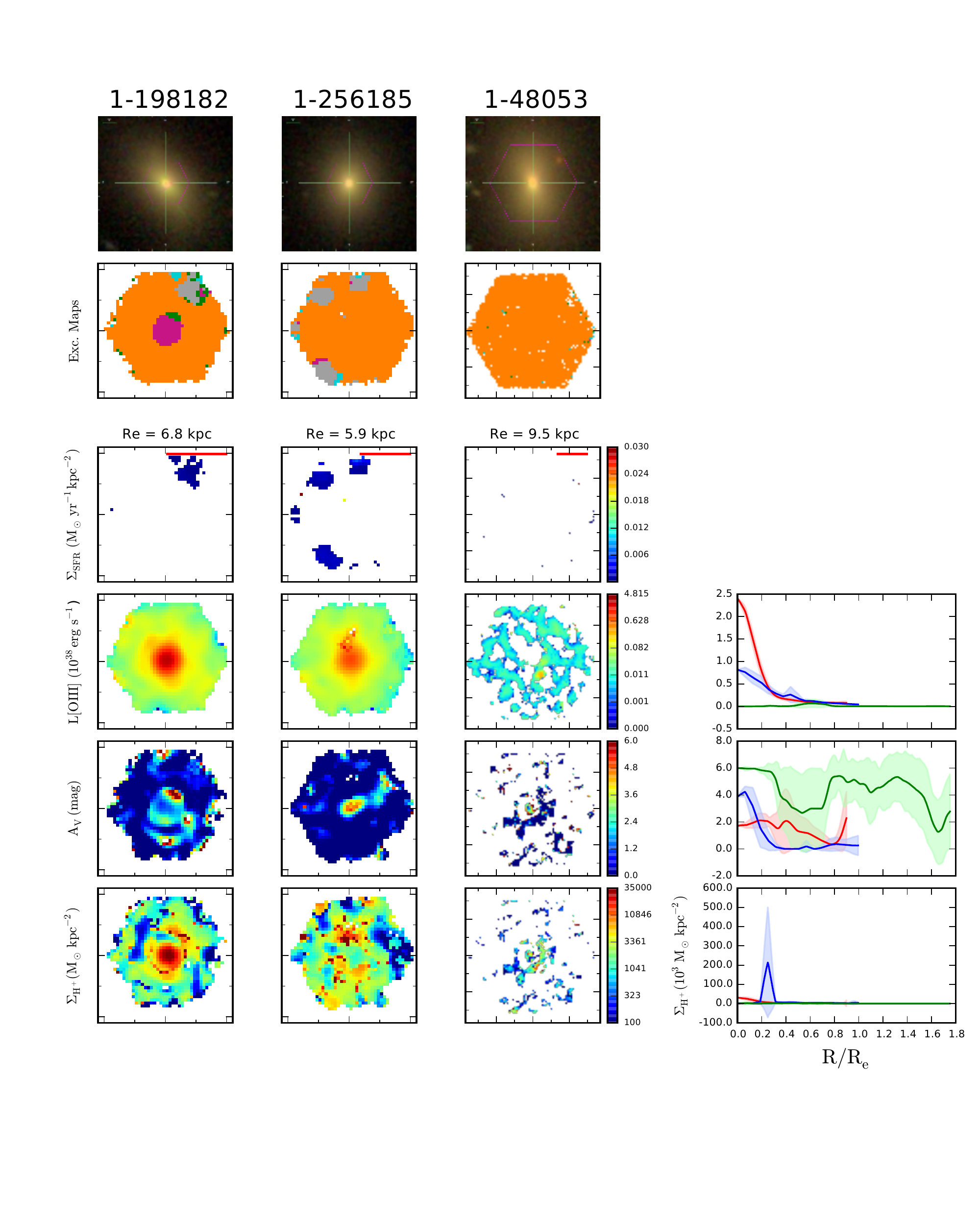}
   \vspace*{-30mm}
   \caption{early-type and weak-luminosity AGN    
   }
\end{figure*}

\begin{figure*}
   \includegraphics[width=2.1\columnwidth]{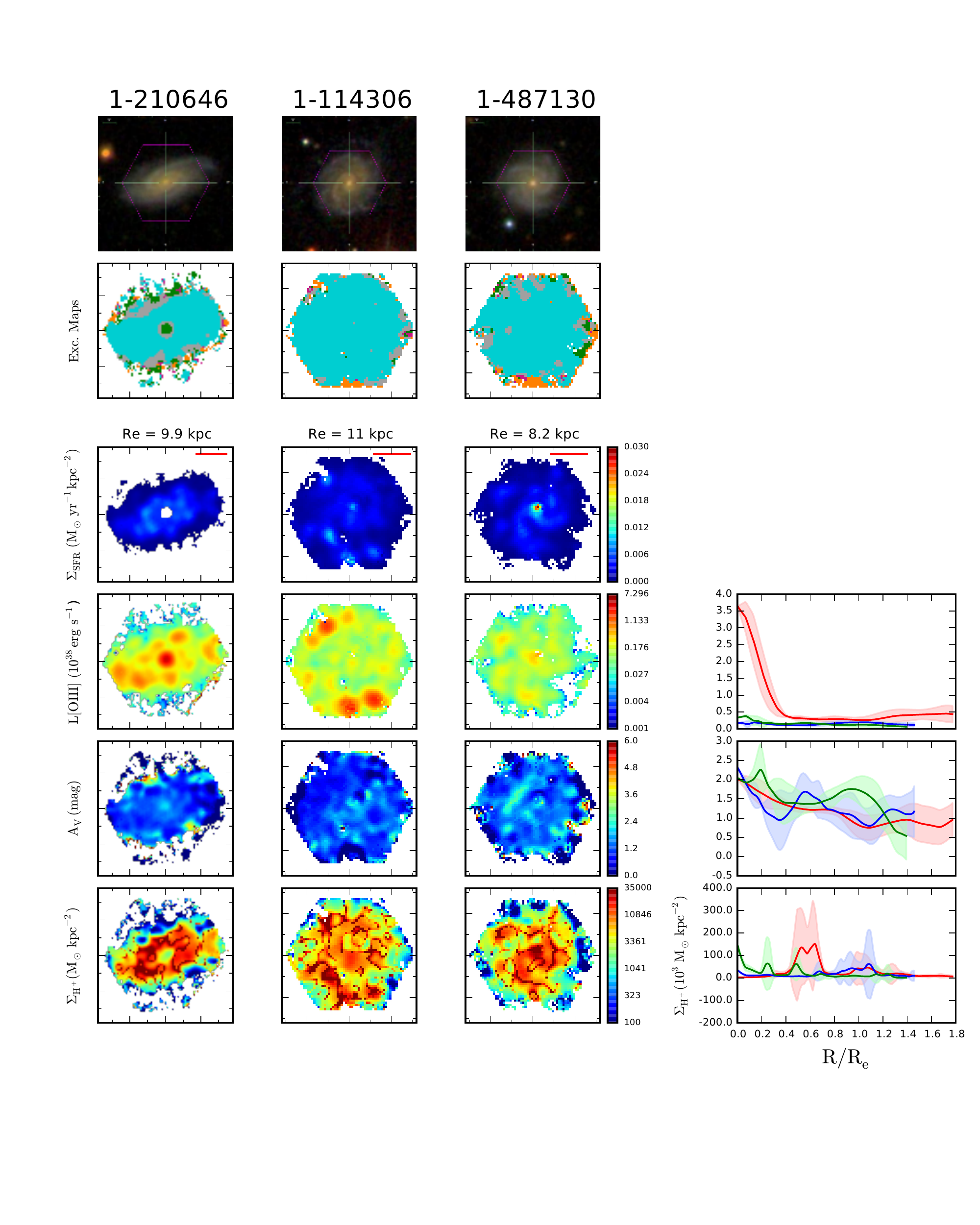}
   \vspace*{-30mm}
   \caption{late type and  weak-luminosity AGN
   }
\end{figure*}

\begin{figure*}
   \includegraphics[width=1.9\columnwidth]{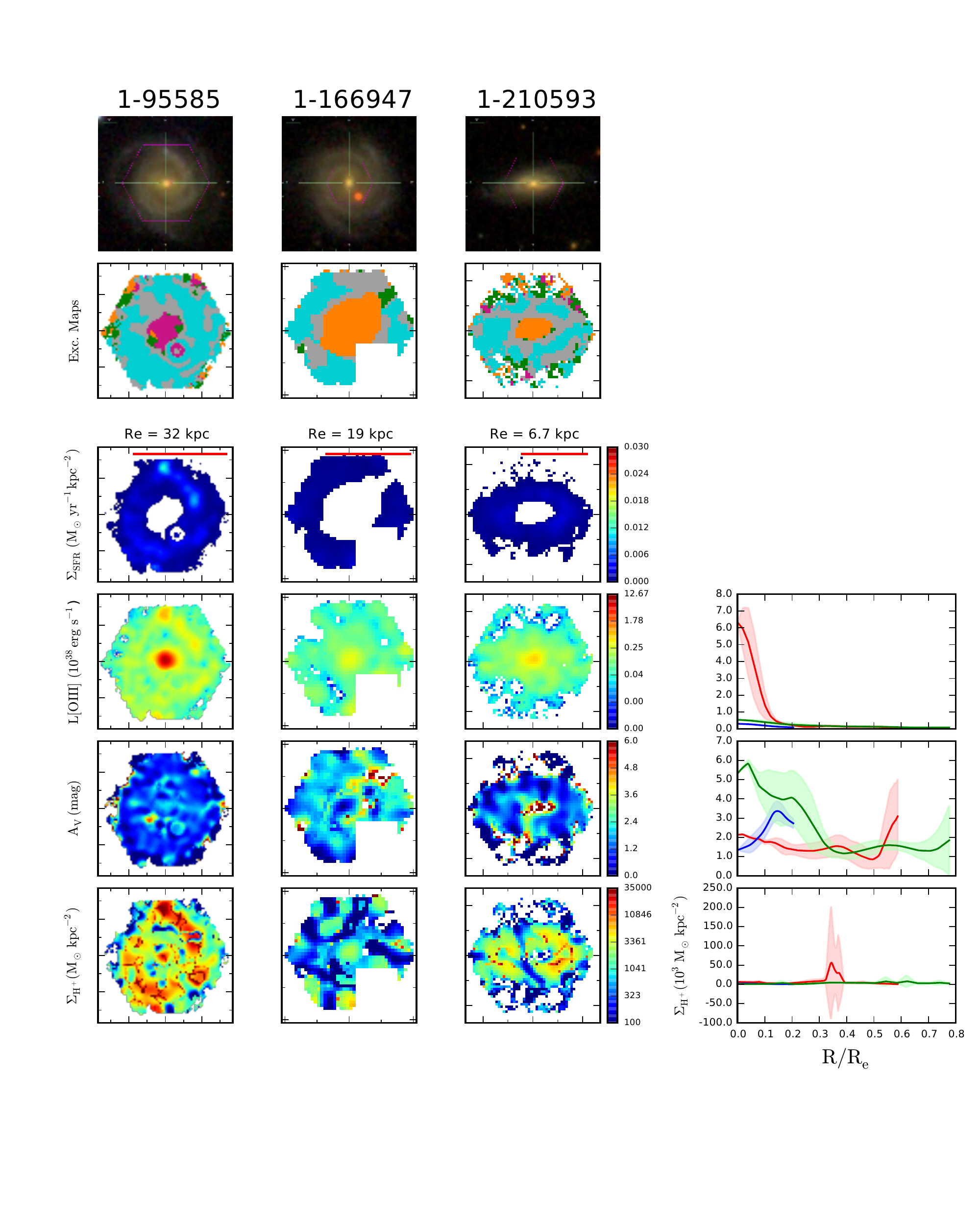}
   \caption{late type and  weak-luminosity AGN
   }
\end{figure*}

\begin{figure*}
   \includegraphics[width=2.1\columnwidth]{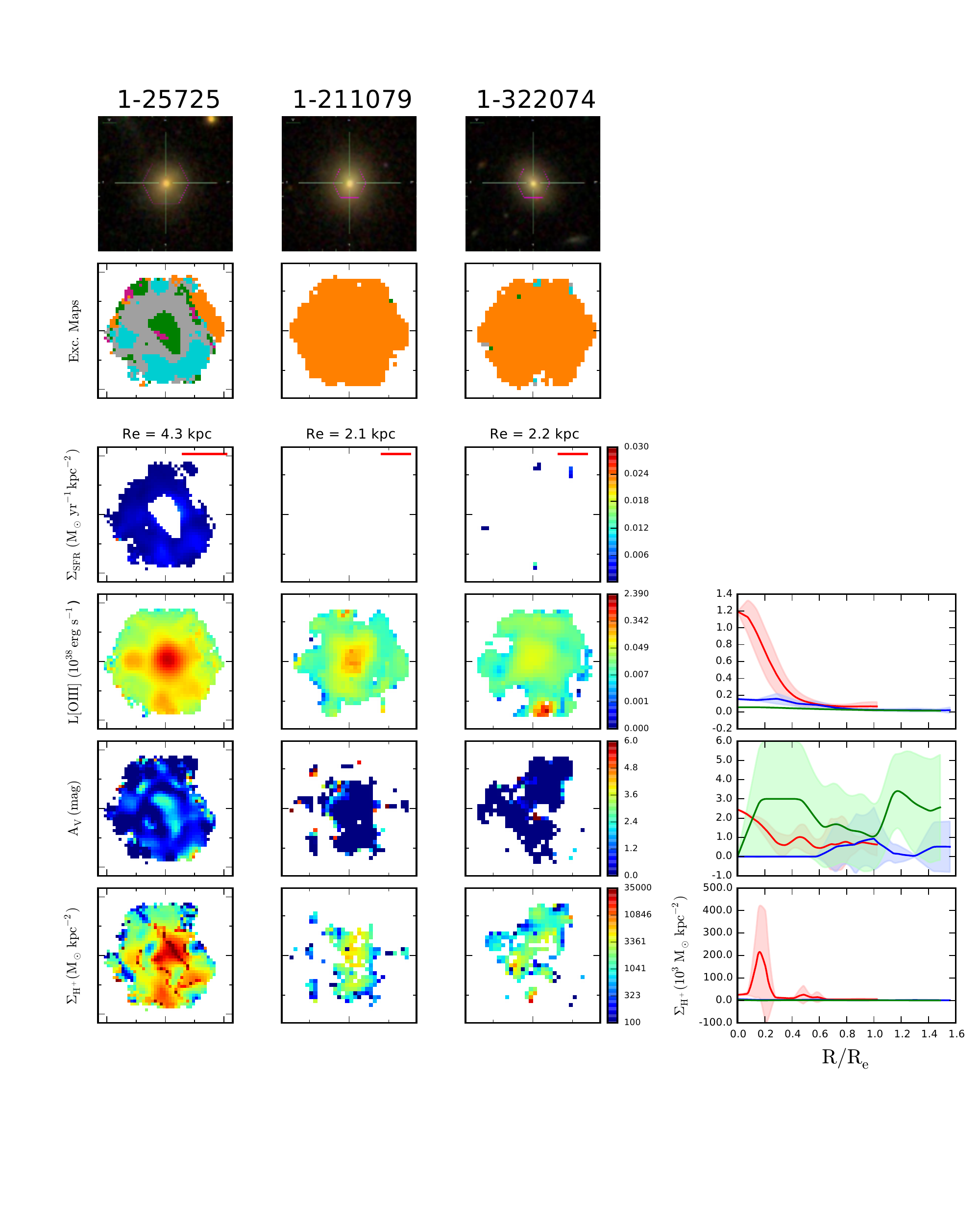}
   \vspace*{-30mm}
   \caption{early-type and weak-luminosity AGN 
   }
\end{figure*}

\begin{figure*}
   \includegraphics[width=2.1\columnwidth]{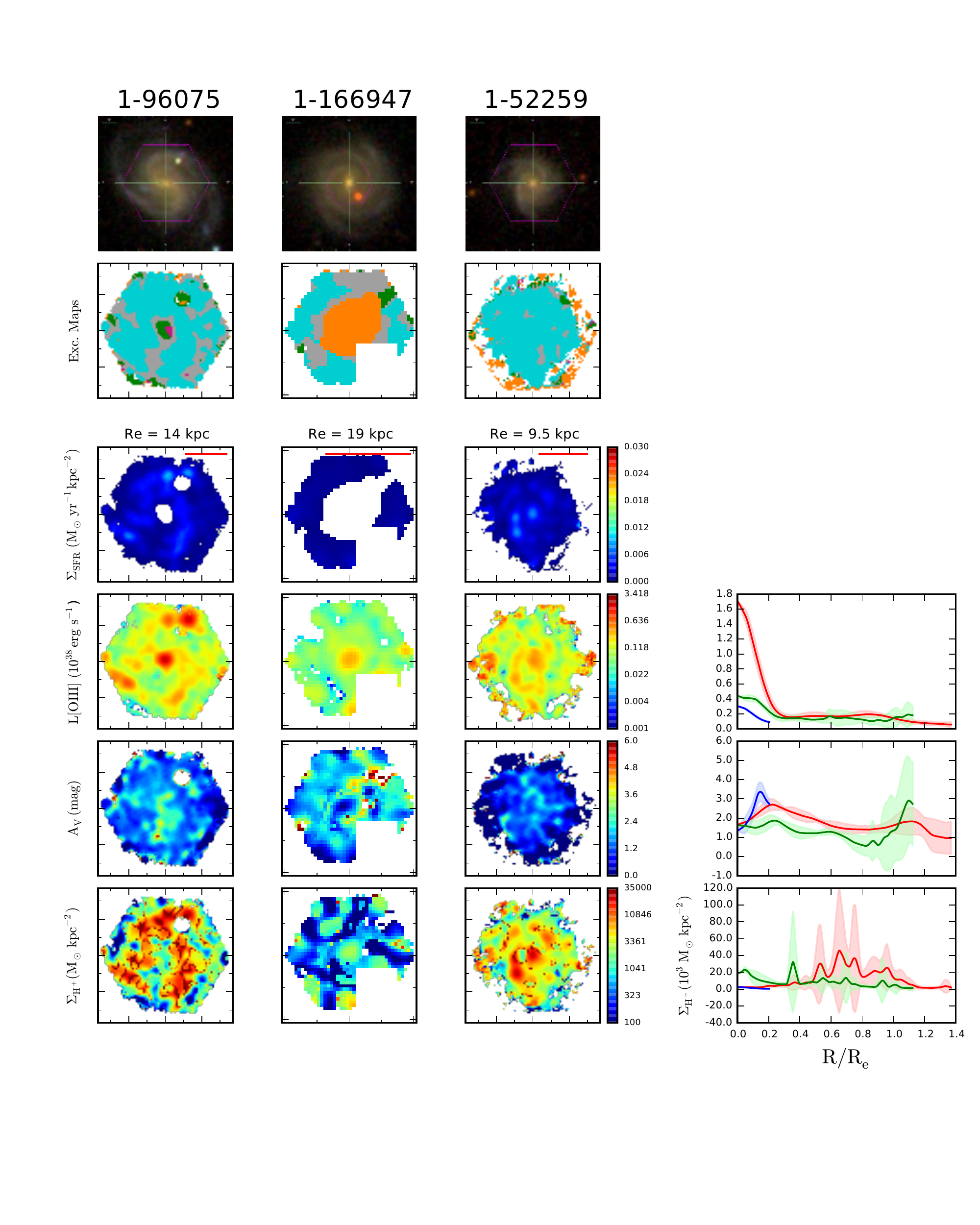}
   \vspace*{-30mm}
   \caption{late type and  weak-luminosity AGN     
   }
\end{figure*}

\clearpage

\begin{figure*}
   \includegraphics[width=2.1\columnwidth]{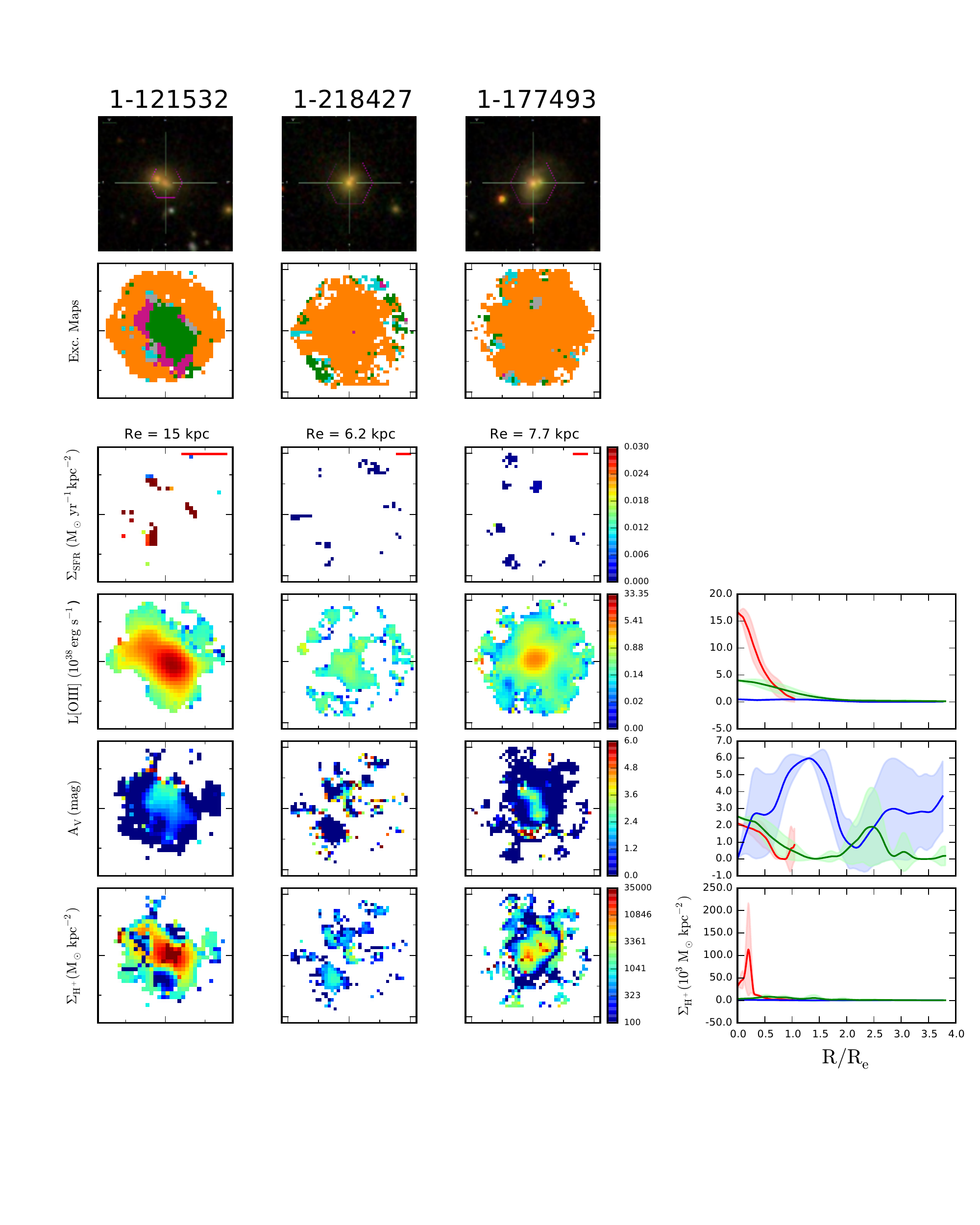}
   \vspace*{-30mm}
   \caption{early-type and strong-luminosity AGN      
  }
\end{figure*}

\begin{figure*}
   \includegraphics[width=2.1\columnwidth]{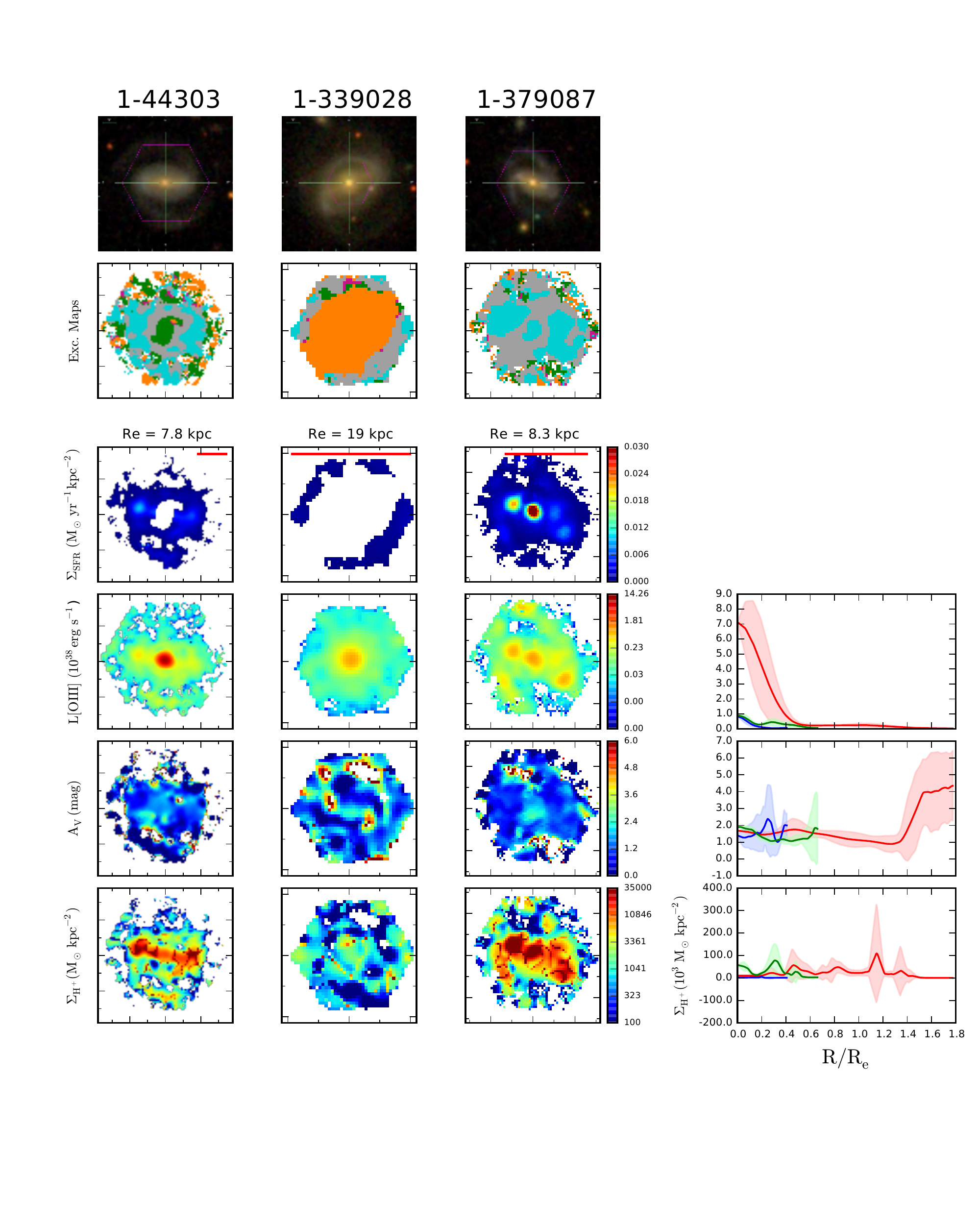}
   \vspace*{-30mm}
   \caption{late-type and strong-luminosity AGN      
   }
\end{figure*}

\begin{figure*}
   \includegraphics[width=2.1\columnwidth]{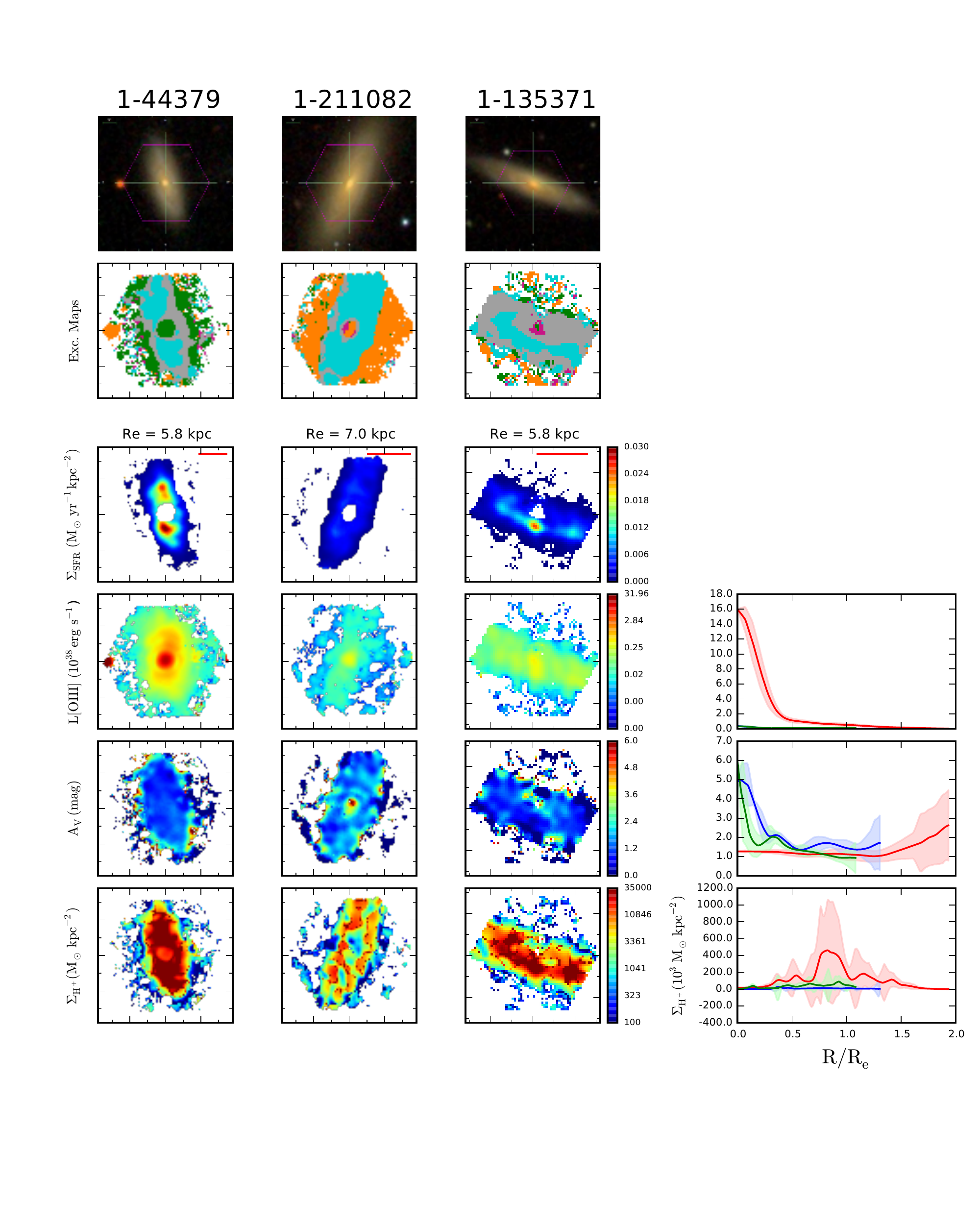}
   \vspace*{-30mm}
   \caption{late-type and strong-luminosity AGN
   }
\end{figure*}

\begin{figure*}
   \includegraphics[width=2.1\columnwidth]{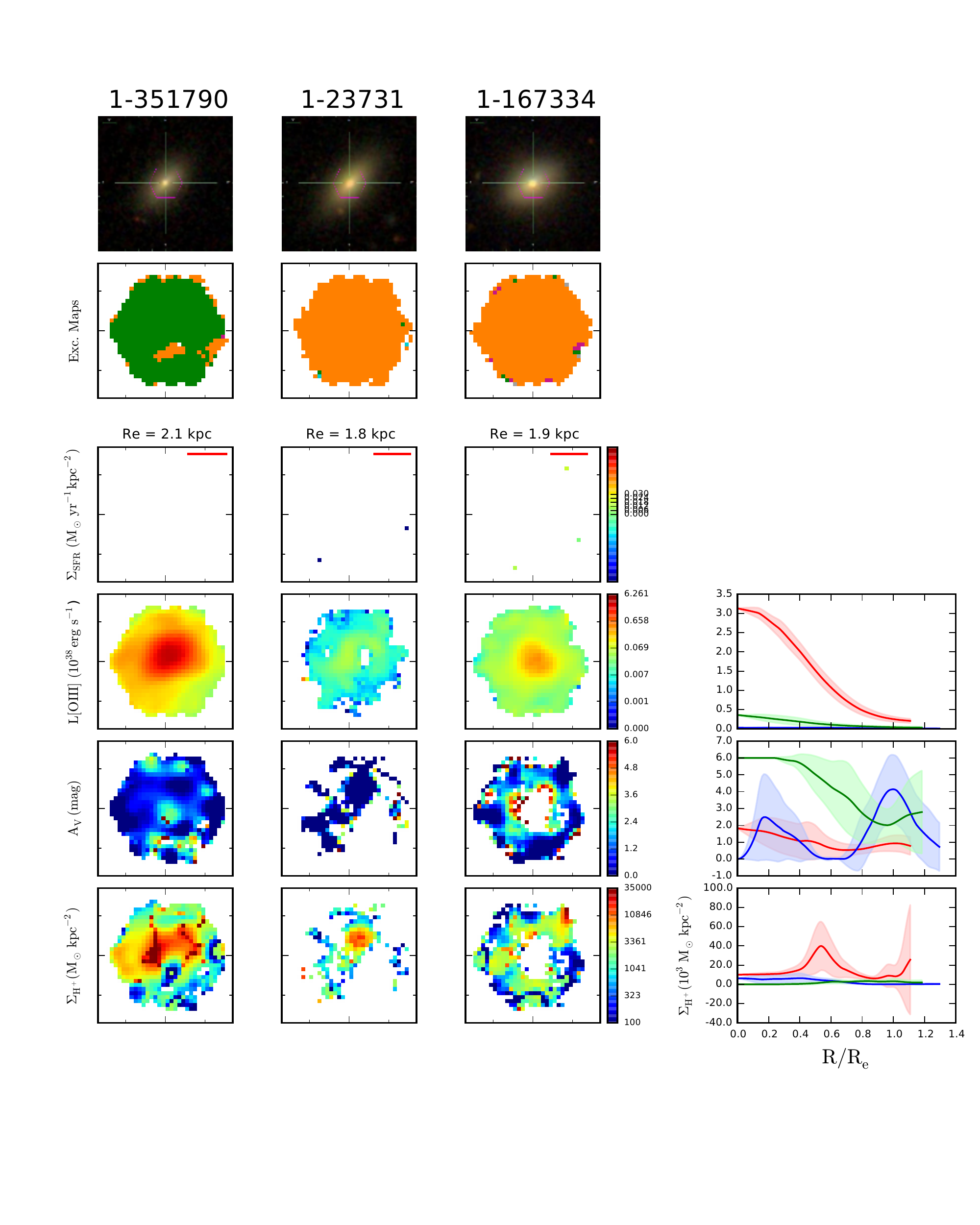}
   \vspace*{-30mm}
   \caption{early-type and weak-luminosity AGN 
   }
\end{figure*}

\begin{figure*}
   \includegraphics[width=2.1\columnwidth]{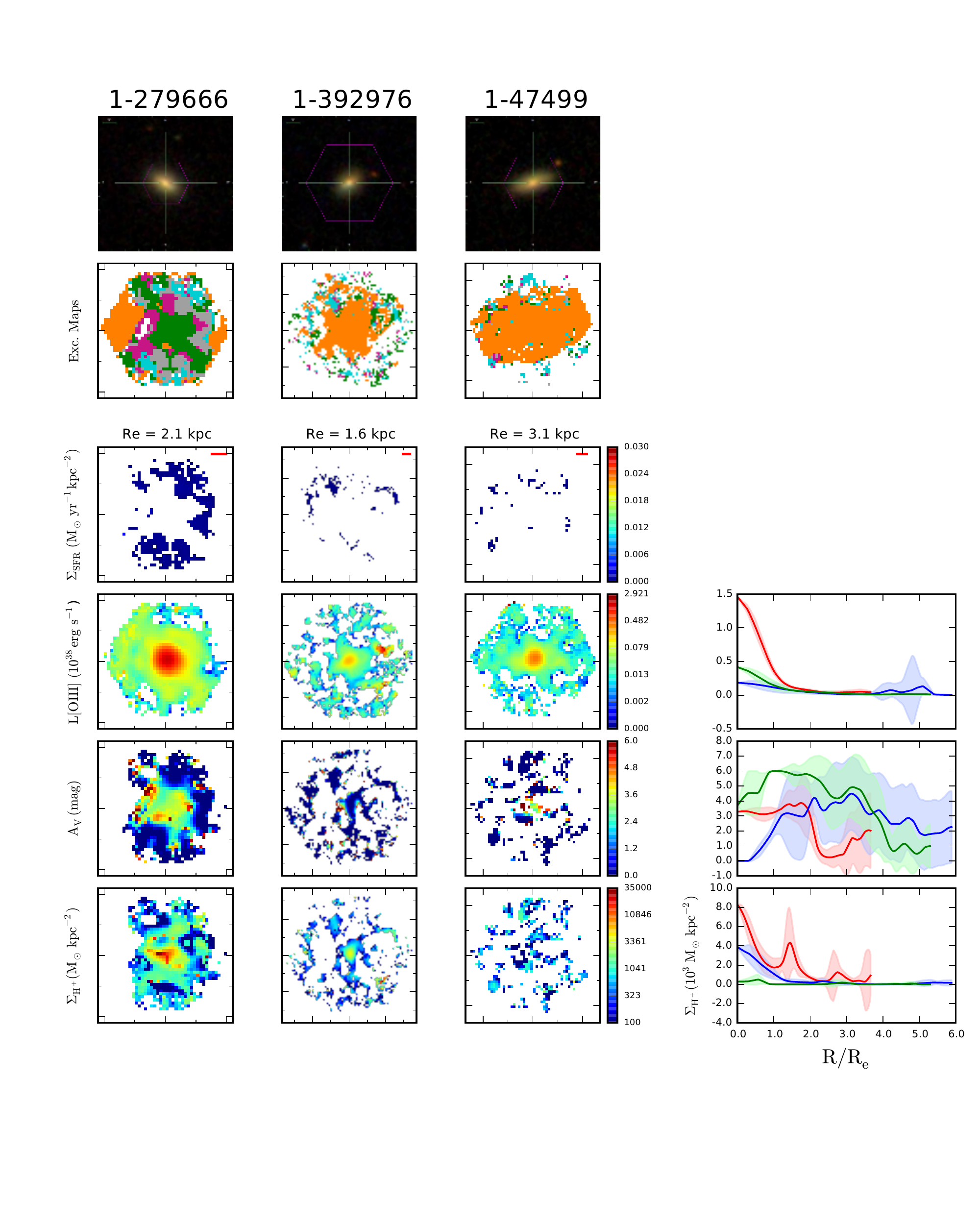}
   \vspace*{-30mm}
   \caption{early-type and weak-luminosity AGN     
   }
\end{figure*}

\begin{figure*}
   \includegraphics[width=2.1\columnwidth]{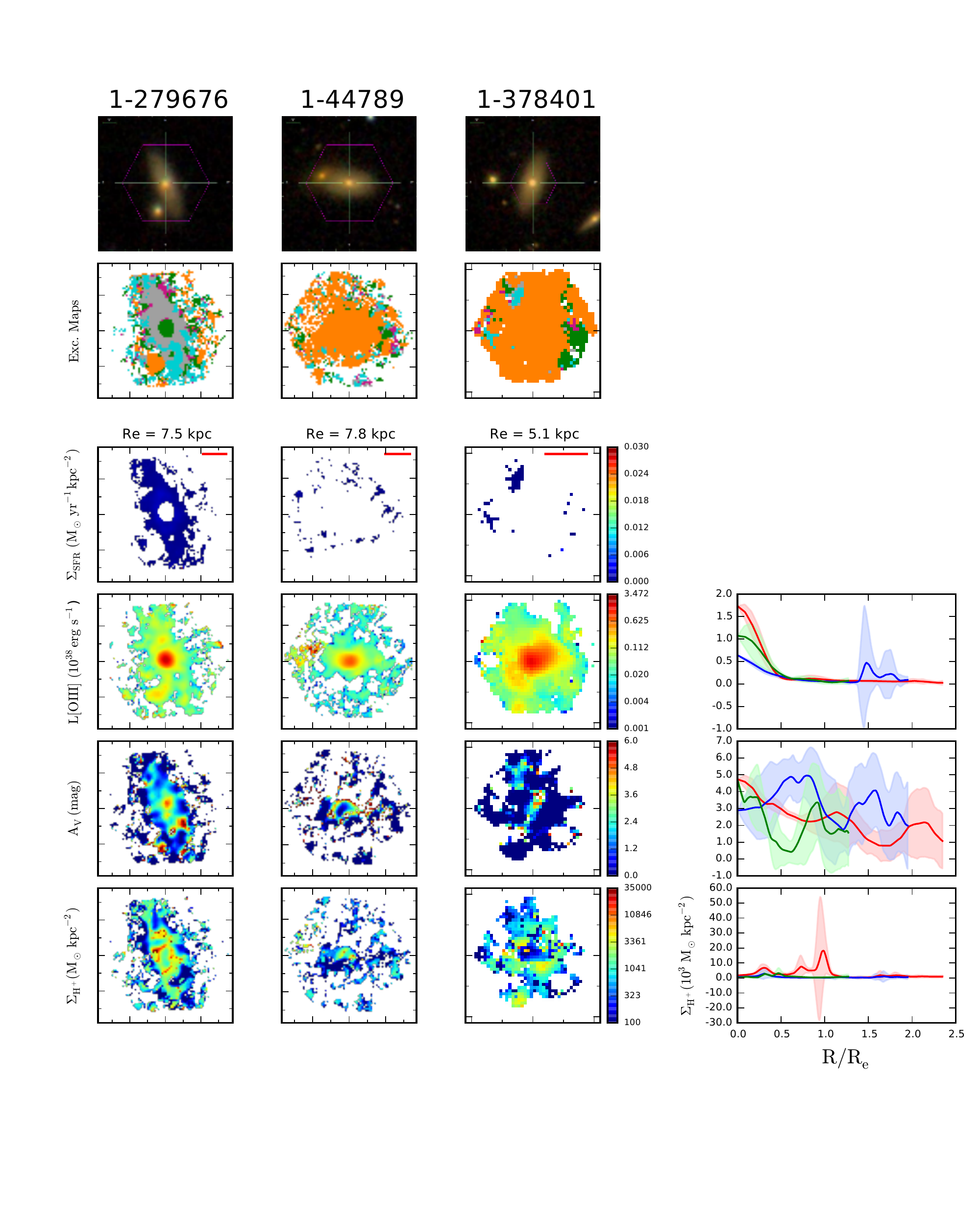}
   \vspace*{-30mm}
   \caption{early-type and weak-luminosity AGN       
   }
\end{figure*}

\begin{figure*}
   \includegraphics[width=2.1\columnwidth]{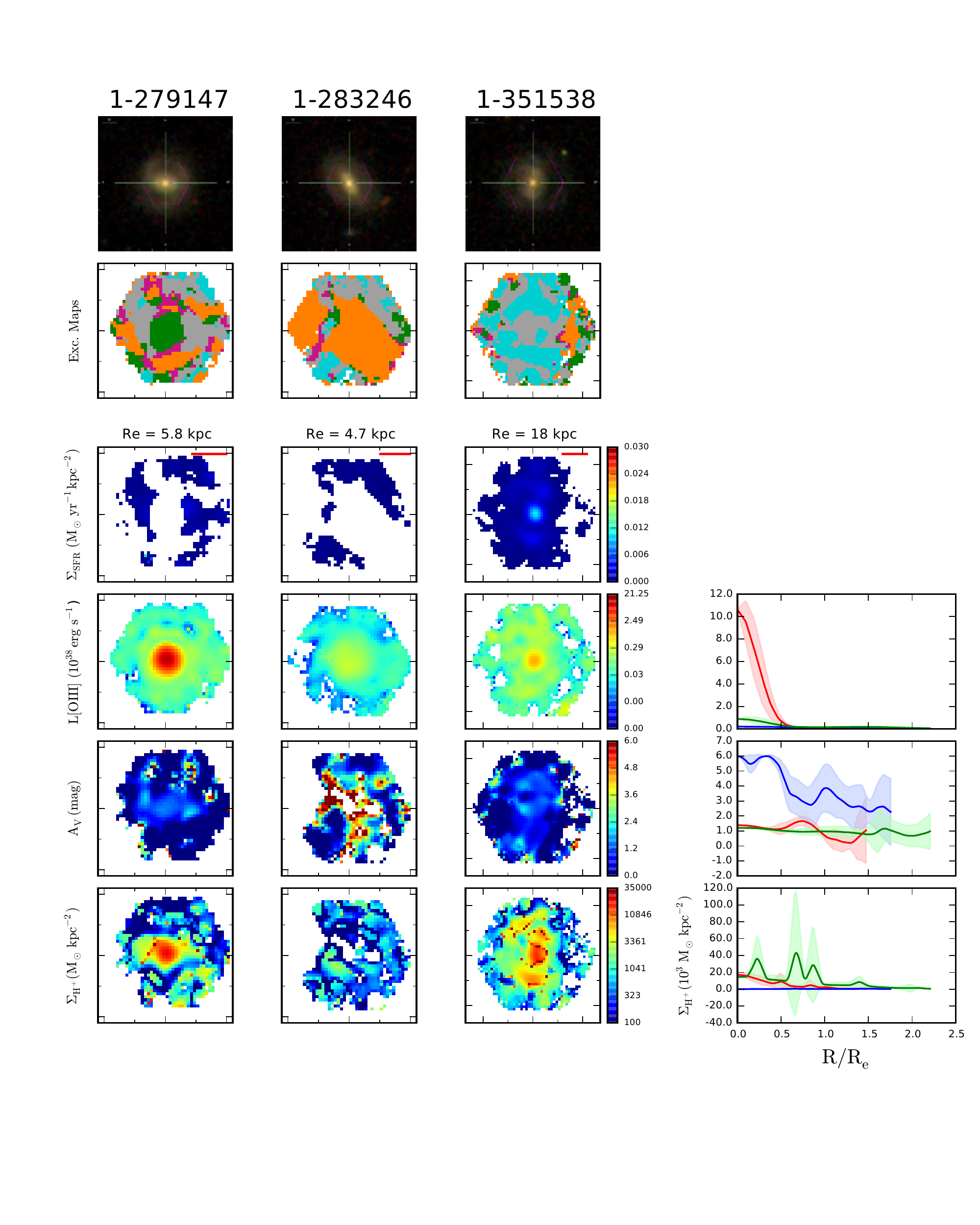}
   \vspace*{-30mm}
   \caption{late-type and strong-luminosity AGN
   }
\end{figure*}

\begin{figure*}
   \includegraphics[width=2.1\columnwidth]{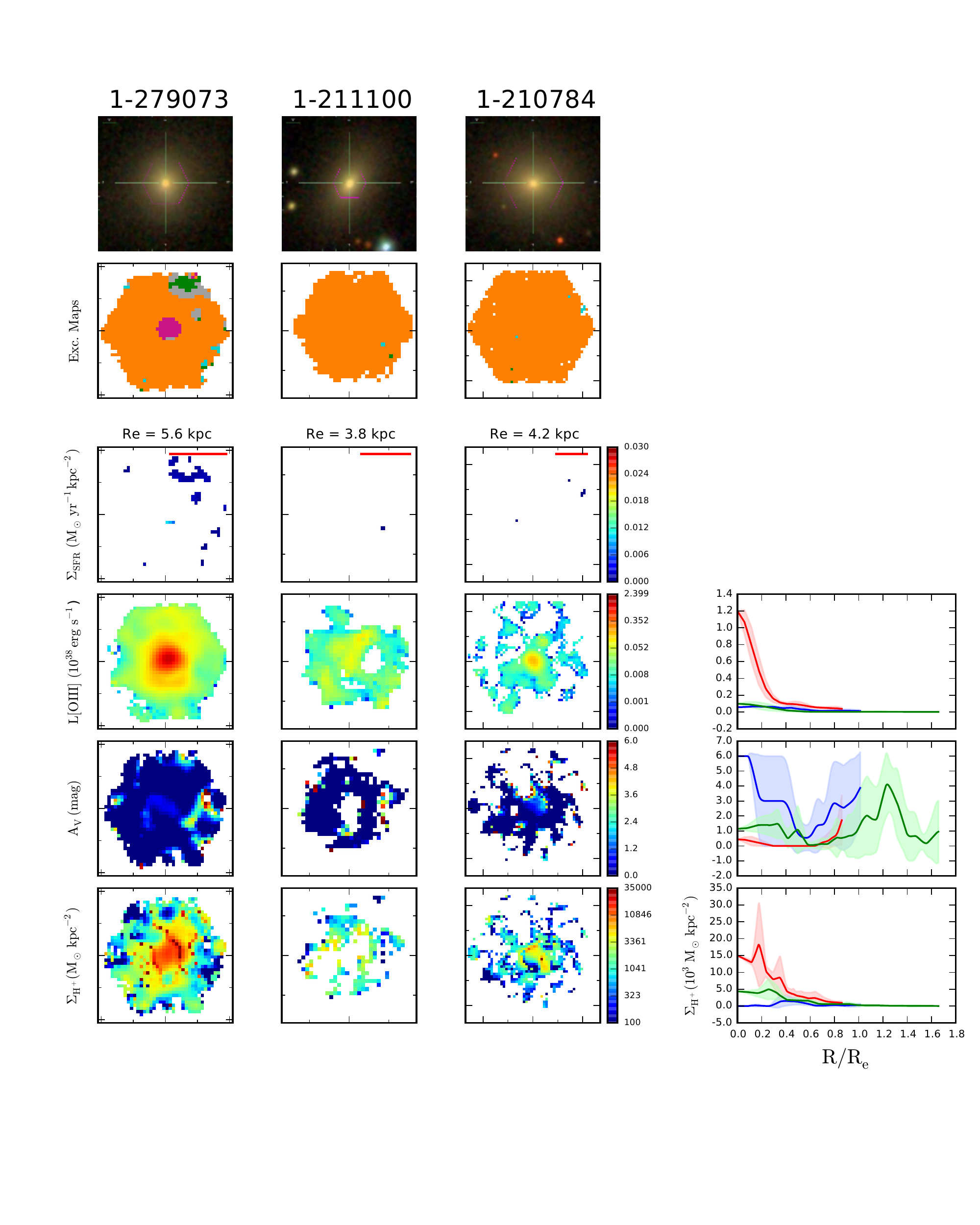}
   \vspace*{-30mm}
   \caption{early-type and weak-luminosity AGN 
   }
\end{figure*}

\begin{figure*}
   \includegraphics[width=2.1\columnwidth]{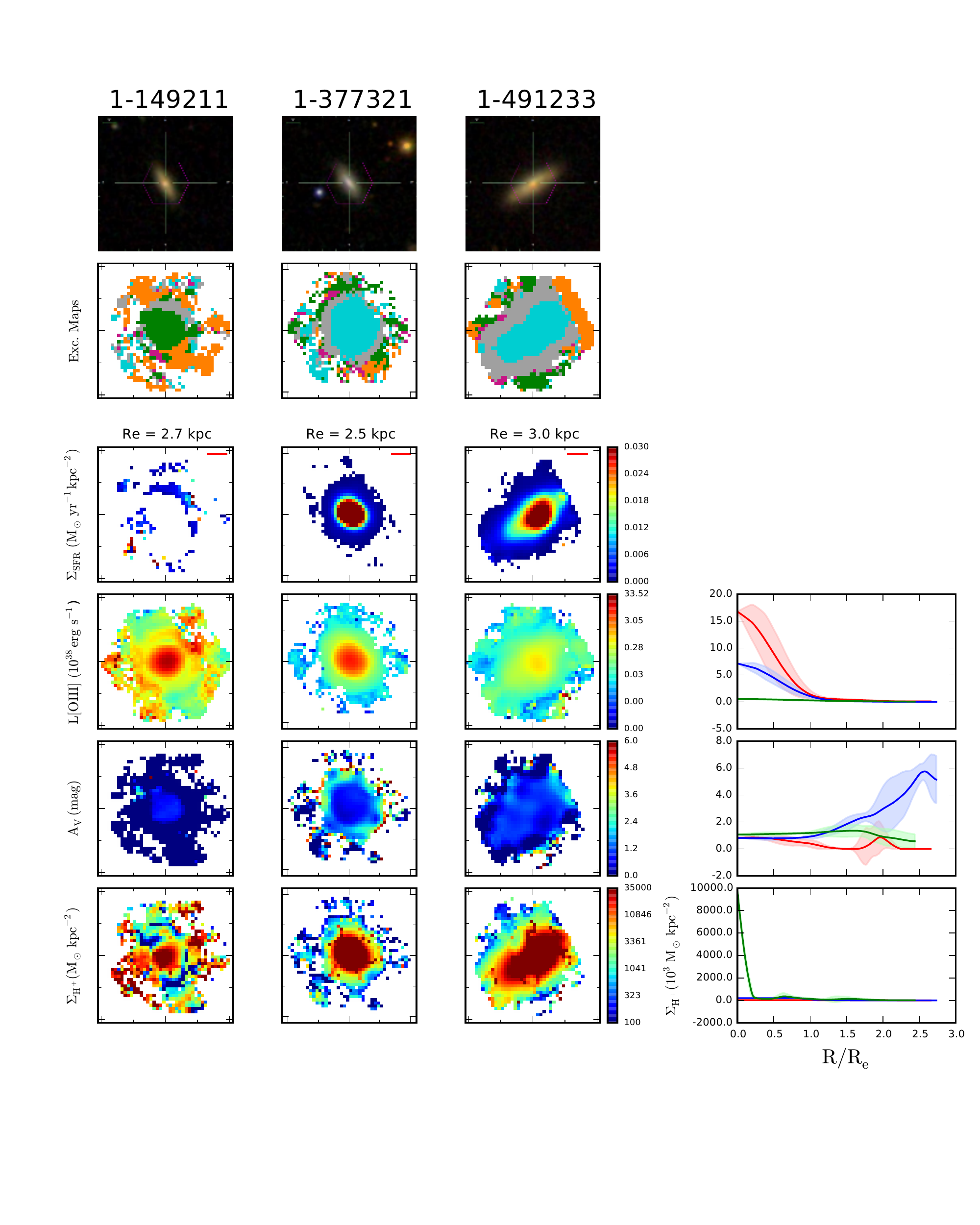}
   \vspace*{-30mm}
   \caption{late-type and strong-luminosity AGN     
   }
\end{figure*}

\begin{figure*}
   \includegraphics[width=2.1\columnwidth]{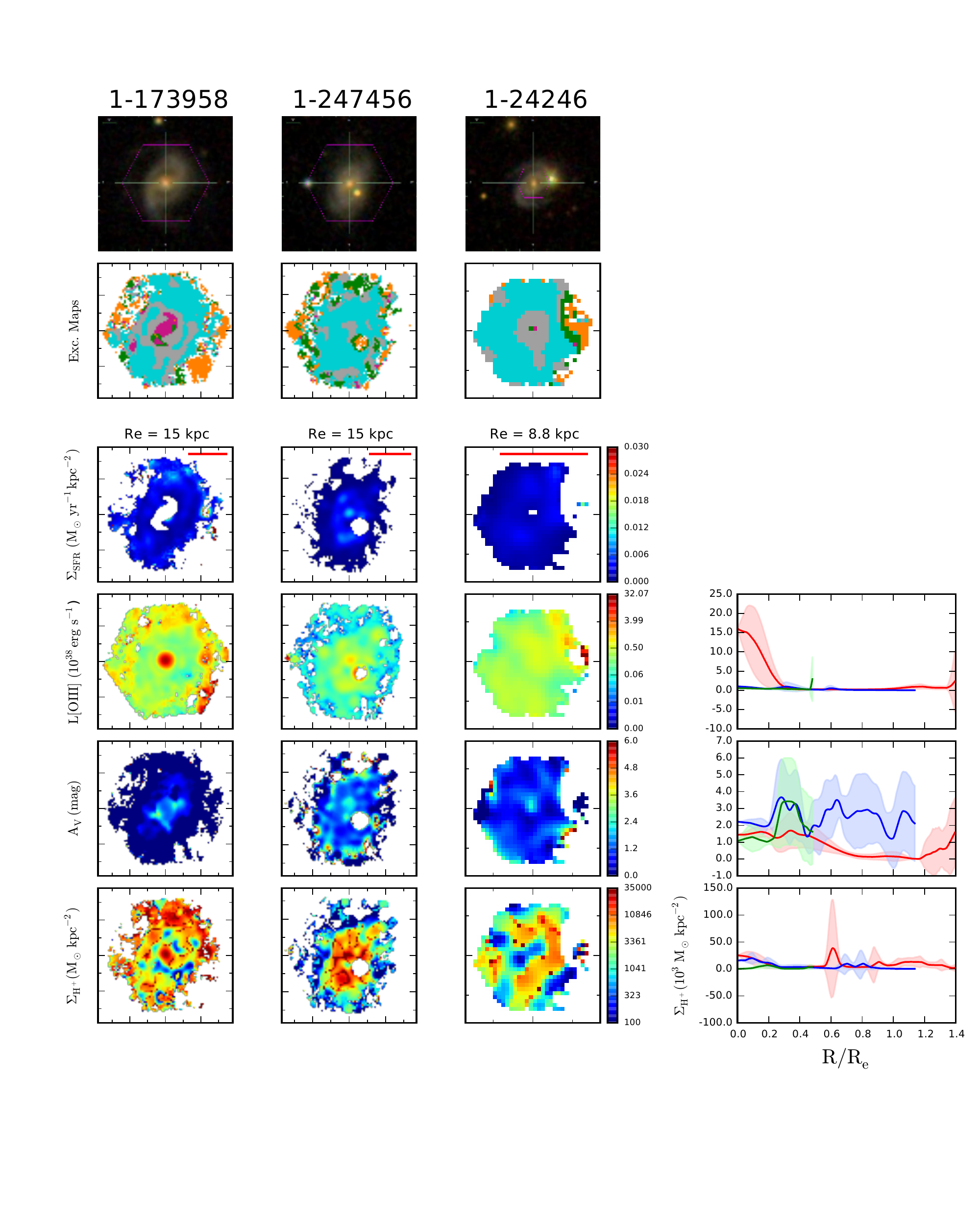}
   \vspace*{-30mm}
   \caption{late-type and strong-luminosity AGN
   }
\end{figure*}

\begin{figure*}
   \includegraphics[width=2.1\columnwidth]{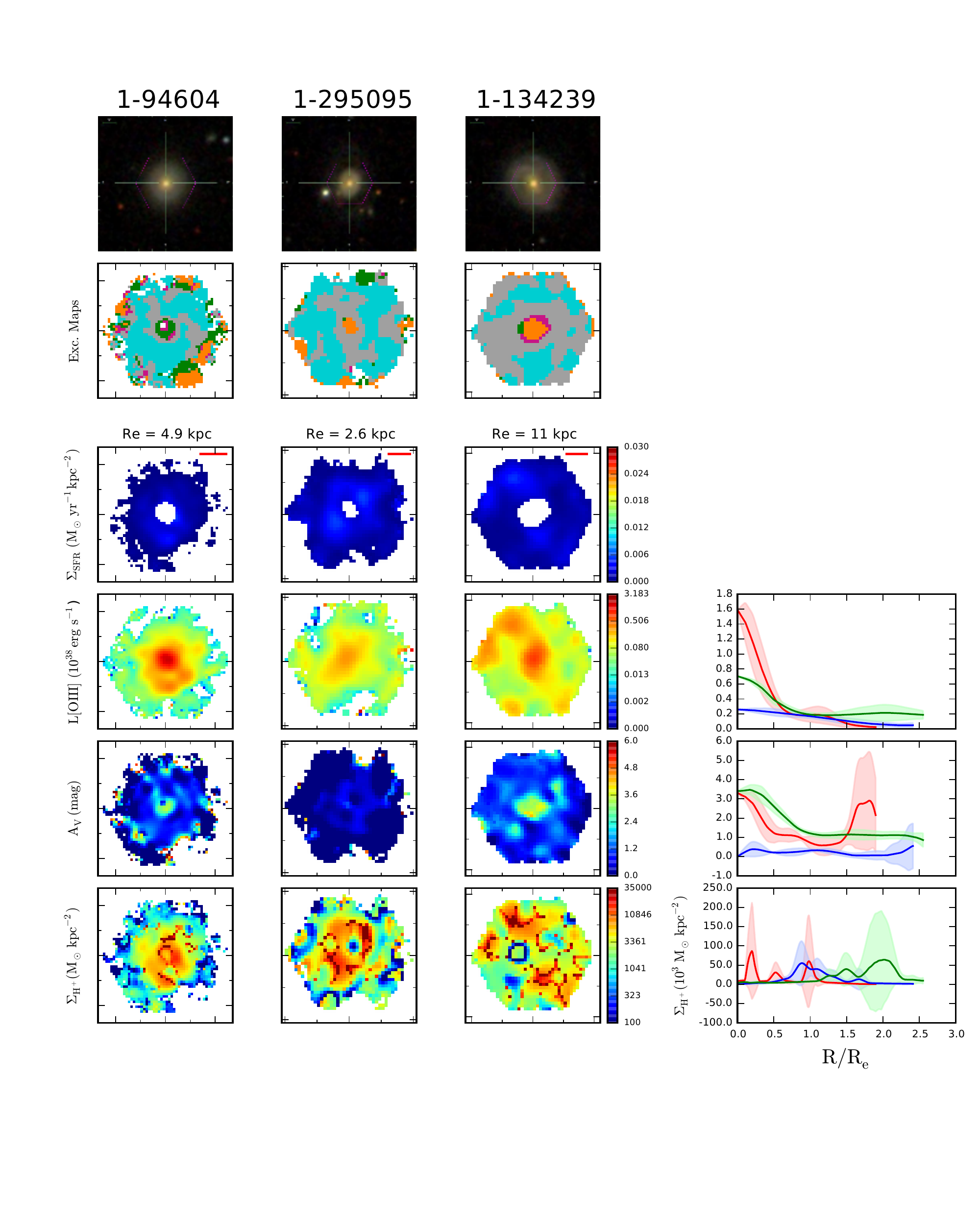}
   \vspace*{-30mm}
   \caption{late type and  weak-luminosity AGN     
   }
\end{figure*}

\begin{figure*}
   \includegraphics[width=2.1\columnwidth]{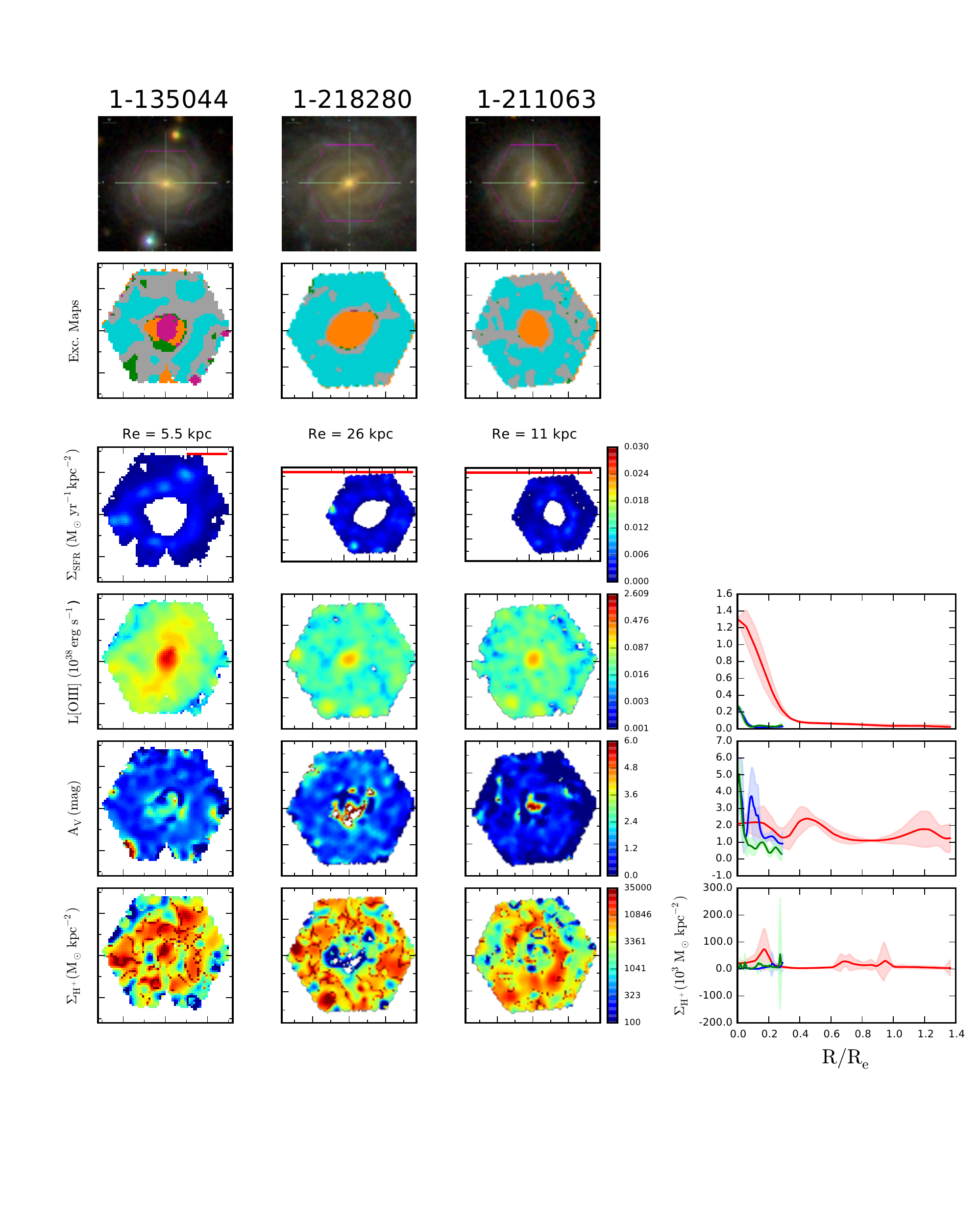}
   \vspace*{-30mm}
   \caption{late type and  weak-luminosity AGN    
   }
\end{figure*}

\begin{figure*}
   \includegraphics[width=2.1\columnwidth]{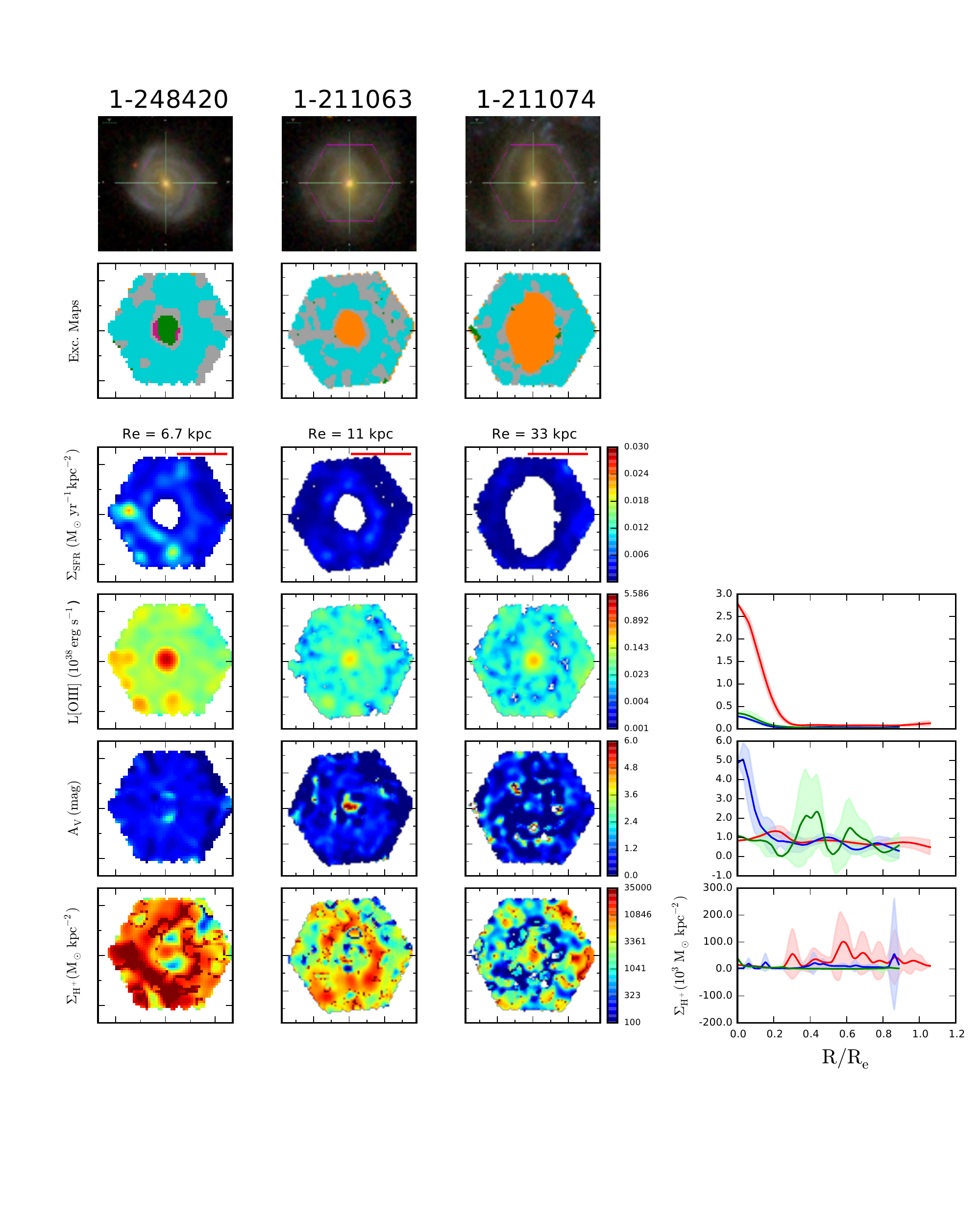}
   \vspace*{-30mm}
   \caption{late type and  weak-luminosity AGN
   }
\end{figure*}

\begin{figure*}
   \includegraphics[width=2.1\columnwidth]{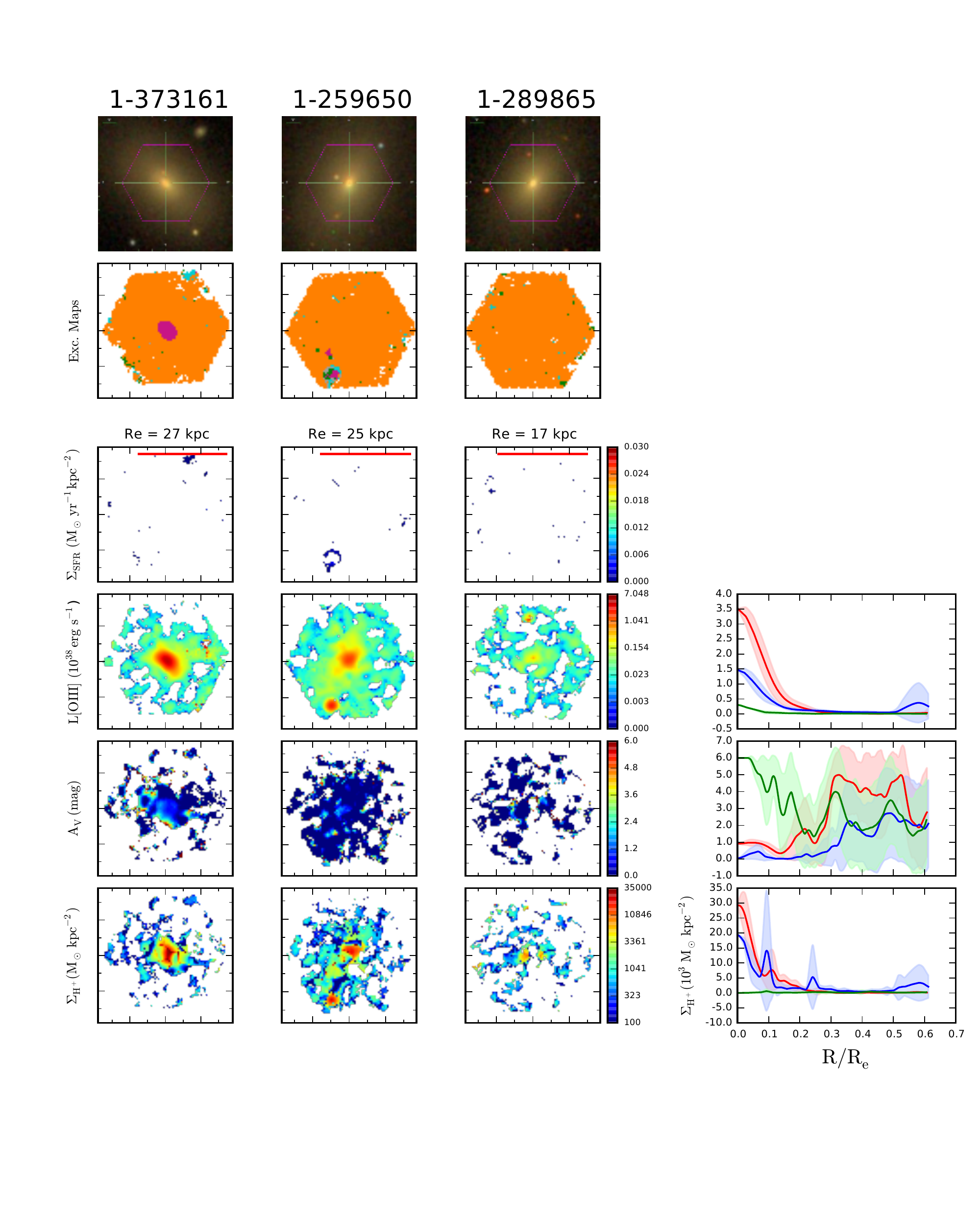}
   \vspace*{-30mm}
   \caption{early-type and weak-luminosity AGN 
   }
\end{figure*}

\begin{figure*}
   \includegraphics[width=2.1\columnwidth]{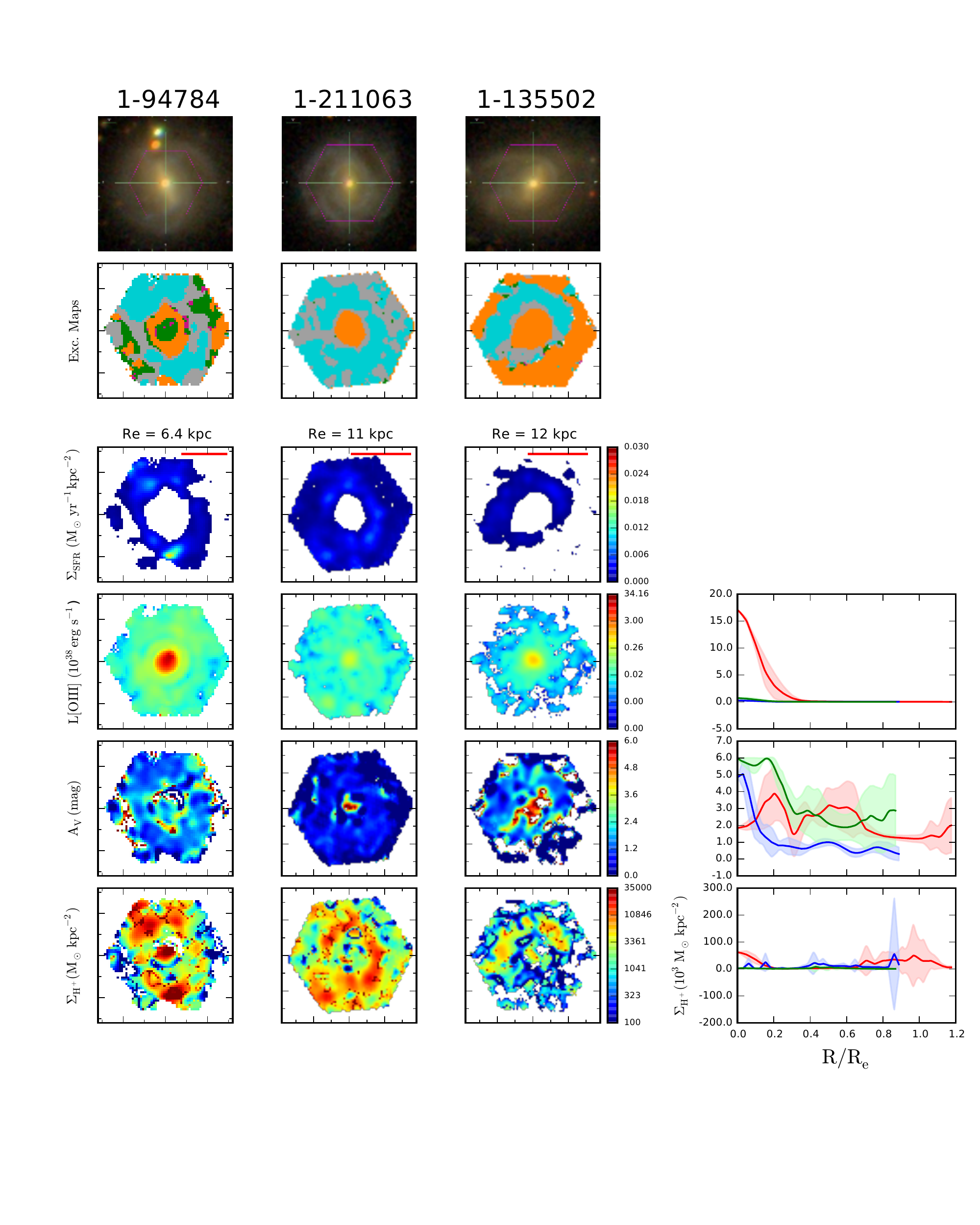}
   \vspace*{-30mm}
   \caption{late-type and strong-luminosity AGN    
   }
\end{figure*}

\begin{figure*}
   \includegraphics[width=2.1\columnwidth]{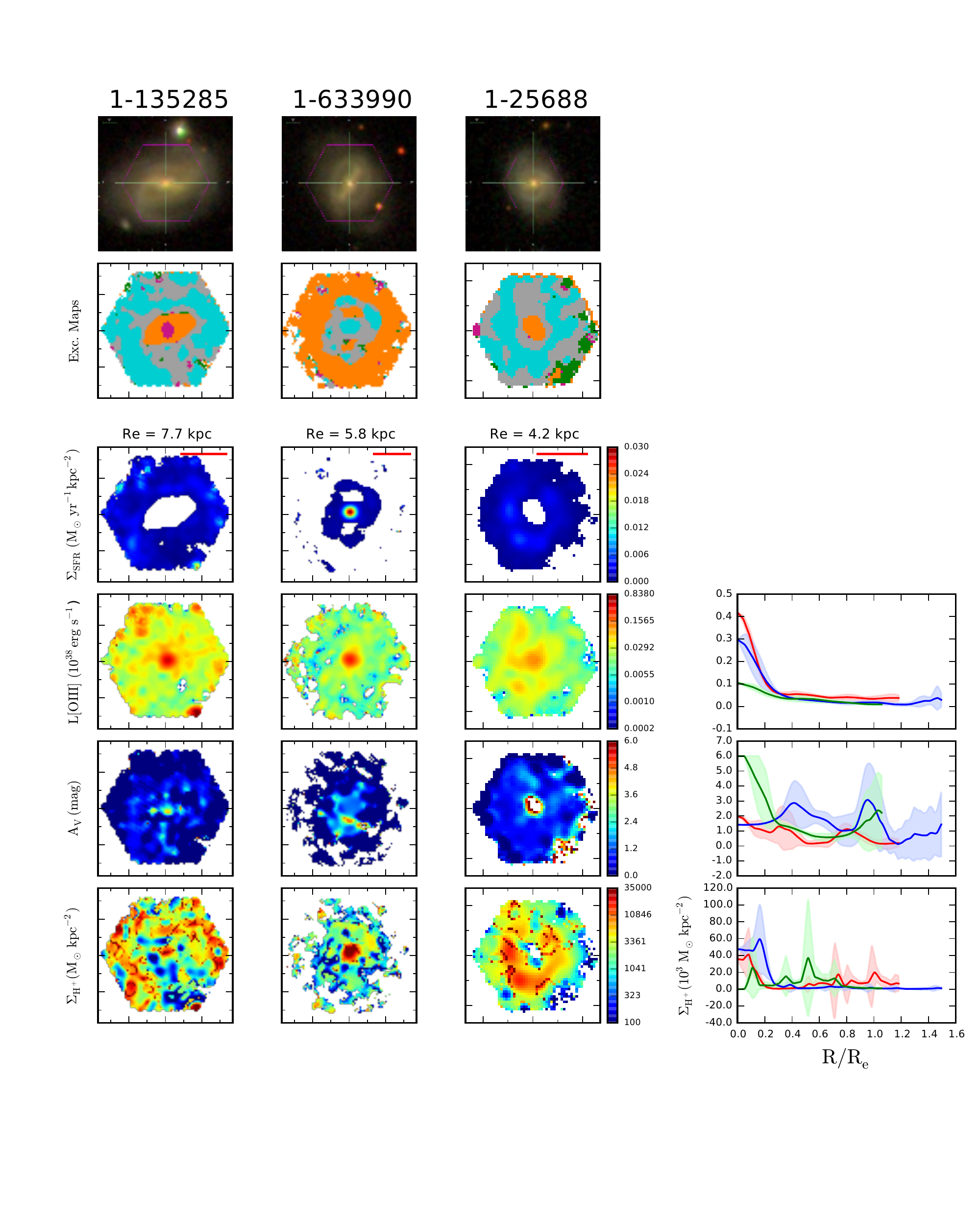}
   \vspace*{-30mm}
   \caption{late type and  weak-luminosity AGN    
   }
\end{figure*}

\bsp	
\label{lastpage}
\end{document}